\documentclass[10pt,journal,onecolumn,compsoc]{IEEEtran}

\ifCLASSOPTIONcompsoc

  \usepackage[nocompress]{cite}
\else

  \usepackage{cite}
\fi

\usepackage{graphicx}
\usepackage{amssymb,amsmath}
\usepackage{color,url}
\usepackage[caption=false]{subfig}

\usepackage[table,xcdraw]{xcolor}
\usepackage{amsmath,amssymb}
\usepackage{color,tikz,verbatim,url}
\usetikzlibrary{positioning}
\usepackage{booktabs}
\usepackage{array}
\usepackage{url}

\usepackage[ruled]{algorithm}
\usepackage[noend]{algpseudocode}

\makeatletter
\algnewcommand{\LineComment}[1]{\Statex \hskip\ALG@thistlm \(\triangleright\) #1}
\makeatother

\newcommand{\RR}{\mathbb{R}}
\newcommand{\Nc}{\mathcal{N}}

\newcolumntype{L}{>{$}l<{$}}

\newtheorem{lem}{Lemma}
\newtheorem{cor}{Corollary}
\newtheorem{prop}{Proposition}
\newtheorem{defn}{Definition}

\newcommand{\AD}[1]{\textcolor{black}{#1}}

\definecolor{SlightGray}{gray}{0.95}
\definecolor{LightGray}{gray}{0.9}
\definecolor{AnotherGray}{gray}{0.8}
\definecolor{ashgrey}{rgb}{0.96,0.96, 0.9}
\definecolor{ashgrey2}{rgb}{0.72, 0.73, 0.69}
\definecolor{beaublue}{rgb}{0.94, 0.97, 0.96}
\definecolor{desertsand}{rgb}{0.95, 0.95, 0.96}

\begin{document}

\title{A Modular Framework for \\ Centrality and Clustering in Complex Networks}

\author{Frederique Oggier,
        Silivanxay Phetsouvanh,
        and~Anwitaman Datta
\IEEEcompsocitemizethanks{\IEEEcompsocthanksitem F. Oggier is with Division of Mathematical Sciences, Nanyang Technological University, Singapore.\protect\\
E-mail: frederique@ntu.edu.sg
\IEEEcompsocthanksitem S. Phetsouvanh and A. Datta was/is with School of Computer Science and Engineering, Nanyang Technological University, Singapore.\protect\\ E-mail: anwitaman@ntu.edu.sg }
}

\IEEEtitleabstractindextext{%
\begin{abstract}

{\textbf{Abstract:} The structure of many complex networks includes edge directionality and weights on top of their topology. Network analysis that can seamlessly consider combination of these properties are desirable. In this paper, we study two important such network analysis techniques, namely, centrality and clustering. An information-flow based model is adopted for clustering, which itself builds upon an information theoretic measure for computing centrality. Our principal contributions include a generalized model of Markov entropic centrality with the flexibility to tune the importance of node degrees, edge weights and directions, with  a closed-form asymptotic analysis. It leads to a novel two-stage graph clustering algorithm. The centrality analysis helps reason about the suitability of our approach to cluster a given graph, and determine `query' nodes, around which to explore local community structures, leading to an agglomerative clustering mechanism. The entropic centrality computations are amortized by our clustering algorithm, making it computationally efficient: compared to prior approaches using Markov entropic centrality for clustering, our experiments demonstrate multiple orders of magnitude of speed-up. Our clustering algorithm naturally inherits the flexibility to accommodate edge directionality, as well as different interpretations and interplay between edge weights and node degrees. Overall, this paper thus not only makes significant theoretical and conceptual contributions, but also translates the findings into artifacts of practical relevance, yielding new, effective and scalable centrality computations and graph clustering algorithms, whose efficacy has been validated through extensive benchmarking experiments. }
{}
\end{abstract}

\begin{IEEEkeywords}
Directed Weighted Graphs, Entropy, Centrality, Graph Clustering, Random Walkers.
\end{IEEEkeywords}}

\maketitle

\IEEEdisplaynontitleabstractindextext

\IEEEpeerreviewmaketitle

\section{Introduction}
\AD{Complex networks \cite{Estrada} are a form of structured data that model interactions among individuals in diverse real-life applications. New instances of complex networks may be studied with existing techniques, or may motivate the introduction of new approaches. An example of a relatively recent and fast evolving complex network is the Bitcoin network, where individuals or nodes are Bitcoin addresses, and edges represent transactions from one address to another; they are thus directed and weighted. There are many reasons to study the Bitcoin network and other cryptocurrency networks as complex networks: understanding their topology and their evolution over time in turn offers economic or societal insights. Centrality analysis and clustering (also known as community detection) are two popular techniques that are used to study network structures and could be applied.}

\begin{figure}
\begin{center}
\includegraphics[scale=.4]{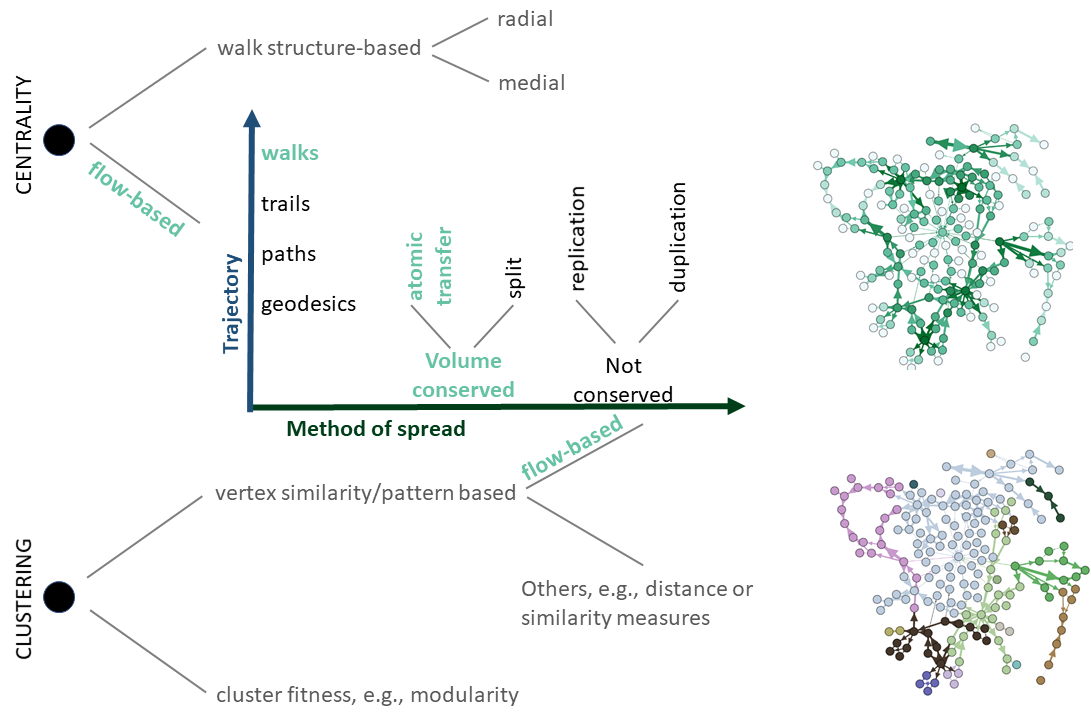}
\caption{\label{fig:summary}
\AD{Positioning our work in the current centrality and clustering landscape.}
}
\end{center}
\end{figure}

\AD{{\bf Centrality:} The identification of key individuals in a complex network is often done using centrality, a family of measures that captures the importance of nodes in a graph. Centrality measures are often categorized (see Figure \ref{fig:summary}) by either flow or walk structure. (i) {\em Flow-based} measures come with a 2-dimensional typology \cite{Borgati}: the trajectory of the flow (categorized as - geodesics, paths where neither nodes nor edges are repeated, trails where nodes are repeated but edges are not, and walks, where both nodes and edges can be repeated) and the method of spread (which includes - transfer of an atomic flow, duplication or replication of the flow or a split of a volume conserved flow). (ii) {\em Walk structure-based} measures are partitioned into radial centralities (counting walks which start/end from a given vertex, e.g. the degree and the eigenvalue centralities) and medial centralities (counting walks which pass through a given vertex, e.g. betweenness centrality).}

\AD{Our focus is on flow-based measures, motivated by the study of phenomena such as the movement of cryptocurrencies across a network from one sender to a receiver, where the volume of the flow is conserved.} \AD{In \cite{T7}, it was argued that the existing flow-based centrality measures back then left a void in the typology because they did not capture well the idea of path-transfer flow. Accordingly entropy, as a measure of uncertainty of dispersal, was advanced as a new centrality measure. The idea was modified in \cite{NRK}, where the path restriction was relaxed from the original idea of \cite{T7}, allowing for the study of the flow as a random walker using a Markov model.}

\AD{In order to distinguish senders from receivers, and consider equivalent or distinct (as the need of a specific analysis might be) the users involved in transactions involving large amounts from users carrying out large number of transactions of small amounts, and other such nuances, it is desirable to have the flexibility to consider the directionality and weights of the edges, which could carry their own semantics. While a notion of weighted degree centrality has been proposed \cite{OAS}, this paper proposes a notion of flow-based centrality which is seamlessly adjustable to undirected, directed and/or weighted graphs.} \AD{Akin to \cite{NRK}, our work uses a random walker starting its walk at any given node, whose centrality is captured by the spread of the walk, measured in terms of entropy, and an absorption probability is used to model the walker stopping at any destination. While \cite{NRK} rendered the usage of entropic centrality more practical than \cite{T7} by introducing the idea of a random walker, it also has limitations both in the theoretical exploration of the concepts as well as the resulting practical implications. These limitations, together with the potential of the ideas themselves, serve as motivations for our current work. Particularly, we provide, among other improvements, closed form expressions to characterize the behaviour of the walker. The specific details are elaborated when we discuss our contributions below. } 

\AD{\textbf{Graph clustering:} A set of nodes that share common or similar features are considered to comprise a cluster in a network. A general understanding of a good cluster is that it is dense (nodes within are well connected), and have few connections with the rest of the network \cite{Schaeffer,MZ}. Beyond that, there is no unique criterion that necessarily captures or reflects the quality of the clusters. Depending on the particular aspects being captured, different approaches may be relevant, and likewise, different outcomes may be expected and desirable.}
\AD{In \cite{Schaeffer}, two broad families of graph clustering methods have been identified (see Figure \ref{fig:summary}) - (i) those based on {\em vertex similarity} in some manner, e.g., distance or similarity measure which may or not take into account structural information (for instance, web-page similarity may consider only content similarity, or may also consider the hyperlinks to determine similarity), adjacency or connectivity based measures; and (ii) those based on {\em cluster fitness} measures, including density and cut-based measures, e.g., modularity. In \cite{MZ}, the former are referred to as {\em pattern based}, since these methods go beyond basic edge density characteristics. They include algorithms relying on random walkers \cite{AL,BNCZ,KSJ}, \AD{where the basic premise is that random walks originating at a given `query' node is more likely to stay confined within a subgraph which forms a local community containing the query node, than to visit nodes outside said community. The probability distribution of a random walker's visit (and absorption) at different nodes in the graph is used as a signal to identify cluster boundaries. }}

\AD{It is observed in \cite{MZ} that while numerous clustering algorithms exist for undirected (weighted) graphs (see \cite{Schaeffer} for a survey of well known clustering techniques for undirected graphs), clustering algorithms for directed graphs are fewer. Most algorithms rely on creating an ``undirected" representation of the directed graph considered, which may result in the loss of semantics and information contained in the directionalities. Examples of notable exceptions where the clustering algorithm is designed specifically for directed graphs are the information flow approaches of \cite{RB}, where clusters are identified as subgraphs where the information flow within is larger compared to outside node groups, and \cite{KSJ}, where an information flow-based cluster is a group of nodes where a random walker (referred as surfer in the work) is more likely to get trapped, rather than moving outside the group. \AD{Tellingly, the other random walker based approaches mentioned above, namely \cite{AL,BNCZ}, because of the additional constraints imposed on the walkers in those algorithms, are only suitable for undirected unweighted graphs.}}

\AD{Since directionality is of importance in many set-ups such as the example of a cryptocurrency network, we will focus on clustering algorithms that can handle both (un)directed and/or (un)weighted graphs, and thus we explore flow circulation based algorithms, which falls within the pattern based (vertex similarity) family of graph clustering algorithms.}
\AD{This choice has a natural coupling with our proposed entropic centrality measure. Under the premise of confinement of flows within a subset of nodes as a way to interpret community structures, one may further consider the flow using any of the many characterization of flows described above in the context of centrality measures, including those supporting the treatment of directed and weighted graphs, each in turn capturing distinct real world phenomena. We would like to emphasize that these distinct approaches of interpreting the flows are not meant to compete with each other (let alone, compete with other clustering algorithms such as those based on local density), and are instead targeted at modelling and capturing distinct behaviors. Given a complex network which is inherently both directed and weighted, e.g., the Bitcoin network, one may need to study its properties considering its directionality, with or without the influence of weights, considering possibly different interpretation of weights, or just its bare topology. It is thus important to have a single clustering method with enough versatility to accommodate and instantiate the spectrum of combinations. Our proposed approach is designed to meet this particular objective.}\\


\AD{
{\bf Organization.} The paper is logically organized in two halves: the first part (Sections \ref{sec:single} and \ref{sec:multiple}) is dedicated to the centrality analysis, while the second (Section \ref{sec:clustering}), which relies on the first part, delves into the clustering algorithm. Relevant experiments are reported within the corresponding sections.} 

\AD{
In Section \ref{sec:single}, we carry out a study of the proposed Markov entropic centrality for unweighted (un)directed graphs. We coin the term `Markov' entropic centrality, as opposed to the original path based model studied in \cite{T7}, because our model leverages the random walks being absorbed in an abstraction called auxiliary nodes with certain probability, and the future steps of the random walker depends solely on its current location, irrespective of its past.  We analyze the behavior of this centrality as a function of (a) the probability that the walker stops at given nodes, and (b) time (or number of steps in the walk). Results are analytically proven in terms of bounds and a closed form expression captures the asymptotic behavior of the random walk probability distribution for unweighted directed graphs (undirected graphs can be readily captured by introducing directed edges for both directions), in contrast to the numerical computations carried out in \cite{NRK}. An immediate practical implication of such a closed form solution instead of reliance on numerical computations is computational efficiency and scalability of the model. Our analysis provides a thorough understanding of the role of an associated parameter termed as the absorption probability, as well as the flexibility to explore the probability being a constant at all nodes, or being dependent on the individual nodes' degrees --- which in turn makes the model versatile to capture different situations, and enables to reason about the choice of the parameter. This is in contrast to the ad-hoc choices used for experiments in \cite{NRK}. Finally, we introduce the notion of Markov entropic centralization which we study analytically in the asymptotic time regime as a way to provide further insights regarding the relative importance of the nodes.}

\AD{
In Section \ref{sec:multiple}, we generalize the theory of entropic centrality for weighted graphs, taking into account the need to accommodate different interpretations of edge weights, and its competition with number of edges, in determining node centrality, e.g., tune the model as per the need of an analysis to let an edge with weight 3 have either more or less or equal contribution to a node's centrality as three edges of weight 1. We refer to \cite{OAS} for a more elaborate discussion on interpreting centrality for weighted graphs. This is furthermore in contrast to the treatment in \cite{NRK}, where weighted graphs are dealt with by changing the transition probabilities of the random walker as per edge weights, which is one specific instance within our tunable model.}

\AD{Overall, Sections \ref{sec:single} and \ref{sec:multiple} result in our \textbf{first set of contributions}, which is vis-\`a-vis entropic centrality. We provide (i) a generalization of the entropic centrality model, which can be tuned to capture a spectrum of combinations in terms of the role of the edges' weights and directionality, accompanied with (ii) a mathematically rigorous analysis which helps understand the role of the model parameters and choose them judiciously, and ultimately resulting in (iii) a computationally, significantly efficient and thus scalable model with respect to prior work.}


Section \ref{sec:clustering} contains the proposed clustering algorithm. 
The above analysis of the random walker probability distributions carried out in the context of entropic centrality, accompanied with an analytical study of the graph centralization measure which captures how tightly \AD{the graph is organized around high centrality nodes, directly lends itself to the design of a novel graph clustering (community detection) algorithm}, which amortizes the computations carried out for computing the node centralities. Foremost, from the study of the (absorption) probability distributions of the random walkers and the resulting entropic centrality scores of the nodes, we observed that: (i) nodes with a low entropic centrality have local clusters where ties are relatively strong, while nodes tend to have a relatively high entropic centrality when they either are embedded within a large cluster, or when they are at the boundary of two clusters (and could arguably belong to either), and moreover, (ii) sharp changes in the random walker absorption probability distribution signal boundaries of local community structures. 

This leads to a two-stage clustering algorithm: \AD{First, observing that nodes at the boundary of clusters often act as hubs and have high centrality, we determine meaningful node centric local communities around nodes with low entropic centrality. This approach inherits the flexibility of the underlying centrality model, and is applicable to the spectrum of combinations (un/directed, un/weighted) of graphs. Furthermore, the strong coupling of the algorithm's design with the aforementioned centrality measure allows to initiate the random walkers in an informed manner, in contrast to previous works based on random walks, such as \cite{AL,BNCZ}, which, in absence of any guidance on where to initiate random walks needed more complex (computationally, as well as in terms of implementation and analysis), constrained random walkers.  Second, the process is reapplied on the created clusters (instead of the nodes) to effectuate a bottom-up, scalable, hierarchical clustering. Working principles behind the heuristic algorithms are supported by formal derivations, and the performance of the algorithms have been tested with experiments in terms of sensitivity and correctness, as well as through rigorous benchmarking using synthetic and real world networks with ground truth.}



\AD{Finally, the existing flow based clustering approaches (and most clustering approaches, in general) do not have an inherent way to reason about the quality of clusters they may yield, given any graph. In contrast} the distribution of the relative values of entropic centrality of nodes and graph centralization also inform us on whether and how much a given graph is amenable to the use of our approach for carrying out clustering.

\AD{In \cite{NRK}, a radically different approach (adapting \cite{girvan2002community}) for community detection is applied, where edges are iteratively removed, and in each iteration one needs to identify the edge such that the average entropic centrality over all the nodes is reduced the most. This requires the computation of entropic centrality of all the nodes multiple times for the different resulting graph snapshots from edge removals, and do so numerically over and over again. Naturally, the approach is computationally intensive and not scalable. For a specific undirected unweighted graph, our algorithm took 1.072 seconds to compute the communities, which is multiple orders of magnitude faster than the 3196.07 seconds needed by the edge removal algorithm \cite{NRK} \AD{(more details on the specific experiments can be found in Subsection \ref{sec:bitcoin178clusters})}. The communities detected by our approach were also qualitatively better for the network where the two approaches were compared head-to-head. Moreover, the authors in \cite{NRK} noted that their approach was not useful for community detection in directed graphs. We demonstrate with experiments that our approach works also with directed graphs. }

To summarize, our \textbf{second set of contributions} are vis-\`a-vis graph clustering. The salient aspects of our work comprise (i) the use of entropic centrality (and underlying analysis) to inform on whether our approach will yield good quality of clusters given a graph, and if so, (ii) how to identify good candidate `query nodes' around which initial sets of local community structures should be explored, which in turn yields (iii) an effective novel two-stage clustering algorithm even while using a single random walker (as opposed to prior studies, where random walkers initiated at random query nodes were shown not to be effective, necessitating more complex and constrained walks by multiple random walkers). Our clustering algorithm (iv) amortizes the previous computation of entropic centralities in computing these communities, and (v) inherits (from the underlying centrality model) the flexibility of interpreting the roles of edge weights and directionality, allowing for capturing different behaviors of a weighted directed graph as per application needs. Our approach is thus \AD{one of the few graph clustering approaches in the literature that handles directed graphs naturally, and additionally, it allows for flexible incorporation and interpretation of edge weights}.

{\bf Experiments and Benchmarking:} 
\AD{Emerging complex networks such as the Bitcoin network motivated the introduction of a framework for centrality and clustering that handles a variety of angles from which the network structures can be studied, considering in combination the direction and weight of edges. The principal objective of the proposed clustering algorithm is to seamlessly accommodate the various combinations in which the network can be interpreted. A second objective is to ensure that our approach also yields meaningful clusters in general, and in particular, that the obtained results are meaningful when applied to standard graphs used for graph clustering benchmarking. Experiments to support our work are designed accordingly.}

\AD{
Clustering of Bitcoin transaction subgraphs \cite{data178,data4571} is performed to explore the versatility of our model vis-à-vis the spectrum of choices in interpreting the directionality and weights of edges. Since no ground truth is known, we compare the obtained results with an existing Bitcoin forensics study \cite{Egret-paper}.
}

\AD{
Evaluating in general the quality of a clustering algorithm is a debated topic \cite{LWG,letopeel}. Complications arise because of multiple reasons, including that different clustering algorithms may identify communities which may reflect different aspects of relationships and thus different results may not necessarily be wrong or worse; there may be multiple interpretations of relationships, resulting in different `ground truths' being valid. We thus report results from a variety of networks, falling into two categories: (a) synthetic networks, and (b) real world graphs with meta-information used to assign a ground truth.}

\AD{
We apply our algorithm on {\em synthetically generated networks}, which eliminates the ambiguity of `ground truth' since it is determined by design. We refer to \cite{LF,LFR} for more details advocating the use of synthetic graphs for benchmarking community detection algorithms. A caveat with using synthetic benchmark graphs is that, they may themselves be created following certain rules, e.g., \cite{LFR} is generated guided by local density measures. That in turn may favor specific families of clustering algorithms. Ranging from small graphs (of 100 to 500 nodes) to larger graphs (of 1000 and 4000 nodes), the proposed algorithm fares well and is comparative with Louvain \cite{blondel2008fast}, with our results showing neither algorithm clearly outperforming the other in determining the clusters, but rather, each algorithm outperforming the other depending on individual instances.}

\AD{
We carry out comparative studies with classical networks and connect these experimental studies back to the conceptual foundations of our approach, which allow us to understand whether our approach is expected to yield good clustering results for a given graph depending on the distribution of the relative values of entropic centralities.
This includes the classical karate club \cite{KarateClubOriginal} and the small cocaine dealing network \cite{cdn} to interpret and explain the meaning of our model and validate the corresponding findings guided by the well understood narrative associated with these graphs, but also the dolphin network \cite{lusseau2003bottlenose} and the American football network \cite{girvan2002community}, for which the performance of our clustering algorithm is compared with well known algorithms, namely Louvain \cite{blondel2008fast}, Infomap \cite{RB} and label propagation \cite{andersen2006local}.
}
Remark: All experiments and implementations presented in this paper were done in Python, using in particular the libraries NetworkX \cite{networkx} and Scikit-Learn \cite{scikit}. All graphs were drawn using either NetworkX or Gephi \cite{gephi}.

\section{Entropic Centrality for Unweighted Graphs}\label{sec:single}

A {\em right stochastic matrix} $P$ is a square matrix, where every coefficient is a non-negative real number and each row sums to 1. A right $n\times n$ stochastic matrix describes a Markov chain $X_t$, $t\geq 1$, over $n$ states. The coefficient $P_{u,v}$ of $P$ is the probability of moving from state $u$ to state $v$, and since the total $\sum_{u=1}^{n}P_{u,v}$ of transition probabilities from state $u$ to all other states must be 1 by definition of probability, the matrix $P$ is indeed a right stochastic matrix, called a transition matrix. To show that $P^k$ is again a right stochastic matrix for any positive integer $k$, take the all 1 column vector ${\bf 1}$ and notice that $P {\bf 1} = {\bf 1}$ for a right stochastic matrix. Then $P^k{\bf 1}=P{\bf 1}={\bf 1}$. 
Let $(P_{u,l})_l$ be the $u$th row of $P$, and $(P_{l,v})_l$ be its $v$th column. Their product $(P_{u,l})_l(P_{l,v})_l = \sum_{l}(P_{u,l})_l(P_{l,v})_l$ determines the probability of going from the state $i$ to the state $j$ in two steps, where the intermediate state is $l$, for any possible state $l$. In general, $P^k$ gives the probability transition of going from any state to any another state in $k$ steps. 

Consider now a connected directed graph $G=(V,E)$ with $n=|V|$ nodes labeled from $1$ to $n$ and $|E|$ directed edges, and a random walker on $G$, which starts a random walk at any given node according to an initial probability distribution. If the walker decides on which node $v$ to walk to, based only on the current node $u$ at which it currently is ($v$ does not have to be distinct from $u$), the random walk is a Markov chain, which can be modeled by a stochastic matrix $P$ where $P_{uv}$ is the probability to go to node $v$ from node $u$. We will assume that every node has a self-loop (a self-loop is needed for the model to encapsulate the case of nodes with zero out-degree.) Let $d_{out}(u)$ denote the out-degree of $u$, which includes the self-loop as per our model. A typical choice that we consider in this section is $P_{uv}=\frac{1}{d_{out}(u)}$ for every node $v$ belonging to the set $\Nc_u$ of out-neighbors of $u$ (this is saying that each neighbor is chosen uniformly at random). 
Other choices are possible, and indeed, subsequently in this paper, we shall consider the case of weighted graphs where the transition probabilities depend on the edge weights.

\subsection{Markov Entropic Centrality}

The notion of entropic centrality $C_H(u)$ for a node $u$ in an undirected graph was originally defined in \cite{T7} as
\begin{equation}\label{eq:CH}
\begin{array}{c}
C_H(u) = -\sum_{v \in V}q_{uv}\log_2(q_{uv})
\end{array}
\end{equation}
where $q_{uv}$ denotes the probability of reaching $v$ from $u$ following any path. This  
formalizes the idea that a node is central if there is a high uncertainty about the destination of a random walker starting at it, assuming that at each hop, the random walker is equally likely to continue to any unvisited node, or to stop there. This is because the Shannon entropy $H(X)=-\sum_x p(X=x)\log_2 p(X=x)$ of a random variable $X$ measures the amount of uncertainty contained in $X$. Entropic centrality given in (\ref{eq:CH}) was then extended to non-atomic flows in \cite{nonatomicflowisita,peerj} and to a Markov model in \cite{NRK} by relaxing the assumption (in \cite{T7}) that a random walker cannot revisit previously traversed nodes. In this paper we generalize the Markov entropic centrality model. 
This notion of centrality captures the influence of ``friends", and of friends of friends (see e.g. \cite{higham} for a discussion of the role of friends in considering centralities).

To define a notion of entropic centrality similar to (\ref{eq:CH}), the random walker actually needs to have the choice of  stopping at any given node. This is not well captured by the transition matrix $P$ described above, even if there is a self-loop at every node. Thus for every node $v$ in $G$, an auxiliary node $v'$ is added \cite{NRK}, together with a single directed edge $(v,v')$, and a probability $p_{vv'}$ to be chosen (the original probabilities $p_{uv}$ are adjusted to the probabilities $\tilde{p}_{uv}$ so that the overall matrix remains stochastic). Once the flow reaches an auxiliary node, it is absorbed because the auxiliary nodes have no outgoing edges. This gives a notion of Markov entropic centrality as defined in \cite{NRK}.

\begin{defn}\label{eq:CHt}
The {\em Markov entropic centrality} of a node $u$ at time $t$ is defined to be	
\begin{equation}\label{eq:Hij}
C_H^{t}(u) = - \sum_{v\in V} (\tilde{p}_{uv}^{(t)}+p_{uv'}^{(t)})\log_2 (\tilde{p}_{uv}^{(t)}+p_{uv'}^{(t)})
\end{equation}

where $\tilde{p}_{uv}^{(t)}$ is the probability to reach $v$ at time $t$ from $u$, for $v$ a node in $V$ where $p_{uv'}$ is the probability to reach an auxiliary node $v'$ from $u$.
\end{defn}

A node $u$ is central if $C_H^{(t)}(u)$ is high: if $C_H^{(t)}(u)$ is high, using the underlying entropy interpretation, this means that for a random walker starting at node $u$, the uncertainty about its destination $v$ after $t$ steps is high, thus $u$ is well connected. 

The time parameter $t$ can also be interpreted as a notion of locality. It describes a diameter around the node $u$, of length $t$ steps. Thus the entropic centrality at $t=1$ emphasizes a node's degree, $t$ being the graph diameter implies that we are considering a period of time by when the whole graph can be first reached, and $t \rightarrow \infty$ describes the asymptotic behavior over time. Thus $C_H^{(t)}(u)$ can be regarded as a measure of influence of $u$ over its close neighborhood for small values of $t$, and over the whole graph asymptotically.

Given an $n\times n$ stochastic matrix $P$ describing the moves of a random walker on a directed graph, let us then introduce $n=|V|$ auxiliary nodes, one for each node $v$, $v\in V$, with corresponding probabilities $p_{vv'}=D_{vv}$ to walk from node $v$ to node $v'$. 
This means the out-degree of $u$ increases by 1 because of the addition of $u'$. Nevertheless, by $d_{out}(u)$, we will actually refer to the degree of $u$ before the addition of $u'$.
This creates the following right stochastic matrix
\[
\hat{P}=
\begin{bmatrix}
\tilde{P} & D \\
{\bf 0}_n   & {\bf I}_n
\end{bmatrix}
\]
where $D=D_{uu}$ is a diagonal matrix, and $\tilde{P}$ is such that $[\tilde{P},D]{\bf 1}={\bf 1}$. 
We assume that $\tilde{P}$ has for $u$-th row $(\tilde{P}_{u,l})_l=(1-p_{uu'})(P_{u,l})_l$. Then $\sum_{l=1}^{2n}(\hat{P})_{ul}=(1-p_{uu'})\sum_{l=1}^np_{ul}+p_{uu'}=1$ for every $u$. This is alternatively written as $\tilde{P}=({\bf I}_n-D)P$.
The identity matrix ${\bf I}_n$ represents the stoppage of the flow at the auxiliary nodes (an absorption of the flow arriving at these nodes). To determine the centrality of a specific node $u$, we assume an initial distribution that gives a probability of 1 to start at $u$, and $0$ elsewhere.

The definition of Markov entropic centrality was used as part of a clustering algorithm in \cite{NRK}, and the probabilities $p_{vv'}$ were explored numerically to optimize the clustering results. Our first contribution is the following closed-form expression for the asymptotic behavior of the transition matrix.

\begin{lem}\label{lem:pt}
For an integer $t\geq 1$ and $D\neq {\bf 0}$, we have  
\begin{equation}\label{eq:P}
\hat{P}^t=
\begin{bmatrix}
\tilde{P}^t & (\sum_{j=0}^{t-1}\tilde{P}^j)D \\
{\bf 0}_n   & {\bf I}_n
\end{bmatrix}.
\end{equation}
In particular
\begin{equation}\label{eq:Pinf}
\hat{P}^t=
\begin{bmatrix}
{\bf 0}_n & ({\bf I}_n-\tilde{P})^{-1} D \\
{\bf 0}_n   & {\bf I}_n
\end{bmatrix}
\end{equation}
when $t\rightarrow \infty$.
\end{lem}
\begin{IEEEproof}
Formula (\ref{eq:P}) follows from an immediate computation. 
Since $\tilde{P},D$ have non-negative real coefficients, so has $\tilde{P}^jD$, thus $(\sum_{j=0}^{l}\tilde{P}^j)D \geq (\sum_{j=0}^{l-1}\tilde{P}^j)D$ for any $l\geq 1$, and equality holds if and only if $\tilde{P}^N=0$ for some $N$. But this $N$ must exist when $t\rightarrow\infty$, because $\hat{P}$ is a stochastic matrix, meaning that the sum of every row must remain 1, while the coefficients of $(\sum_{j=0}^{t-1}\tilde{P}^j)D$ increases at every increment of $t\leq N$. Then
\[
\hat{P}^N=
\begin{bmatrix}
{\bf 0}_n & (\sum_{j=0}^{N-1}\tilde{P}^j)D \\
{\bf 0}_n   & {\bf I}_n
\end{bmatrix}
\]
but since $(\tilde{P}-{\bf I}_n)(\sum_{j=0}^{N-1}\tilde{P}^j)=\tilde{P}^N-{\bf I}_n = -{\bf I}_n$, the matrix $(\tilde{P}-{\bf I}_n)$ is invertible,
$
\sum_{j=0}^{N-1}\tilde{P}^j = -(\tilde{P}-{\bf I}_n)^{-1}$,  
yielding (\ref{eq:Pinf}).
\end{IEEEproof}

The matrix $\hat{P}^t$ in (\ref{eq:P}) contains a first block $\tilde{P}^t$ whose coefficients $\tilde{p}_{uv}^{(t)}$ are the probabilities to go from $u$ to $v$ in $t$ steps. The second block $(\sum_{j=0}^{t-1}\tilde{P}^j)D$ contains as coefficients the probabilities to go from $u$ to $v'$ in $t$ steps, which we denote by $p_{uv'}^{(t)}$. Therefore
the probabilities $\tilde{p}_{uv}^{(t)}+p_{uv'}^{(t)}$ in (\ref{eq:Hij}) are obtained from the $u$th row of the matrix $\hat{P}^t$, by summing the coefficients in the columns $v$ and $v'$. When $t\rightarrow \infty$, the term in column $v$ becomes $0$, and we are left with the term in column $v'$,

of the form
\[
\pi_{uv}:=
p_{uv'}^{(\infty)}=
\left\{
\begin{array}{ll}
\sum_{t\geq 1}\tilde{p}_{uv}^{(t)}p_{vv'}& u \neq v \\
(\sum_{t\geq 1}\tilde{p}_{uv}^{(t)}+1)p_{vv'}& u = v.\\
\end{array}
\right. 
\]
This means that we are looking at the probability to start at $u$ and reach $v$ in $t \geq 0$ steps, and then to get absorbed at $v$ (that is reaching $v'$, and then not leaving $v'$). Asymptotically, the entropic centrality defined in (\ref{eq:Hij}) becomes
\[
C_H^{\infty}(u) = -\sum_{v\in V}\pi_{uv}\log_2(\pi_{uv}).
\]

We discuss next how the choice of the probabilities $p_{uu'}$ to reach an auxiliary node $u'$ from $u$ influences the random walker at time $t$. 

\begin{lem}\label{lem:ptilde}
The probability $\tilde{p}_{uv}^{(t)} $ to start a walk at $u$ and to reach $v$ in $t$ steps is bounded as follows: 
\[
(1-p_{uu'}) (\max_{w\in V}(1-p_{ww'}))^{t-1}p_{uv}^{(t)} \geq \tilde{p}_{uv}^{(t)} \geq (1-p_{uu'}) (\min_{w\in V}(1-p_{ww'}))^{t-1}p_{uv}^{(t)}.
\]
\end{lem}
\begin{IEEEproof}
Since $\tilde{p}_{uv}=(1-p_{uu'})p_{uv}$, we have
\begin{eqnarray*}
\tilde{p}_{uv}^{(2)} 
&=& \sum_{w\in out(u)\cap in(v)}\tilde{p}_{uw}\tilde{p}_{wv} 
= \sum_{w }(1-p_{uu'})p_{uw}(1-p_{ww'})p_{wv} \\
& \geq &(1-p_{uu'})\min_{w}(1-p_{ww'}) \sum_w p_{uw}p_{wv}
= (1-p_{uu'})\min_{w}(1-p_{ww'}) p^{(2)}_{uv}, \\
\tilde{p}_{uv}^{(3)} 
&=& \sum_{x}\tilde{p}_{ux}\tilde{p}_{xv}^{(2)} 
\geq \sum_{x}(1-p_{uu'})p_{ux}(1-p_{xx'})\min_{w\in out(x)\cap in(v)}(1-p_{ww'}) p^{(2)}_{xv} \\
&\geq & (1-p_{uu'})\min_{x\in out(u)\cap in(w)}(1-p_{xx'}) \min_{w}(1-p_{ww'})p_{uv}^{(3)}  
\end{eqnarray*}
and we observe that the minimization is taken over all walks from $u$ to $v$ in $t$ steps, or more precisely over all nodes involved in such walks, except $u$ and $v$. Certainly for $x$ in such a walk, $\min_x(1-p_{xx'})\geq \min_{w\in V}(1-p_{ww'})$. The other inequality can be established identically, and thus
\[
(1-p_{uu'}) (\max_{w\in V}(1-p_{ww'}))^{t-1}p_{uv}^{(t)} \geq \tilde{p}_{uv}^{(t)} \geq (1-p_{uu'}) (\min_{w\in V}(1-p_{ww'}))^{t-1}p_{uv}^{(t)}.
\]
For $t$ small or for $u$ an isolated node, better bounds are obtained by replacing the minimum/maximum over all nodes by those in the different walks from $u$ to $v$, but for $t$ large, starting from $t\geq n-1$, it is likely that for most well connected nodes, the bounds are getting tight.
\end{IEEEproof}
\begin{cor}
In particular:
\begin{enumerate}
\item If $p_{uu'}=a<1$ for all $u\in V$, we have $\tilde{p}_{uv}^{(t)} = (1-a)^tp_{uv}^{(t)}$.
\item
If $p_{uv}=\frac{1}{d_{out}(u)}$ and $p_{uu'}=\frac{1}{d_{out}(u)+1}$, then $\tilde{p}_{uv}=\frac{1}{d_{out}(u)+1}$, 
and
\[
(1-\tfrac{1}{d_{out}(u)+1}) (\max_{w\in V}\frac{d_{out}(w)}{d_{out}(w)+1})^{t-1}p_{uv}^{(t)} \geq \tilde{p}_{uv}^{(t)} \geq (1-\tfrac{1}{d_{out}(u)+1}) (\min_{w\in V}\frac{d_{out}(w)}{d_{out}(w)+1})^{t-1}p_{uv}^{(t)}.
\]
\end{enumerate}
\end{cor}
\begin{IEEEproof}
\begin{enumerate}
\item
All inequalities in the proof of the lemma are equalities in this case.
\item
By definition, $\tilde{p}_{uv}=(1-\frac{1}{d_{out}(u)+1})\frac{1}{d_{out}(u)}=\frac{1}{d_{out}(u)+1}$. 
\end{enumerate}
\end{IEEEproof}

The case when $p_{uu'}=a<1$ is reminiscent of the notion of Katz centrality. A factor $(1-a)^t$ is introduced so that the longer the path, the lower the probability. If $a$ is chosen close to 1, e.g. $a=0.9$, then $(1-a)^t$ (e.g. $\tfrac{1}{10^t}$) becomes quickly negligible. If instead $a$ is chosen close to $0$, e.g. $a=0.1$, then it takes longer for probabilities to become negligible (e.g. $(\tfrac{9}{10})^{50}\approx 0.0051$). 

The case $p_{uu'}=\frac{1}{d_{out}(u)+1}$ instead uses (inverse) proportionality to the number of outgoing edges. Both the upper and lower bounds given in the proof of the above corollary depend on the degree of the nodes included in walks from $u$ to $v$, and when the walk length grows, the number of distinct nodes is likely to increase, giving the bounds stated in the corollary. These bounds depend on the function $\tfrac{x}{x+1}$, which closely converges to 1, in fact for $d_{out}(v)=9$, we already get $0.9$. Therefore, $\tilde{p}_{uv}^{(t)}$ is mostly behaving as $p_{uv}^{(t)}$, except if $d_{out}(u)$ is small enough (say less than 8). 

In summary, the emphasis of the case $p_{uu'}=a<1$ is on the length of the walk, not on the walk itself, while for $p_{uu'}=\frac{1}{d_{out}(u)+1}$ it is actually on the nodes traversed during the walk, with the ability to separate the nodes of low entropic centrality from the others.   

The bounds of Lemma \ref{lem:ptilde} can be applied to the asymptotic case. 

\begin{lem}\label{lem:piuv}
The probability $\pi_{uw}=\sum_{t\geq 1}\tilde{p}_{uw}^{(t)}p_{ww'} $ to start at $u$ and to be absorbed at $w\neq u$ over time is bounded as follows:  
\[
(1-p_{uu'})(\sum_{t\geq 1}(\min_{x\in V}(1-p_{xx'}))^{t-1}p_{uw}^{(t)})p_{ww'}
\leq \pi_{uw} 
\leq 
(1-p_{uu'})(\sum_{t\geq 1}(\max_{x\in V}(1-p_{xx'}))^{t-1}p_{uw}^{(t)})p_{ww'}.
\]
\end{lem}
\begin{IEEEproof}
Recall that $\pi_{uw}=\sum_{t\geq 1}\tilde{p}_{uw}^{(t)}p_{ww'} $ is the probability to start at $u$ and to be absorbed at $w$ over time. Then
\begin{eqnarray}
\pi_{uw}=\sum_{t\geq 1}\tilde{p}_{uw}^{(t)}p_{ww'}  
 &=&  (\tilde{p}_{uw}+\sum_{t\geq 2}\sum_{i=1}^{t-1}\tilde{p}_{uv}^{(i)}\tilde{p}_{vw}^{(t-i)} + \sum_{t\geq 2}\sum_{i=1}^{t-1}\sum_{y\neq v}\tilde{p}_{uy}^{(i)}\tilde{p}_{yw}^{(t-i)} )p_{ww'} \label{eq:pabs}
\end{eqnarray}
and use Lemma \ref{lem:ptilde}:
\[
\tilde{p}_{uv}^{(i)} \geq (1-p_{uu'})(\min_{x\in V}(1-p_{xx'}))^{i-1} p_{uv}^{(i)},~ 
\tilde{p}_{vw}^{(t-i)}\geq (1-p_{vv'})(\min_{x\in V}(1-p_{xx'}))^{t-i-1} p_{vw}^{(t-i)}
\]
to get
\begin{eqnarray*}
\sum_{t\geq 2}\sum_{i=1}^{t-1}\tilde{p}_{uv}^{(i)}\tilde{p}_{vw}^{(t-i)}
&\geq & (1-p_{uu'})(1-p_{vv'})
\sum_{t\geq 2}(\min_{x\in V}(1-p_{xx'}))^{t-2}\sum_{i=1}^{t-1}p_{uv}^{(i)}p_{vw}^{(t-i)} \\
\sum_{t\geq 2}\sum_{i=1}^{t-1}\sum_{y\neq v}\tilde{p}_{uy}^{(i)}\tilde{p}_{yw}^{(t-i)}
&\geq & (1-p_{uu'})
\sum_{t\geq 2}(\min_{x\in V}(1-p_{xx'}))^{t-2}\sum_{i=1}^{t-1}\sum_{y\neq v}(1-p_{yy'})p_{uy}^{(i)}p_{yw}^{(t-i)} \\
& \geq & (1-p_{uu'})
\sum_{t\geq 2}(\min_{x\in V}(1-p_{xx'}))^{t-1}\sum_{i=1}^{t-1}\sum_{y\neq v}p_{uy}^{(i)}p_{yw}^{(t-i)}.\end{eqnarray*}
Therefore
\begin{eqnarray}
\pi_{uw} 
&\geq &
(1-p_{uu'})[p_{uw}+(1-p_{vv'})
\sum_{t\geq 2}(\min_{x\in V}(1-p_{xx'}))^{t-2}\sum_{i=1}^{t-1}p_{uv}^{(i)}p_{vw}^{(t-i)} \nonumber \\ 
&&+
\sum_{t\geq 2}(\min_{x\in V}(1-p_{xx'}))^{t-1}\sum_{i=1}^{t-1}\sum_{y\neq v}p_{uy}^{(i)}p_{yw}^{(t-i)})]p_{ww'}, \label{eq:lowertrans}
\end{eqnarray}
for $v$ an intermediate node between $u$ and $w$.

The bound can be made coarser, using that $(1-p_{vv'}) \geq \min_{x\in V}(1-p_{xx'})$:  

\begin{eqnarray*}
\pi_{uw} & \geq & 
(1-p_{uu'})[p_{uw}+\sum_{t\geq 2}(\min_{x\in V}(1-p_{xx'}))^{t-1}
\sum_{i=1}^{t-1}(p_{uv}^{(i)}p_{vw}^{(t-i)}+\sum_{y\neq v}p_{uy}^{(i)}p_{yw}^{(t-i)})]p_{ww'} \\
& = & (1-p_{uu'})[p_{uw}+\sum_{t\geq 2}(\min_{x\in V}(1-p_{xx'}))^{t-1}
p_{uv}^{(t)}]p_{ww'} \\
& = & (1-p_{uu'})\sum_{t\geq 1}(\min_{x\in V}(1-p_{xx'}))^{t-1}
p_{uv}^{(t)}p_{ww'}. 
\end{eqnarray*}

\end{IEEEproof}

The bounds are well matching the intuition: three components mostly influence $\pi_{uw}$: the likelihood of leaving $u$ (the term $1-p_{uu'}$), that of reaching $w$ from $u$, and that of getting trapped at $w$ (the term $p_{ww'}$). When $p_{uu'}=a<1$ for all $u \in V$, the bounds on $\pi_{uw}$ are met with equality and we have that
\[
\pi_{uw} = (1-a)\sum_{t\geq 1}(1-a)^{(t-1)}p_{uw}^{(t)}a,~
\pi_{uu} = (1-a)\sum_{t\geq 1}(1-a)^{(t-1)}p_{uu'}^{(t)}a +a.
\]
Therefore $\pi_{uw}$ is weighted by $(1-a)$ and $a$ irrespectively of the choice of $u,w$, and so is the corresponding entropic centrality: when $a$ grows, $\pi_{uu}$ grows, thus $\pi_{uw}$ for $w\neq u$ must decrease since probabilities sum up to 1 and
\begin{eqnarray*}
\sum_w \pi_{uw}\log_2(\tfrac{1}{\pi_{uw}}) 
& = & \pi_{uu}\log_2(\tfrac{1}{\pi_{uu}}) + \sum_{w \neq u}\pi_{uw}\log_2(\tfrac{1}{\pi_{uw}}),\\
& \leq & \pi_{uu}\log_2(\tfrac{1}{\pi_{uu}}) + (1-a)\log_2(\tfrac{n-1}{1-a}) 
\end{eqnarray*}
since the second sum contains $n-1$ terms whose sum is at most $1-a$, and whose maximum is reached when all probabilities are $\tfrac{1-a}{n-1}$. Also the function $-x\log_2(x)=-\tfrac{1}{\ln 2}x\ln x$ is concave and has a global maximum at $x$ such that $-\tfrac{1}{\ln 2}(\ln x+1)=0$, that is $x=\tfrac{1}{3}\approx 0.36788$, for which $-x\log_2(x) \approx 0.53074$. Thus
\begin{equation}\label{eq:paupper}
C_H^{\infty}(u) \leq 0.53074 + (1-a)\log_2(\tfrac{n-1}{1-a}) 
\end{equation}
and for a given $n$, this upper bound becomes small when $a$ grows. The same argument repeated for $p_{uu'}=\frac{1}{d_{out}(u)+1}$ shows (part of) the role of the out-degree of $u$ in its centrality.

In what follows, we will provide some experiments to illustrate the consequences of choosing $p_{uu'}=a<1$, but we will then focus on the case $p_{uu'}=\frac{1}{d_{out}(u)+1}$ for the reasons just discussed: (1) choosing $\tilde{p}_{uv}$ is influenced by the choice of the walk rather than its length, and (2) $\pi_{uv}$ depends on the degrees of $u$ and $v$ rather than on the constant $a$. 

\begin{lem}\label{lem:centralization}
Set $n=|V|$. The asymptotic Markov entropic centralization (see \cite{freeman}) defined by
\[
C_H^{\infty}(G) = \frac{\sum_{v\in V}C_H^{\infty}(\hat{v})-C_H^{\infty}(v)}
{\max \sum_{v\in V}C_H^{\infty}(\hat{v})-C_H^{\infty}(v)}
\]
where $\hat{v}$ is the node with the highest Markov entropic centrality in $V$ is 
given by
\[
C_H^{\infty}(G) = \frac{\sum_{v\in V}C_H^{\infty}(\hat{v})-C_H^{\infty}(v)}{n\log_2(n)}
\]
when $p_{uu'}=\frac{1}{d_{out}(u)+1}$.
\end{lem}
\begin{IEEEproof}
The maximum at the denominator is taken over all possible graphs with the same number of vertices. 
A graph that maximizes the denominator would have one node with the maximum centrality, and all the other nodes with a centrality 0 (thus minimizing the terms contributing negatively). This graph is the star graph on $|V|$ vertices, since the middle node has $|V|$ outgoing edges, and the $|V|-1$ leaves have none. Therefore all leaves have an entropic centrality of 0, and the center $\hat{v}$ of the star has maximum entropic centrality. Formally, with $|V|=n$ and putting $\hat{v}$ in the first row:
\[
\hat{P}=
\begin{bmatrix}
\tilde{P} & D \\
{\bf 0}_n & {\bf I}_n
\end{bmatrix},~
\tilde{P}=
\begin{bmatrix}
\tfrac{1}{n+1} & \tfrac{1}{n+1} & \ldots & \tfrac{1}{n+1} \\
{\bf 0}_{n-1,1} & \tfrac{1}{2}{\bf I}_{n-1,n-1} \\
\end{bmatrix},~
D=\begin{bmatrix}
\tfrac{1}{n+1} & {\bf 0}_{1,n-1} \\
{\bf 0}_{n-1,1} & \tfrac{1}{2}{\bf I}_{n-1,n-1} \\
\end{bmatrix}.
\]
Then
\[
{\bf I}_n-\tilde{P} = \begin{bmatrix}
\tfrac{n}{n+1} & -\tfrac{1}{n+1} & \ldots & -\tfrac{1}{n+1} \\
{\bf 0}_{n-1,1} & \tfrac{1}{2}{\bf I}_{n-1,n-1} \\
\end{bmatrix},
\]
and using Schur complement:
\[
({\bf I}_n-\tilde{P})^{-1} = 
\begin{bmatrix}
\tfrac{n+1}{n} & \tfrac{2}{n} & \ldots & \tfrac{2}{n} \\
{\bf 0}_{n-1,1} & 2{\bf I}_{n-1,n-1} \\
\end{bmatrix},
\]
so
\[
({\bf I}_n-\tilde{P})^{-1}D 
=
\begin{bmatrix}
\tfrac{1}{n} & \tfrac{1}{n} & \ldots & \tfrac{1}{n} \\
{\bf 0}_{n-1,1} & {\bf I}_{n-1,n-1} \\
\end{bmatrix}.
\]
Lemma \ref{lem:pt} tells us that $\pi_{\hat{v}v}=\frac{1}{n}$ for all $v$ and thus its Markov entropic centrality is $\log_2(n)$, while $C_H^{\infty}(v)$ is 0 for $v\neq \hat{v}$.
\end{IEEEproof}

\begin{defn}\label{def:seq} 
Given a graph $G$, label its vertices in increasing order with respect to their Markov entropic centralities, that is $C_H^{\infty}(v_i) \leq C_H^{\infty}(v_{i+1})$ for $i=1,\ldots,n-1$ and $p_{uu'}=\tfrac{1}{d_{out}(u)+1}$. 
We define {\em the centralization sequence} of $G$ to be the ordered sequence 
\[
(\frac{\sum_{v\in V}C_H^{\infty}(v_i)-C_H^{\infty}(v)}{\log_2(n)})_{i=1,\ldots,n}.
\]
\end{defn}

It is an increasing sequence, with values ranging from -1 to 1. We already know that the maximum is 1 by Lemma \ref{lem:centralization}. The minimum is achieved when $v_1$ has centrality 0, and every other node has centrality $\log_2(n)/n$. This is the case if we consider a graph on $n$ vertices defined as follows: build a complete graph on $n-1$ vertices, that is each of the $n-1$ vertices have $n-1$ outgoing edges (including to themselves). Then add one additional vertex ($v_1$), and an outgoing edge from every of the previous $n-1$ vertices of the complete graph to this new vertex. The advantage of the centralization sequence with say considering the sequences of ordered centralities is that it captures for every node how central it is with respect to other nodes, normalized by a factor that takes into account the size of the graph, which allows comparisons among different graphs.

\subsection{A Markov Entropic Centrality Analysis of the Karate Club Network}
\label{sec:karatecentrality}
\begin{figure}
	\centering
	\includegraphics[scale=0.65]{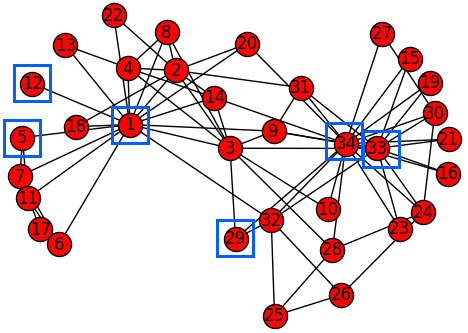}
	\caption{\label{fig:kcg}
		The karate club network: we show in Figure~\ref{fig:kcgplot} the path and Markov entropic centralities (for different values of $D$ and $t$) for the nodes 1, 5, 12, 29, 33 and 34. The degrees are respectively $16,3,1,3,12,17$. 
	}
\end{figure}

Consider the karate club network \cite{KarateClubOriginal} (used as an example in \cite{NRK}) shown in Figure~\ref{fig:kcg}. We use this small social network comprising 34 members (nodes) as a toy example to illustrate and validate some of the ideas explored in this paper. The 78 edges represent the interactions between members outside the club, which eventually led to a split of the club into two, and are used to predict which members will join which group.
This is an undirected unweighted graph, which is treated as a particular case of directed graph ($d_{out}(u)$ is the degree of $u$). 
Let $P$ denote the transition matrix such that $P_{uv}=\frac{1}{d_{out}(u)}$ for every $u\in V$ and every neighbor $v$ of $u$. The work \cite{NRK} explores the choice of $D$ (and $t$) in the context of clustering. Here, we start by investigating the role of $D$ in terms of the resulting Markov entropic centrality, for $t$ finite ($t=1,\ldots,6$ since the karate club network has a diameter of 5) and asymptotic (using Lemma \ref{lem:pt}).

\begin{figure}
	\centering
	\includegraphics[scale=0.6]{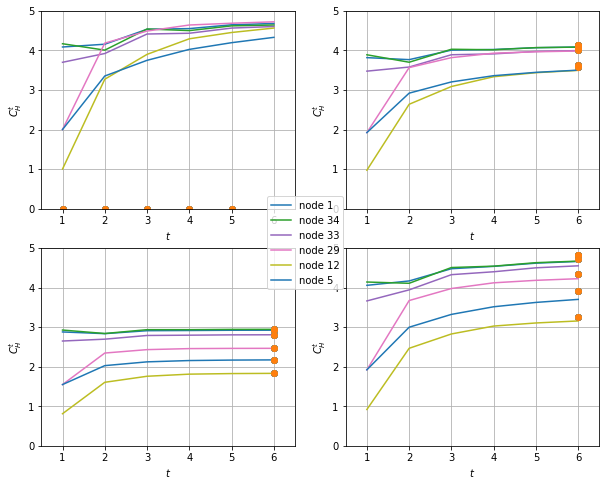}
	\caption{\label{fig:kcgplot}
		The Markov entropic centrality $C_H^t(u)$ for $u=1,34,33,29,12,5$ (from Figure~\ref{fig:kcg}) for $t=1,2,3,4,5,6$: on the upper left, $D=0.001{\bf I}_{34}$, on the upper right, $D=\tfrac{1}{5}{\bf I}_{34}$, on the lower left, $D=\tfrac{1}{2}{\bf I}_{34}$ and for $D$ such that $D_{uu}=\tfrac{1}{d_{out}(u)+1}$ on the lower right. The dots show the asymptotic values $C_H^{\inf}(u)$.
	}
\end{figure}

\begin{paragraph}{Influence of $D$.}
Figure~\ref{fig:kcgplot} illustrates the Markov entropic centrality $C_H^t(u)$ of the nodes $u \in \{1,34,33,$ $29,12,5\}$\footnote{We choose these specific nodes, since these were studied explicitly in \cite{NRK}.} for different values of $D$: for $D=a{\bf I}_{34}$,\footnote{Note that ${\bf I}_{M}$ represents the identity matrix of dimension $M\times M$.} with $a=0.001$\footnote{We cannot use $D= {\bf 0}$ since this means no absorption probability. Also the matrix $({\bf I}_n-\tilde{P})$ in Lemma \ref{lem:pt} would have no reason to be invertible.}, $0.2, 0.5$, we observe that $C_H^t(u)$ is decreasing and flattening, for all nodes and for every value of $t$. This is expected, since when the probabilities at the auxiliary nodes are increasing as a constant, the overall uncertainty about the random walk is reducing (as computed in (\ref{eq:paupper})). Thus the higher the absorption probability, the greater the attenuation of the entropic centralities for all nodes. More precisely, (\ref{eq:paupper}) upper bounds $C_H^{\infty}(u)$ by
\[
0.53074+(1-a)\log_2(\tfrac{33}{1-a})\approx 5.5715,4.8238,3.5529~for~a=0.001,0.2,0.5,
\]
which is consistent with the numerical values obtained ($\approx 4.8232$, $4.1319$, $2.9475$).

For $D$ such that $D_{uu}=\tfrac{1}{d_{out}(u)+1}$ (shown on the lower right subfigure), the Markov entropic centralities are more separated than previously: indeed, a node with small degree then has a high absorption probability, which induces a large attenuation on its entropic centrality (as discussed after (\ref{eq:paupper})). The net effect is a wide gulf in the centrality scores between nodes with low and high degrees.
\end{paragraph}

\begin{paragraph}{Influence of $t$.}
We notice how the centrality of a node is influenced over time by its neighbors. 
Consider for example the upper left corner figure for nodes 12 and 5. Node 12 starts (at $t=1$) with an entropic centrality significantly lower than node 5  - indeed node 5 has three neighbors ($d_{out}(5)=2$ neighbors and the self-loop), more than 12 which has only one (thus $C_H^1(12)=1(1/2)\log_2(2)$ including the self-loop).
Yet node 12 reaches a ``hub" (a node of itself high centrality), namely node 1, at $t=2$, and thus for $t\geq 2$, its entropic centrality grows, and eventually ends up being almost as high as that of node 1 itself. In contrast, even though node 5 reaches the same hub, it also belongs to a local community within the graph, inside which a significant amount of its influence (flow) stays confined. This explains why node 5 ends up having a lower entropic centrality than node 12 in particular. 
In the upper right and lower left plots, we have assigned a significant volume of the flow to be absorbed at the auxiliary nodes, which has a net effect of attenuating the absolute values of entropic centrality for all nodes, i.e., a downward shift and flattening. 
This happens for nodes 5, 12, 29 on the lower right plot, since $\tfrac{1}{d_{out}(u)+1}$ is significant if the node has a low degree like 5, 12, and 29.
Taken together with the initial gap among the centralities of nodes 5 and 12, we thus do not observe the overtake in these experiments, unlike in the case where the absorption probability was negligible.  

Another less prominent behavior that we notice for node 34, and to a certain extent for node 1, is that there is a dip in their entropic centralities at the beginning, before they rise up to their long term values. We hypothesize that this is because some of the immediate neighbors of these nodes form local groupings (for instance, 34's neighbor node 15 has only 33 and 34 as neighbors, node 27 has only nodes 30 and 34 as neighbors), and thus after the initial dispersal of influence among a relatively large number of these neighbors (hence the initial high centrality), their influence gets concentrated within the local cluster with the initiating node itself for a while (thus the dip), before reaching out more evenly to the rest of the network, leading to another rise in entropic centrality.  
\end{paragraph}

\begin{paragraph}{Centralization sequence.}
The min and the max of the centralities are 3.0859 and 4.7216, while the mean and the median are 4.07636 and 3.9111. 
The min, median, mean and max of the centralization sequence as defined in Definition \ref{def:seq} are [-0.19467,-0.03248,0,0.12682].
The values for the above studied nodes of interest are 0.1226 for 1, 0.1241 for 34, 0.10195 for 33, 
 0.03485 for 29, -0.17471 for 12 and -0.06289 for 5.
\end{paragraph}

\begin{paragraph}{Hypothesis.}
The above analysis suggests that changes in the Markov entropic centralities over time are indicative of local communities in the graph, with changes in gradient corresponding to traversal of boundaries from one community to another. Nodes with low centralization have a low centrality with respect to both other nodes in the graph and graphs of the same size, since they are likely to be either isolated or to belong to a small community (e.g. node 5). While the reverse argument suggests that nodes with high centralization (e.g. nodes 1, 33 and 34) are likely to be either hubs or close to hubs.
We will explore and exploit this observation to design a clustering algorithm in Section \ref{sec:clustering}.
\end{paragraph}

\begin{paragraph}{Comparison with known centralities.}
Table \ref{table:centralities} compares different centralities. In addition to the previously mentioned prominent centrality measures, we also consider load centrality \cite{loadcentrality} in our experiments, because it captures the fraction of traffic that flows through individual nodes (load), considering all pairwise communications to be through corresponding shortest paths. Both the path and the asymptotic Markov centralities give the same ranking, it is similar to the ranking given by the degree centrality $C_D$ (with some difference regarding node 5 and 12, which are explained by the above discussion). The betweenness and load centrality agree on their ranking, which is different from the other ones. These results are not surprising: for a small graph, the degree heavily influences short paths/walks, explaining the agreement among $C_H,C_H^{\infty}$ and $C_D$, but it has little to do with the betweenness and the load. The same ranking observed between betweenness and load centrality is expected, since the latter was proposed to be a different interpretation of betweenness centrality \cite{loadcentrality}, even though some minor differences have been identified subsequently \cite{loadbetween}.
\end{paragraph}   


\begin{table}
\begin{center}
\begin{tabular}{c|c|c|c|c|c}
\rowcolor{LightGray}node & $C_H$ & $C_H^{\infty}$ & $C_D$   & $C_B$ & $C_L$\\
node 34 &4.83992 & 4.82504  & 0.5151  & 0.3040  & 0.2984\\
node 1  &4.83041 & 4.81999  & 0.4848  & 0.4376  & 0.4346 \\   
node 33 &4.76892 & 4.72539  & 0.3636  & 0.1452  & 0.1476\\
node 29 &4.50092 & 4.34323  & 0.0909  & 0.0017  & 0.0017\\
node 5  &4.07244 & 3.90674  & 0.0909  & 0.0006  & 0.0006\\
node 12 &3.39469 & 3.26763  & 0.0303  & 0       & 0\\
\end{tabular}
\caption{\label{table:centralities}
Different centrality measures: $C_H$ and $C_H^{\infty}$  are the path (\ref{eq:CH}) and asymptotic Markov entropic centralities (\ref{eq:Hij}),  $C_D$, $C_B$, $C_L$ are respectively the degree, betweenness and load centralities.
}
\end{center}
\end{table}

\section{Entropic Centrality for Weighted Graphs}\label{sec:multiple}

Consider now a weighted directed graph $G_w=(V,E)$, where the weight function $w:E \rightarrow \RR_{\geq 0}$ attaches a non-negative weight $w(u,v)$ to every edge $(u,v)\in E$. For a node $u\in V$, let $\Nc_u=\{ v\in V,~(u,v)\in E\}$ be the set of out-neighbors of $u$, $d_{out}(u)=|\Nc_u|$ be the out-degree of $u$, and $d_{w,out}(u)=\sum_{v\in\Nc_u}w(u,v)$ be the weighted out-degree of $u$.

A natural way to define a transition matrix $P_w$ to describe a random walk over $G_w$ taking into account the weight function $w$ would be to set $P_{w,uv}=\frac{w(u,w)}{d_{w,out}(u)}$. Now at $t=1$, a node $u_1$ whose out-degree is 2 with outgoing edges of weight 1 is considered less important than a node $u_2$ with out-degree 6 and all 6 outgoing edges of weight 1, because $u_1$ contributes a probability $\frac{1}{2}$ instead of $\frac{1}{6}$ for $u_2$, therefore reducing the uncertainty from $\log_2 6 \approx 2.585$ to $\log_2 2 = 1$. 
But if a node $u_3$ has two outgoing edges, one with weight 2 and one with weight 3, then its entropic centrality is $-\frac{2}{5}\log_2\frac{2}{5}-\frac{3}{5}\log_2\frac{3}{5} \approx 0.97 < 1$. Thus the node $u_3$ with weighted outgoing edges has lower entropic centrality than the node $u_1$ with unweighted outgoing edges. This is meaningful from an entropy view point, since it captures uncertainty, and the uniform distribution over the outgoing edges gives the most uncertainty about the random walk. However this may not be desired if the given scenario requires nodes with high weighted out-degree to be more central than others, e.g., when using the centrality measure as a proxy for determining influence. It is well known that centrality measures for unweighted graphs can be adapted to weighted graphs, but at the risk of changing their meaning, and that one way to remedy this is by the introduction of a weight parameter, see e.g. \cite{OAS}.

\subsection{Weighted Markov Entropic Centrality}

In order to capture the weights of the outgoing edges in the current framework of Markov entropic centrality, we introduce two tuning parameters, a conversion function $\alpha:w(e)\rightarrow \alpha(w(e))$ and a node weight function $\mu:v \rightarrow \mu(v)$.

The conversion function $\alpha:w(e)\rightarrow \alpha(w(e))$ adjusts the transition matrix $P_{\alpha(w)}$ such that $P_{\alpha(w),uv}=\frac{\alpha(w(u,w))}{d_{\alpha(w),out}(u)}$ (with $d_{\alpha(w),out}(u)=\sum_{v\in\Nc_u}\alpha(w(u,v))$) depending on the importance that weights are supposed to play compared to edges. The two obvious choices for $\alpha$ are $\alpha(w(e))=1$ for all edges $e$, which treats the weighted graph as unweighted, and $\alpha(w(e))=w(e)$ which considers the graph with its given weights, more generally one could define $ \alpha(w(e))=w(e)^\beta$ for some parameter $\beta$.
This formulation has the flexibility to give more or less importance to weights with respect to edges. It is enough here to consider the case where weighted edges are more important than unweighted ones. 

The conversion function $\alpha$ allows the weights to play a role in the transition matrix $P_{\alpha(w)}$, based on the scenario needs, but it does not address the issue of defining entropic centrality for weighted graphs.
To do so, we need a tuning parameter (e.g., \cite{OAS}) within the definition of entropic centrality, that maintains the semantics and meaning of the notion of entropy. We use the node weight function $\mu:v \rightarrow \mu(v)$ and propose the notion of weighted Markov entropic centrality, which is inspired by the concept of weighted entropy \cite{G71}.
\begin{defn}
The {\em weighted Markov entropic centrality} $C_{\alpha(w),H}^{t}(u)$ of a node $u$ at time $t$ is defined to be	
\begin{equation}\label{eq:wHij}
- \sum_{v\in V} \mu(v)(\tilde{p}_{\alpha(w),uv}^{(t)}+p_{uv'}^{(t)})\log_2 (\tilde{p}_{\alpha(w),uv}^{(t)}+p_{uv'}^{(t)}),
\end{equation}
where $\tilde{p}_{\alpha(w),uv}^{(t)}$ is the probability to reach $v$ at time $t$ from $u$, for $v$ in $V$, taking into account the weights $\alpha(w(e))$ for every edge in $E$. 
Auxiliary nodes defined for the unweighted case are still present
and $p_{uv'}^{(t)}$ is the probability to reach an auxiliary node $v'$ from $u$ at time $t$, which depends on the choice of the absorption probability matrix $D$, as in the unweighted case.  The probabilities $\tilde{p}_{\alpha(w),uv}^{(t)}+p_{uv'}^{(t)}$ are obtained from the matrix $\hat{P}^t_{\alpha(w)}$ (use $P_{\alpha(w)}$ instead of $P$ in Lemma \ref{lem:pt}).
\end{defn}

Before discussing the choice of $\mu(v)$, we recall that when $t\rightarrow \infty$, as for the unweighted case, the terms in column $v$ of $\hat{P}^{t}_{\alpha(w)}$ becomes $0$, and we are left with the term in column $v'$, of the form 
\[
\pi_{\alpha(w),uv}:=
\left\{
\begin{array}{ll}
\sum_{t\geq 1}\tilde{p}_{\alpha(w),uv}^{(t)}p_{vv'}& u \neq v \\
(\sum_{t\geq 1}\tilde{p}_{\alpha(w),uv}^{(t)}+1)p_{vv'}& u = v,\\
\end{array}
\right. 
\]
asymptotically yielding the entropic centrality 
\[
C_{\alpha(w),H}^{\infty}(u) = -\sum_{v\in V}\mu(v)\pi_{\alpha(w),uv}\log_2(\pi_{\alpha(w),uv}).
\]
Lemma \ref{lem:piuv} holds, therefore similarly to the unweighted case, we have that if the probability of absorption is $p_{uu'}=a$, then repeating the arguments leading to (\ref{eq:paupper}) yields: 
\begin{eqnarray}
\sum_w \mu(w)\pi_{uw}\log_2(\tfrac{1}{\pi_{uw}}) 
& = & \mu(u)\pi_{uu}\log_2(\tfrac{1}{\pi_{uu}}) + \sum_{w \neq u}\mu(w)\pi_{uw}\log_2(\tfrac{1}{\pi_{uw}}),\nonumber\\
& \leq & \max_{v\in V}\mu(v)(0.53074 + (1-a)\log_2(\tfrac{n-1}{1-a})) 
\label{eq:paw}
\end{eqnarray}
which shows that while $\max_{v\in V}\mu(v)$ may increase the overall centrality, it remains true, as for the unweighted case, that increasing $a$ just reduces the overall centrality. Therefore we will continue to use $p_{uu'}=\tfrac{1}{d_{out}(u)+1}$ as absorption probability.

The computation of $C_{\alpha(w),H}^{t}(u)$ involves a sum over all nodes $v\in V$, including  $v=u$, and the probability to go from $u$ to every $v$. The weight $\mu(v)$ captures the influence of the reached node $v$: if we reach influential nodes $v$ of weight $\mu(v)$ in $t$ time, then $u$ should be more central than if the only nodes we reach from it have no influence themselves. 
We define $\mu(v)$ to be $\mu(v)=\left(\tfrac{d_{w,out}(v)}{d_{out}(v)}\right)^\gamma$ if $d_{out}(v)\neq 0$. When $d_{out}(v) = 0$, $d_{w,out}(v)=0$ and we set $\mu(v)=1$. The weighted Markov entropic centrality (\ref{eq:wHij}) gives an influence which is proportional to the ratio $\left(\tfrac{d_{w,out}(v)}{d_{out}(v)}\right)^\gamma$, while $\gamma$ gives a way to amplify or reduce this influence. 

If all weights are 1, then the ratio simplifies to 1, and we are back to the non-weighted definition of Markov entropic centrality (as also with $\gamma=0$). Other possible choices for $\mu(v)$ could be explored, such as $\mu(v)=\tfrac{\log_2d_{w,out}(v)}{\log_2d_{out}(v)}$. Indeed, $\log_2d_{w,out}(v)$ would be the influence that $v$ would have had over its neighbors in terms of entropic centrality if it had $d_{w,out}(v)$ neighbors of weight 1, while $\log_2d_{out}(v)$ is the entropic centrality over the neighbors of $v$, ignoring the weights.

If the weights are normalized such that the lowest weight is mapped to 1, the entropic centrality values obtained when using a non-trivial function for $\mu(v)$ will necessarily be at least as much as obtained with $\mu(v)=1$.

The centralization as computed in Lemma \ref{lem:centralization} can be generalized to the weighted case. In the unweighted case, comparison was done among graphs with the same number of vertices. In the weighted case, comparison is done among graphs with the same number of vertices, with given weights $\mu(v)$.
A graph that would give the maximum centrality consists again of a star graph, with center $v_1$, assuming that the leaves have a self-loop which can be weighted, in which case $d_{out}(v_i)=1$ for $i\neq 1$, and  $\mu(v_i)$ is any weight. The centrality of the leaves will be zero, but that of $v_1$ will be 
$\sum_{i=1}^n \mu(v_i)p_i\log_2(1/p_i)$. However maximizing this quantity over $p_i$ does not have a closed form expression. Loose bounds can be computed though: $\frac{1}{n}\log(n)\sum_{i=1}^n \mu(v_i)\leq\max_{p_1,\ldots,p_n}\sum_{i=1}^n \mu(v_i)p_i\log_2(1/p_i) \leq 0.53074 \sum_{i=1}^n \mu(v_i)$ (see before (\ref{eq:paupper}) for a bound on $p\log_2(1/p)$).

\subsection{A Weighted Markov Entropic Centrality Analysis of a Cocaine Dealing Network}

We consider the cocaine dealing network \cite{cdn}, a small directed weighted graph coming from an investigation into a large cocaine traffic in New York City. It involves 28 persons (nodes), and 40 directed weighted edges indicate communication exchanges obtained from police wiretapping.

\begin{figure}
\centering
\includegraphics[scale=0.25]{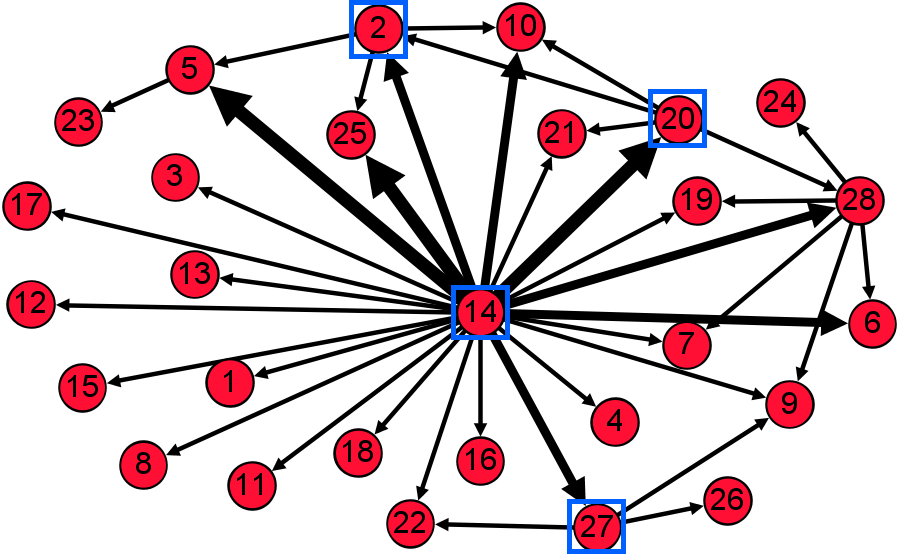}
\caption{\label{fig:gcd}
The cocaine dealing network \cite{cdn}: weighted edges are drawn with quantized girth. Figure~\ref{fig:gcdplot} shows the (weighted) Markov entropic centrality of the nodes 2, 27, 20 and 14. 
}
\end{figure}

The (weighted) Markov entropic centrality depends on the choice of the absorption matrix $D$, we thus start by looking at how a change in $D$ influences the entropic centralities. We keep the same choices for $D$ as for the unweighted case, namely $D=0.001{\bf I}_{28}$, $D=0.2{\bf I}_{28}$, $D=0.5{\bf I}$ and $D$ such that $D_{uu}=\tfrac{1}{d_{w,out}(u)+1}$, since when $w(e)=1$ for all edges, then $d_{w,out}(u)=d_{out}(u)$ for all nodes.

\begin{figure}
\centering
\includegraphics[scale=0.49]{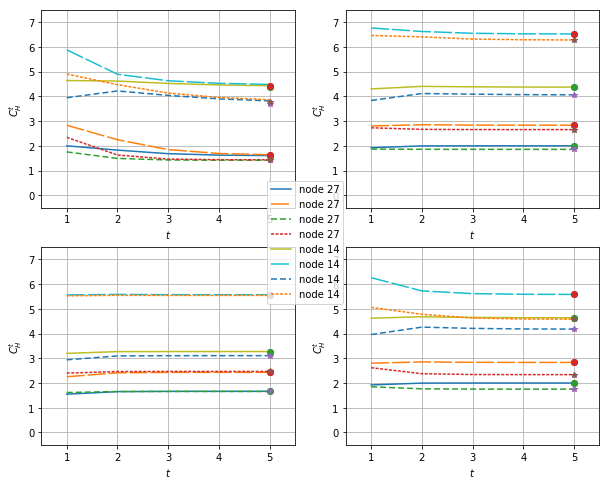}
\caption{\label{fig:gcdplot}
The Markov entropic centrality $C_{\alpha(w),H}^t(u)$ for $u=27,14$ (from Figure~\ref{fig:gcd}) for $t=1,\ldots,5$: on the upper left, $D=0.001{\bf I}_{28}$, on the upper right, $D=0.2{\bf I}_{28}$, on the lower left, $D=0.5{\bf I}$ and $D$ with $D_{uu}=\tfrac{1}{d_{w,out}(u)+1}$ on the lower right.
}
\end{figure}
 
In Figure~\ref{fig:gcdplot}, we plot the (weighted) Markov entropic centrality $C_{\alpha(w),H}^t(u)$ for $u=27,14$ for $t=1,2,3,4,5,6$, for the 4 choices of absorption probabilities $D$. For $u=27,14$, 4 variations of Markov entropic centralities are considered: $\alpha(w)=1$ and $\mu(v)=1$ (straight lines), corresponding to the unweighted case, $\alpha(w)=1$ and $\mu(v)=\left(\tfrac{d_{w,out}(v)}{d_{out}(v)}\right)$ (long dash lines)  for the case where the transition matrix $P$ is used, $\alpha(w)=w$ and $\mu(v)=1$ (short dash lines)  to show how centrality is computed based purely on $P_w$, and finally $\alpha(w)=w$ with $\mu(v)=\left(\tfrac{d_{w,out}(v)}{d_{out}(v)}\right)$ (dotted lines), for which $P_w$ is used for the transition matrix, together with the weighted entropic centrality.
Nodes 27 and 14 are different in nature, in that node 14 is a hub, with high degree compared to the rest of the other nodes, while node 27 has only one incoming edge and two outgoing edges.
They are chosen as representatives of nodes of respectively high and low degree (even though the degree is not the only contributing factor in the node centrality, it still plays an important role for small values of $t$). For each of the 4 plots, the 4 upper lines characterize the behavior of node 14, and the 4 lower lines the one of node 27. We observe that the behavior of the entropic centralities when $\mu(v)=1$ are similar to what was already observed for the karate club network: for $D=0.2{\bf I}_{28}$ and $D=0.5{\bf I}$, the centralities are flattened because the path uncertainty is reduced when the absorption probabilities are increasing (as shown in (\ref{eq:paw})), while for $D$ such that $D_{uu}=\tfrac{1}{d_{w,out}(u)+1}$, the gap among the centralities is (slightly) increasing. When $\mu(v)=\tfrac{d_{w,out}(v)}{d_{out}(v)}$, both centralities are amplified. Since node 14 has many outgoing edges, with weights including 10, 11, 14 18, 19, the introduction of $\mu(v)$ creates a higher amplification for node 14 than for node 27, whose edge weights are all less than 5. Zooming at time $t=2$ for node 14 and $D=0.001{\bf I}_{28}$, we notice that the short dashed line has a peak, before going down. This is explained by the fact that when $\alpha(w)=w$ and $\mu(v)=1$, the entropic centrality is that of a node with degree amplified by the weights, thus creating an initial jump in the entropy. However at the next time, several reached nodes have in turn very few (or no) neighbors, and edges of low weights, and this leads to a saturation effect.  This behavior is less prominent for other choices of $D$, where the absorption probability is high, and hence the effect of the edge weights is attenuated.

\begin{figure}
\centering
\includegraphics[scale=0.49]{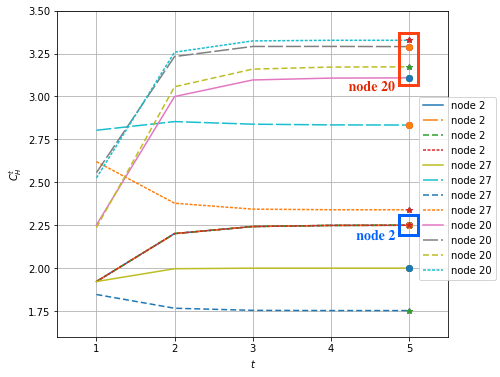}
\caption{\label{fig:gcdw1plot}
The Markov entropic centrality $C_{\alpha(w),H}^t(u)$ with $D_{uu}=\tfrac{1}{d_{w,out}(u)+1}$: for $u=2,27,20$ (from Figure~\ref{fig:gcd}), for $t=1,\ldots,5$ and for $(\alpha(w),\mu(v))=(1,1)$ (straight lines), $(1,\tfrac{d_{w,out}(u)}{d_{out}(u)})$ (long dash lines), $(w,1)$ (short dash lines) and $(w,\tfrac{d_{w,out}(u)}{d_{out}(u)})$ (dotted lines). 
}
\end{figure}

We then focus on the case where $D_{uu}=\tfrac{1}{d_{w,out}(u)+1}$, which depends only on the network setting, instead of other choices of $D$ which are parameters whose range can be explored one by one. With this choice of $D$, we consider the (weighted) Markov entropic centrality for $u=2,27,20$, as displayed in Figure~\ref{fig:gcdw1plot}. These nodes have outgoing edges with weights $1,1,1$, for $u=2$, weights $4,3,1$ for $u=27$, and weights $2,1,1,1$ for $u=20$. These nodes are chosen because they have similar out-degrees, but different weighted out-degrees. The same 4 scenarios as above are considered. For node 2, its edge to 10 stops at 10, its edge to 5 has only one connection to node 23 which has weight 1, and its last edge goes to 25, which has no connection either, therefore the 4 centralities are actually of the same quantity, and there all the 4 lines are coincident. For node 27, since it has the same out-degree as node 2, in the unweighted case, it starts at the same centrality as node 2. However it is even less influential since none of its own neighbors have neighbors. In the 3rd scenario, its entropic centrality is even lower than in the unweighted case, which is normal, since in this case, the walk distribution is not uniform anymore. In the two cases where $\mu(v)=\tfrac{d_{w,out}(v)}{d_{out}(v)}$, we see a jump in the centrality, explained by the weights of the outgoing edges which are higher than for node 2. However, the centrality of node 27 remains below that of 20, whose weighted out-degree is less than that of 27, but its out-degree is actually more.   

\subsection{Interpreting Entropic Centrality with a Bitcoin Subgraph}
\label{subsec:bitcoin178}

We next consider a small subgraph extracted from the bitcoin network comprising 178 nodes and 250 edges. Nodes are bitcoin addresses, and there is an edge between one node to another if a bitcoin payment has been made. Since the proposed Markov entropic centrality measure is suitable for undirected, directed and weighted graphs, we look at the chosen bitcoin subgraph as an undirected unweighted graph by ignoring the direction of transactions, as a directed unweighted graph by just considering whether any transactions exists, and as a  directed weighted graph (with $\alpha(w(e))=w(e)$,  $\mu(v)=\tfrac{d_{w,out}(v)}{d_{out}(v)}$) to capture the effect of transaction amount. 

\begin{figure*}
	\centering
        \subfloat[]{\label{fig:undirCH} 
		\includegraphics[width=0.3\textwidth]{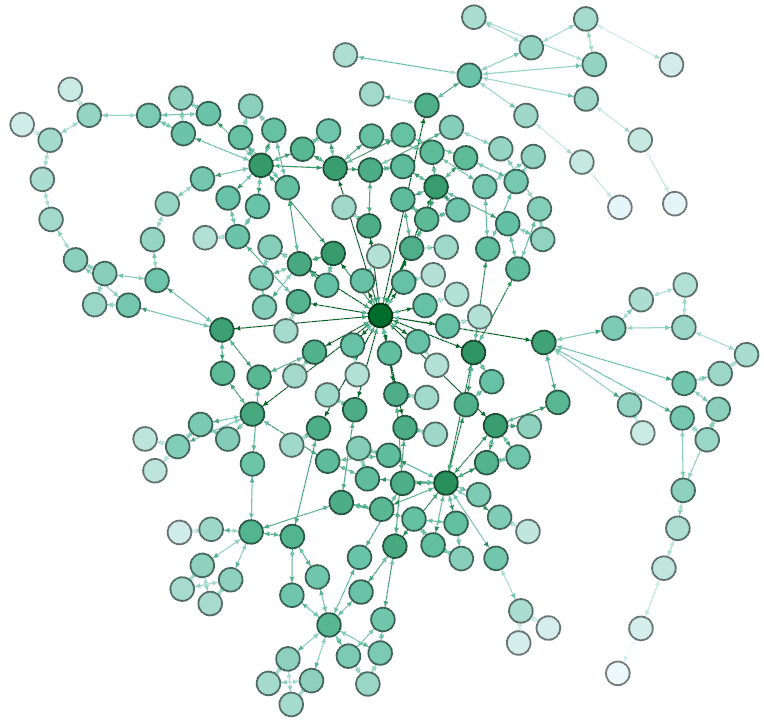}}
        ~ 
	\subfloat[]{\label{fig:dirCH} 
		\includegraphics[width=0.3\textwidth]{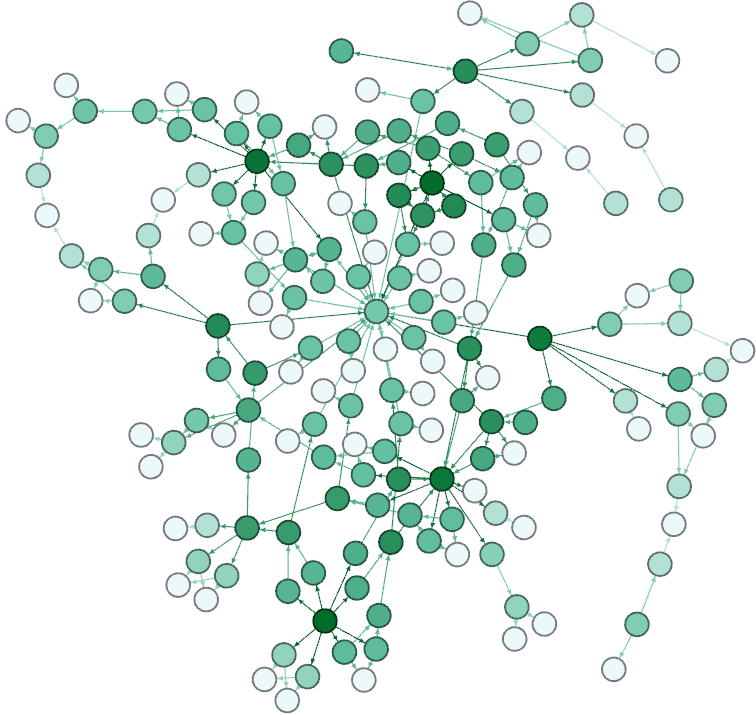}}
        ~
	\subfloat[]{\label{fig:weightCH}
		\includegraphics[width=0.3\textwidth]{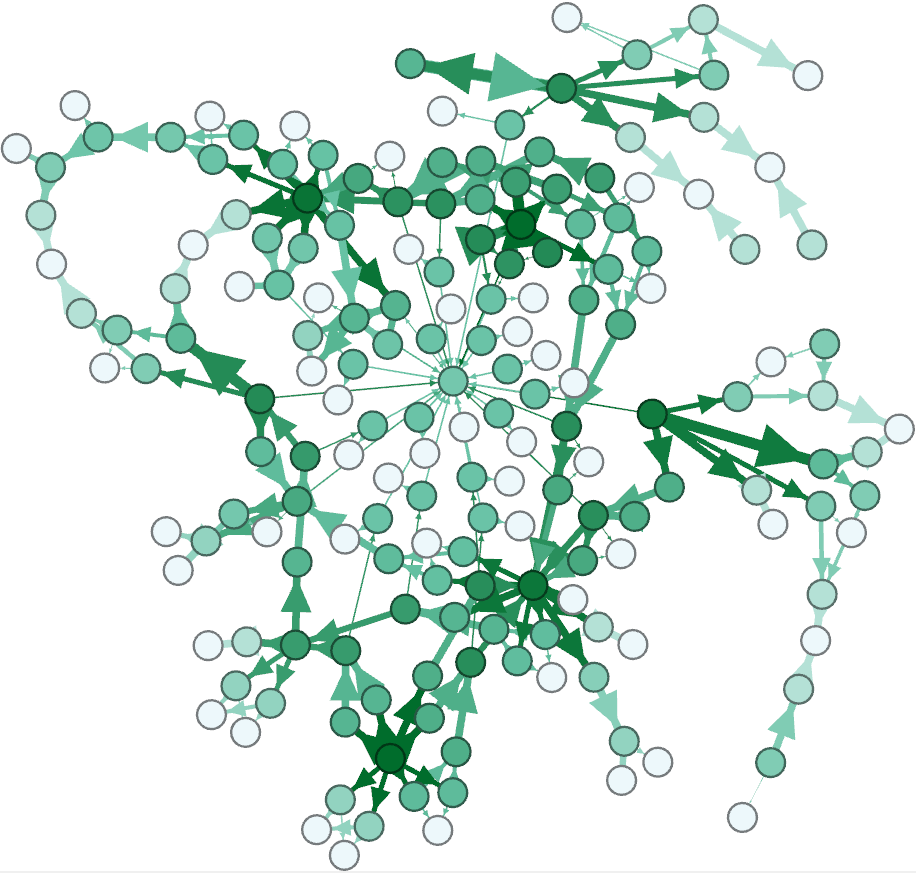}}
	\caption{\label{fig:entropybitcoin}
		Asymptotic Markov entropic centralities for a 178 node subgraph of the bitcoin network: undirected and unweighted on the left, directed and unweighted in the middle, directed and weighted on the right. The top 10\%, next 20\%, 30\% and remaining 40\% nodes in terms of their entropic centrality are shown in progressively lighter colors.} 
\end{figure*}

In Figure \ref{fig:entropybitcoin} we show the three variations with nodes colored according to their Markov entropic centrality at $t=\infty$ (asymptotic). The darker the node color, the higher the Markov entropic centrality.
In subfigure \ref{fig:undirCH}, one hub stands apart, with the highest entropic centrality, which is a node which is highly connected to other nodes. If we look at the same node in subfigure \ref{fig:dirCH}, we see that it is not central anymore: this is because in the directed graph, this node is actually seen to receive bitcoin amounts from many other nodes, so as far as sending money is concerned, it has very little actual influence (in the same graph but with edge directions reversed, this node would have had the highest centrality). On the other hand, nodes that were not so prominent in the undirected graph appear now as important. The graph in subfigure \ref{fig:dirCH} highlights nodes which are effectuating many bitcoin transactions. Now one node may do several transactions with little bitcoin amount, we may want a node that does few but high amount transactions to be more visible, which is illustrated in subfigure \ref{fig:weightCH}.
It turns out that for the subgraph we studied, high centrality nodes in the unweighted case remain high centrality nodes in the weighted case, meaning that nodes performing many bitcoin transactions are also those sending high amount transactions.

For the sake of completeness, we provide Kendall-$\tau$ rank correlation coefficients among different variations of (weighted) Markov entropic centralities for the directed graph, for $t=\infty$: for $(\alpha(w(e))=1,\mu(v)=1)$ (shown on subfigure \ref{fig:dirCH}), for $(\alpha(w(e))=w(e),\mu(v)=\tfrac{d_{w,out}(v)}{d_{out}(v)})$ (shown on subfigure \ref{fig:weightCH}), but also for $(\alpha(w(e))=w(e),\mu(v)=1)$, and for $(\alpha(w(e))=1,\mu(v)=\tfrac{d_{w,out}(v)}{d_{out}(v)})$. The out-degree centrality is also tested. A Kendall-$\tau$ coefficient is a measure of rank correlation: the higher the measure, the higher the pairwise mismatch of ranks across two lists.
Results are shown in Table \ref{tab:kendalltau}. Above the diagonal, coefficients are obtained using all 178 nodes. Below the diagonal, only 67 nodes are used, which are in the union of the top 20\% nodes for each of the 5 aforementioned centrality measures. In the given graph, many (peripheral) nodes have the same (low) entropic centrality, so when all nodes are considered for ranking correlation, the average is misleadingly low. Since high centrality nodes are often of interest, ranking variations among the top nodes is pertinent, and we observe that (i) out-degree is not a good proxy for most entropic centrality variants, and (ii) the different variations yield significantly distinct sets of high centrality nodes, corroborating the need for our flexible entropic centrality framework.

\begin{table}[]
	\centering
	\begin{tabular}{p{1.4cm}lllp{1.3cm}p{1.4cm}}
		\cellcolor{AnotherGray}\textbf{$\alpha$,$\mu$} & \cellcolor{ashgrey}1,1 & \cellcolor{ashgrey} $d_{out}$ & \cellcolor{ashgrey} $w$,1 & \cellcolor{ashgrey} 1,$\tfrac{d_{w,out}(v)}{d_{out}(v)}$ & $w$,$\tfrac{d_{w,out}(v)}{d_{out}(v)}$ \cellcolor{ashgrey} \\
		\cellcolor{ashgrey} 1,1 &  \cellcolor{LightGray}   0                     & \cellcolor{beaublue}0.159 & \cellcolor{beaublue}0.075 & \cellcolor{beaublue}0.084  & \cellcolor{beaublue}0.086 \\
		\cellcolor{ashgrey} $d_{out}$ 
		& \cellcolor{desertsand}0.419&   \cellcolor{LightGray}   0                    & \cellcolor{beaublue}0.193 & \cellcolor{beaublue}0.214 & \cellcolor{beaublue}0.218 \\
		\cellcolor{ashgrey} $w$,1 & \cellcolor{desertsand}0.315 & \cellcolor{desertsand}0.60 &      \cellcolor{LightGray}   0                 & \cellcolor{beaublue}0.120 & \cellcolor{beaublue}0.050 \\
		\cellcolor{ashgrey} 1,$\tfrac{d_{w,out}(v)}{d_{out}(v)}$ & \cellcolor{desertsand}0.186 & \cellcolor{desertsand}0.538 & \cellcolor{desertsand}0.305 &    \cellcolor{LightGray}  0                    & \cellcolor{beaublue}0.089 \\
		\cellcolor{ashgrey} $w$,$\tfrac{d_{w,out}(v)}{d_{out}(v)}$ & \cellcolor{desertsand}0.346 & \cellcolor{desertsand}0.671 & \cellcolor{desertsand}0.097 & \cellcolor{desertsand}0.283 &  \cellcolor{LightGray}  0                     
	\end{tabular}
	\caption{\label{tab:kendalltau}
		Kendall-$\tau$ coefficients of 5 centrality measures: 4 of them are the Markov entropic centralities for different values of $\alpha(w(e))$ and $\mu(v)$; $d_{out}$ is the out-degree centrality. Above the diagonal, coefficients are obtained using all 178 nodes. Below the diagonal, only 67 nodes are used, which are in the union of the top 20\% nodes for each of these five centrality measures.}
\end{table}

\section{Entropic Centrality Based Clustering}\label{sec:clustering}

We next explore whether and how the local communities inferred from the gradients observed in the evolution of the entropic centralities as a function of $t$ (see Section \ref{sec:single}) can be exploited to realize effective clustering. 

The formulas (\ref{eq:Hij}) and (\ref{eq:wHij}) for the (weighted) Markov entropic centralities involve the sum of probabilities $p_{uv}^{(t)}+p_{uv'}^{(t)}$ and $p_{\alpha(w),uv}^{(t)}+p_{\alpha(w),uv'}^{(t)}$ respectively. We will use $\hat{P}$ to denote the matrix whose $u$th row contains the coefficient $p_{uv}^{(t)}+p_{uv'}^{(t)}$ in the unweighted case, and $p_{\alpha(w),uv}^{(t)}+p_{\alpha(w),uv'}^{(t)}$ in the weighted case, in column $v$. Since the algorithm that we describe next uses $\hat{P}$ in the same manner, irrespectively of $t$ (though we will focus on the case $t=\infty$) and of whether there is a weight function, we keep the same notation $\hat{P}$.

The proposed algorithm works in two stages: first, we create local clusters around `query nodes', and then, we carry out a hierarchical agglomeration. The choice of query nodes in our approach is informed by the entropic centrality, which we describe first (Subsection \ref{sec:querynodechoice}), before we elaborate the algorithm (in Section \ref{sec:clusteringalgo}).

\subsection{Query Node Selection}
\label{sec:querynodechoice}

In the initial step of the algorithm, we look for a cluster around and inclusive of a specific query node, similarly to \cite{BNCZ}. In \cite{BNCZ} though, it was not possible a priori to determine suitable query nodes (e.g., a node could be at the boundary of two clusters, and thus yield the union of subsets of these clusters as a resultant cluster), and hence multiple constrained random walkers were deployed to increase the confinement probability of the random walks within a single cluster. In contrast, our choice of query nodes is informed by their Markov entropic centrality.

Let us first formalize some of the information captured by the notion of entropy.

\begin{prop}\label{lem:H}
Given the probabilities $p_1\leq p_2 \leq \ldots \leq p_m < p_{m+1}=\frac{1}{n}=\ldots = p_{m+s} < p_{m+s+1} \leq \ldots \leq p_n$, $p_i \geq 0$, $\sum_{i=1}^n p_i =1$, and $\gamma > 1$ such that $\sum_{i=1}^n p_i \log_2(1/p_i)=\frac{\log_2(n)}{\gamma}$, we have:
\begin{enumerate}
\item
$s \leq \tfrac{n}{\gamma}$, and $n-m-s \geq n\tfrac{\tau-1}{\tau}\tfrac{\gamma-1}{\gamma}$ for $\tau > 1$,
\item
$\sum_{i=m+s+1}^n p_i > \tfrac{\tau-1}{\tau}\tfrac{(\gamma-1)}{\gamma} \iff \sum_{i=1}^m p_i < 1 - \frac{s}{n} - \tfrac{\tau-1}{\tau}\tfrac{(\gamma-1)}{\gamma}$.
\end{enumerate}
\end{prop}
\begin{IEEEproof}
Since we ordered the $n$ probabilities by increasing order, we write
\[
\sum_{i=1}^n p_i \log_2(1/p_i)=  \sum_{i=1}^m p_i \log_2(1/p_i)+ \frac{s}{n} \log_2(n)  + 
\sum_{i=m+s+1}^n p_i \log_2(1/p_i),
\]
and $\sum_{i=1}^n p_i \log_2(1/p_i)=\frac{\log_2(n)}{\gamma}$ for some $\gamma > 1$ certainly implies 
$\frac{s}{n} \log_2(n) \leq \frac{\log_2(n)}{\gamma}$. 
Thus there must be $s \leq \tfrac{n}{\gamma}$ probabilities of the form $\tfrac{1}{n}$ (in particular, if $\gamma > n$, $s=0$, and likewise, when $\gamma > 1$ , $s\leq n-1$). In turn, we must have $n-s$ which is at least $\tfrac{n(\gamma-1)}{\gamma}$ probabilities shared between $p_1,\ldots,p_m$ and $p_{m+s+1},\ldots,p_n$, and they must sum up to $1-\tfrac{s}{n}=\tfrac{n-s}{n}$. 

It is not possible that all $n-s$ probabilities belong to the group $p_1,\ldots,p_m$ (i.e., $< \frac{1}{n}$), because if all of them were strictly less than $\frac{1}{n}$, we would have at most $\frac{s}{n}+(n-s)p_m < \frac{s}{n}+\frac{n-s}{n}<1$ (a similar argument holds for not having all the probabilities in the group $p_{m+s+1},\ldots,p_n$). More generally, this is saying that deficiency from $1/n$ among the $p_1,\ldots,p_m$ has to be compensated by the $p_{m+s+1},\ldots,p_n$:
\[
\sum_{i=1}^m p_i = 1 -\frac{s}{n}-\sum_{i=m+s+1}^n p_i,
\]
where $0 < \sum_{i=1}^m p_i < \frac{n-s}{n}$, and accordingly $\sum_{i=m+s+1}^n p_i$ varies from being strictly less than $1-\frac{s}{n}$ to strictly more than $0$. 

We can refine these ranges using the number of probabilities we are allowed in each sum. Suppose that among the $n-s$ probabilities that are left to be assigned, $\frac{1}{\tau}(n-s)$ are the probabilities $p_1,\ldots,p_m$ for some $\tau > 1$, and the rest, namely $\tfrac{\tau-1}{\tau}(n-s) \geq \tfrac{\tau-1}{\tau}\tfrac{n(\gamma-1)}{\gamma}$ are the probabilities $p_{m+s} < p_{m+s+1} \leq \ldots \leq p_n$.

For at least $\tfrac{\tau-1}{\tau}\tfrac{n(\gamma-1)}{\gamma}$ probabilities strictly more than $1/n$, we have:
\[
\sum_{i=m+s+1}^n p_i > \tfrac{\tau-1}{\tau}\tfrac{(\gamma-1)}{\gamma}
\iff \sum_{i=1}^m p_i < 1 - \frac{s}{n} - \tfrac{\tau-1}{\tau}\tfrac{(\gamma-1)}{\gamma}.
\]
\end{IEEEproof}

We apply this proposition to $\pi_{uv}$, for a given node $u$, and for $v$ ranging over possible nodes in $V$. Let $\tau>1$ be a parameter that decides the proportion of probabilities less than $\frac{1}{|V|}$ (the larger $\tau$, the smaller the proportion), and let $\gamma>1$ be a parameter that characterizes how small the Markov entropic centrality of a node $u$ is (the larger $\gamma$, the lower the entropic centrality). The result says that given $\gamma$, there is a tension between agglomeration of small probabilities on the one hand and agglomeration of large probabilities on the other hand: the larger the agglomeration of large probabilities (as a function of $\tau$ and $\gamma$), the lower the agglomeration of small probabilities, and conversely. Furthermore, as $\gamma$ grows, it creates an agglomeration of large probabilities which is detached from that with small probabilities. This suggests that nodes with a low entropic centrality have local clusters where ties are relatively strong, while

nodes tend to have a relatively high entropic centrality when they either are embedded within a large cluster, or when they are at the boundary of two clusters (and could arguably belong to either). Accordingly, we choose to start identifying local communities by choosing the low entropic centrality nodes as query nodes. 

\begin{figure}
\centering
\subfloat[A high centrality query node.]{\label{fig:highcentral}
		\includegraphics[width=0.4\textwidth]{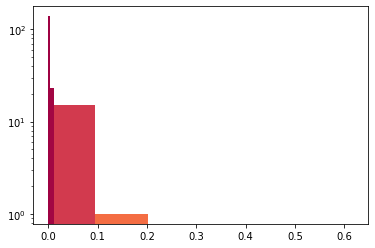}}
		~
\subfloat[A low centrality query node.]{\label{fig:lowcentral}
		\includegraphics[width=0.4\textwidth]{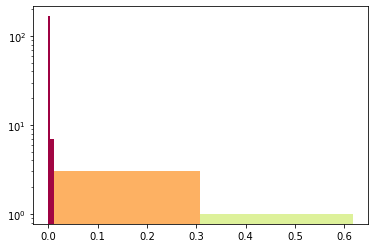}}
		
\caption{Histogram showing number of nodes ($y$-axis in log-scale) at which a random walker is absorbed for given (range of) values of absorption probability ($x$-axis) for the 178 node Bitcoin subgraph, treated as unweighted and undirected.\label{CvsP}
}
\end{figure}

This is illustrated in Figure \ref{CvsP} for the 178 node Bitcoin subgraph, treated as unweighted and undirected. We consider a node with high centrality on the left hand-side, it has centrality $\approx 5.08297$, and one with low centrality on the right hand-side (with centrality $\approx 1.73539$). By high and low, we mean that both belong to the top 10 nodes in terms of respective high/low centrality. The histograms show the probability that a random walker is (asymptotically) absorbed at a node. The $y$-axis is in logscale. The bins are placed at 0, $1/2|V|$, $2/|V|$, and then at half the maximum probability (as observed across the two cases) and the maximum probability. In the first bin, for both cases, there are more than 100 nodes which have a negligible probability of the random walker being absorbed. As predicted by Proposition \ref{lem:H}, for the node with high entropic centrality, they are less spread apart (and there are more probabilities close to $1/|V|$) than for the low entropic centrality node, which has, furthermore one high probability (between 0.3 and 0.6). Such higher probabilities of absorption at a few nodes in the graph are precisely what we use as a signal for a local community structure.

\begin{figure}
\begin{center}
\includegraphics[width=0.42\textwidth]{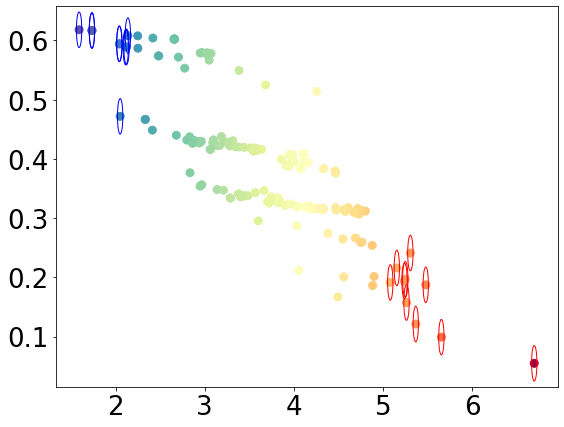}
\caption{\label{scatter}
On the $x$-axis, the centrality of nodes, on the $y$-axis, the value of the highest probability of absorption at any node, for a random walker starting at that node (in the 178 node Bitcoin subgraph, treated as unweighted and undirected). The two vertical bars demarcate the top 30\% and 40\% of nodes in terms of entropic centrality. Nodes that are circled have highest/lowest centralization.
}
\end{center}
\end{figure}

This finding is consistent with the hypothesis formulated at the end of Section \ref{sec:karatecentrality}, namely that if there is no significant change in the entropic centrality over time, then the node belongs to a small community, while larger entropic centralities indicate nodes that are well connected (possibly indirectly) to many nodes rather than being strongly embedded in a particular well defined community. In Figure \ref{scatter} we show a scatter diagram illustrating, again for the 178 Bitcoin subgraph, the relations between the entropic centrality of a node and the highest probability of absorption at any node for a random walker starting at the respective node. We see that nodes with high centralities have as highest absorption probability, values below 0.3, while low entropic centrality nodes have highest absorption probability of more than 0.4. For this subgraph, the centralization sequence has a minimum of -0.27962, a maximum of 0.40350, a mean of 0 and a median of 0.00560. The nodes that are ``circled" are those with highest centralization in red (centralization values more than 0.18), and those with lowest centralization in blue (centralization values less than -0.2).

The Markov entropic centrality score acts as a good summary of the detailed probability distributions of absorption for a random walker starting at different query nodes. Specifically, we see that using low entropic centrality nodes as query nodes would yield meaningful local clusters precisely by identifying the nodes at which the random walker is absorbed with high probability to constitute a local community. Such communities can then be further coalesced into larger (and fewer) clusters or re-clustered into smaller (and more numerous) communities, as deemed suitable, in a hierarchical manner. We will see, in our experiments, what is a suitable `low' or `high' entropic centrality, particularly, that this would depend on the distribution of the centrality scores. The distribution of the relative Markov centrality scores for a given graph is thus also a good indicator of whether the proposed approach would be effective for clustering that graph instance, and helps us reason about whether to use the proposed approach. This will be illustrated in our experiments, in particular in Subsection \ref{sec:groundtrutheva}. 

\subsection{Entropic Centrality Based Clustering Algorithms}
\label{sec:clusteringalgo}

{Keeping in mind the above discussion, we describe the \textbf{\emph{first stage}} (without the iterative hierarchical process) of clustering in Algorithms \ref{alg:querynodecentric} and \ref{alg:processrawcluster}. 

In Algorithm \ref{alg:querynodecentric}, we maintain $S_{cluster}$ as a current global view of clusters. We start from the lowest entropic centrality node as a query node, and repeat the process as long as there are nodes that do not already belong to some existing cluster (listed in $S_{cluster}$).

\begin{algorithm}[th]
	\caption{Probability distribution based graph clustering }\label{alg:querynodecentric}
	\begin{algorithmic}[1]
		\Procedure{ProbDistClustering}{$G=(V,E),N,t$}
		\LineComment \textcolor{gray}{$N << |V|$ stands for the top-$N$ most central nodes}
		\State Compute $\hat{P}$ and the entropic centrality of all $v\in V$. 
		\State{Assign $S_{HE}=\{$the top-$N$ entropic centrality nodes$\}$.}
		\LineComment \textcolor{red}{Initialization}
		\State{Set $Q$: nodes in ascending order of entropic centrality.}
		\State{Set $S_{cluster}=\varnothing$.}\Comment\textcolor{gray}{Current clusters}
		\LineComment \textcolor{red}{Global clustering}
		\While{$Q$ is not empty}
		\State Take the query node $v_q$ from $Q$'s head (remove it).
		\LineComment \textcolor{red}{Obtain query node centric local cluster}
		\State Apply a(ny) clustering algorithm on the row $(\hat{P}_{v_q,v})_{v\in V}$ of $\hat{P}$ to form the set $S_{ini,v_q}$ of clusters.
		\State Choose $S_{v_q}$ from $S_{ini,v_q}$ with the highest average transition probability (include $v_q$). \Comment\textcolor{gray}{$v_q$'s raw cluster}
		\LineComment \textcolor{red}{Prune the raw cluster $S_{v_q}$ using Algorithm \ref{alg:processrawcluster}.}
		\State ProcessRawCluster($S_{v_q},S_{HE},S_{cluster}$)
		\State $\forall v \in S_{v_q}$, remove $v$ from $Q$.
		\LineComment \textcolor{red}{Integrate the local result with the global view.}
		\If {$S_{v_q}$ intersects with any cluster(s) in $S_{cluster}$}              \State Merge them (update $S_{cluster}$ accordingly). 
		\Else ~Add $S_{v_q}$ to $S_{cluster}$.
		\EndIf
		\EndWhile
		\State \textbf{return} $S_{cluster}$
		\EndProcedure
	\end{algorithmic}
\end{algorithm}

We consider the transition probabilities from a query node $v_q$ to all the other nodes as per $\hat{P}$, and carry out a clustering of these (one-dimensional, scalar) probability values. 
Lemma \ref{lem:H} shows the existence of (up to) ``three clusters", formed by probabilities of values around $1/|V|$, and of values away from $1/|V|$, either by being smaller or larger enough. How clearly defined these clusters are depends on $\gamma$: if $\gamma$ is small, close to 1, probability values can still be close to uniform, on the other extreme, if $\gamma$ is large, there may be very few or no value around $1/|V|$, resulting in mostly two clusters.   
In our implementation, we use the Python scikit-learn agglomerative clustering to look for this initial set of (up to) three clusters $S_{ini,v_q}$. Among these clusters, we choose the cluster with the highest average transition probability from $v_q$. We define $S_{v_q}$ to be the nodes corresponding to} this cluster along with $v_q$ itself since, (i) the comprising nodes have similar probabilities for random walks to end up there when originating from $v_q$ (this follows from the clustering of the probability values), and crucially, these nodes are considered to comprise the immediate local community because (ii) this is the largest (in expectation) such probability. 

We consider the absorption probabilities for a random walker starting at $v_q$ to be absorbed at any of the nodes in $S_{v_q}$, and define the minimum of these values as $\sigma$ i.e.  $\sigma = \min_{v\in S_{v_q}}\hat{P}_{v_q,v}$. Thus, $\sigma$ can be understood as an effective threshold (deduced a posteriori) above which the probability of being absorbed in the local cluster of $v_q$ is high enough. Proposition \ref{prop:trans} and its corollary below show that if $v$ belongs to the local cluster of $v_q$, but $w$ should also belong to the local cluster of $v$, then provided that the absorption probability $p_{vv'}$ at $v$ is not too large ($p_{vv'}\leq \sigma$), $w$ will also belong to $S_{v_q}$, together with $v$.

\begin{prop}\label{prop:trans}
The probability $\pi_{uw}=\sum_{t\geq 1}\tilde{p}_{uw}^{(t)}p_{ww'} $ to start at $u$ and to be absorbed at $w\neq u$ over time is lower bounded by:
\[
\tilde{p}_{uw}p_{ww'}+\pi_{uv}\frac{\pi_{vw}}{p_{vv'}}, 
\]
\end{prop}
\begin{IEEEproof}
We have that $\pi_{uw}=\sum_{t\geq 1}\tilde{p}_{uw}^{(t)}p_{ww'} $ is the probability to start at $u$ and to be absorbed at $w$ over time. 
We start again with (\ref{eq:pabs}): 
\begin{eqnarray*}
\pi_{uw} 
 &=&  (\tilde{p}_{uw}+\sum_{t\geq 2}\sum_{i=1}^{t-1}\tilde{p}_{uv}^{(i)}\tilde{p}_{vw}^{(t-i)} + \sum_{t\geq 2}\sum_{i=1}^{t-1}\sum_{y\neq v}\tilde{p}_{uy}^{(i)}\tilde{p}_{yw}^{(t-i)} )p_{ww'} \\
& \geq &  (\tilde{p}_{uw}+\sum_{t\geq 2}\sum_{i=1}^{t-1}\tilde{p}_{uv}^{(i)}\tilde{p}_{vw}^{(t-i)} )p_{ww'} \\
& = &  (\tilde{p}_{uw}+\sum_{s\geq 1}\tilde{p}_{vw}^{(s)} \sum_{i\geq 1}\tilde{p}_{uv}^{(i)})p_{ww'} =  \tilde{p}_{uw}p_{ww'}+ \frac{\pi_{uv}}{p_{vv'}}\pi_{vw}
\end{eqnarray*}
since we recognize a Cauchy product, and we invoked Merten's Theorem where $\sum_{i\geq 1}\tilde{p}_{uv}^{(i)} \rightarrow \pi_{uv}$, $\sum_{s\geq 1}\tilde{p}_{vw}^{(s)}\rightarrow \pi_{vw}$, 
and both sequences $(\tilde{p}_{uv}^{(i)})_i,(\tilde{p}_{vw}^{(s)})_s$ are absolutely converging to 0.
\end{IEEEproof}
\begin{cor}
Given that the probability $\pi_{uv}$ to start at $u$ and to be absorbed at $v$ is more than a threshold $\sigma$, and that the probability to start at $v$ and to be absorbed at $w$ is also more than $\sigma$, then if $p_{vv'}\leq \sigma$, $\pi_{uw}\geq \sigma$.
\end{cor}
\begin{IEEEproof}
Suppose $\pi_{uv},\pi_{vw}\geq \sigma$, then by the above proposition:
\[
\pi_{uw} 
\geq \tilde{p}_{uw}p_{ww'}+\pi_{uv}\frac{\pi_{vw} }{p_{vv'}}
\geq \pi_{uv}\frac{\pi_{vw} }{p_{vv'}}
\geq \frac{\sigma^2}{p_{vv'}}
\]
so $p_{vv'} \leq \sigma$ implies $\pi_{uw}\geq \sigma$. Note that this reasoning is iterative, namely if now if we consider that from $u$, we got absorbed at $w$,  but also from $w$, we got absorbed at $y$, with $\pi_{yw}\geq \sigma$, then
\[
\pi_{uy} \geq \tilde{p}_{uy}p_{yy'}+\frac{\sigma^2}{p_{ww'}}
\]
and again $p_{ww'} \leq \sigma$ implies $\pi_{uy}\geq \sigma$.
\end{IEEEproof}

If the query node $v_q$ is a low entropic centrality node, it is expected that once a cluster $S_{v_q}$ is formed, nodes belonging to $S_{v_q}$ are unlikely to be high centrality nodes themselves (otherwise $v_q$ would likely have inherited the influence and would not be a low centrality node itself). However, if the relative difference between what are considered low or high entropic centrality is not high, then several well connected nodes may get clubbed together in the clustering process, further bringing in many other nodes, coalescing a large group of nodes in the early stage of the clustering. Likewise, if the query node $v_q$ belongs to a pre-designated group of high entropic centrality nodes $S_{HE}$, then there is a risk that it may inadvertently merge multiple clusters which one may want to be separate. These motivate the addition of a pruning algorithm described in Algorithm \ref{alg:processrawcluster}, which works as follows.

\begin{algorithm}[tb]
	\caption{Pruning of the raw cluster $S_{v_q}$.}
	\label{alg:processrawcluster}
	\begin{algorithmic}[1]
		\Procedure{ProcessRawCluster}{$S_{v_q},S_{HE},S_{cluster}$}
		\LineComment \textcolor{gray}{most central node list $S_{HE}$, current cluster list $S_{cluster}$}
		\If {$v_q \in S_{HE}$}
		
		\State Set $\{C_1,\ldots,C_r\}=\arg\max_{C\in S_{cluster}}|C\cap S_{v_q}|$.
                \If {$r > 1$}
                 $C'=rand{\{C_1,\ldots,C_r\}}$
                \Else~ $C'=C_1$
                \EndIf
                \State $S_{v_q} = S_{v_q}\backslash(\cup_{C\in S_{cluster}}(C \cap S_{v_q})\backslash C')$.
		\EndIf
		\If {$|S_{v_q}\cap (S_{HE}\backslash v_q)| > 1$}
		\LineComment \textcolor{gray}{$S_{v_q}$ contains multiple high entropy nodes beside $v_q$} 
		\State Among nodes in $S_{HE}\backslash v_q$, keep only the node(s) which have the highest transition probability from $v_q$ \LineComment \textcolor{gray}{nodes not in $S_{HE}$ are not affected}
		\EndIf
		\State \textbf{return} $S_{v_q}$
		\EndProcedure
	\end{algorithmic}
\end{algorithm}

If the query node $v_q$ is a high entropic centrality node, then we check the intersection of $S_{v_q}$ against existing clusters from $S_{cluster}$, and in case there are non-trivial intersections, and yet there is no unique cluster with which there is a largest intersection, we retain only one of the largest clusters as $S_{v_q}$ chosen randomly, since it otherwise risks merging clusters which ought to be distinct. Otherwise, we retain as $S_{v_q}$ the nodes from the largest intersection, as well as nodes that were in no intersection, but discard the rest. Irrespective of the query node, if there are multiple pre-designated high entropic centrality nodes in $S_{v_q}$ (other than $v_q$), we retain among these only the ones to which the transition probability from $v_q$ is highest. This pruned list is given to Algorithm \ref{alg:querynodecentric}, where it is checked against existing clusters in $S_{cluster}$, and if there is any intersection, then they are merged, otherwise, $S_{v_q}$ is added as a new cluster in $S_{cluster}$. Note that the pruning mechanism introduces an element of randomization in our algorithm. As such, all the reported results in this paper are based on ten experiment runs. For all but one of the graphs used in our experiments, the conditional statement which introduces the randomization was in fact never triggered, and thus the results are produced in a consistent fashion. Only some variations of the 178 nodes Bitcoin subgraph (results shown in Figure \ref{fig:178}) triggered the conditional statement: When it was treated as undirected and unweighted, the conditional statement was triggered but nevertheless resulted in same clusters across the ten experiment runs. For the same graph treated as directed but unweighted, the randomization yielded distinct but very similar cluster results (F-score 0.982 among the distinct results), and one of these result instances is shown.

Having identified localized query-node centric community structures, in the \textbf{\emph{second stage}}, we agglomerate these to identify clusters at different degrees of granularity. A single stage of agglomeration is almost identical to the initial clustering process described above, with the following subtle changes. The cluster results from the previous step are considered as the new nodes. We still only use the matrix $\hat{P}$, and hence we do not (need to) explicitly define edges connecting the clustered nodes. The new coalesced nodes are assigned an entropic centrality value corresponding to the average of the entropic centrality of their constituent nodes. For transition probabilities across clustered nodes (say $\bar C$ and $\tilde C$), we considered the minimum, mean and maximum transition probabilities amongst all node pairs $u\in \bar C,v\in \tilde C$ as per $\hat{P}$. Finally, we discard a specific agglomeration in case the resulting agglomerated cluster would not result in a (weakly) connected graph. Our experiments indicate that the best clustering results are obtained using the minimum transition probabilities, we only report the corresponding results next. 

\begin{figure*}
	\centering
	\subfloat[Edge removal (20 iterations).]{\label{fig:karateremoval} 
		\includegraphics[width=0.3\textwidth]{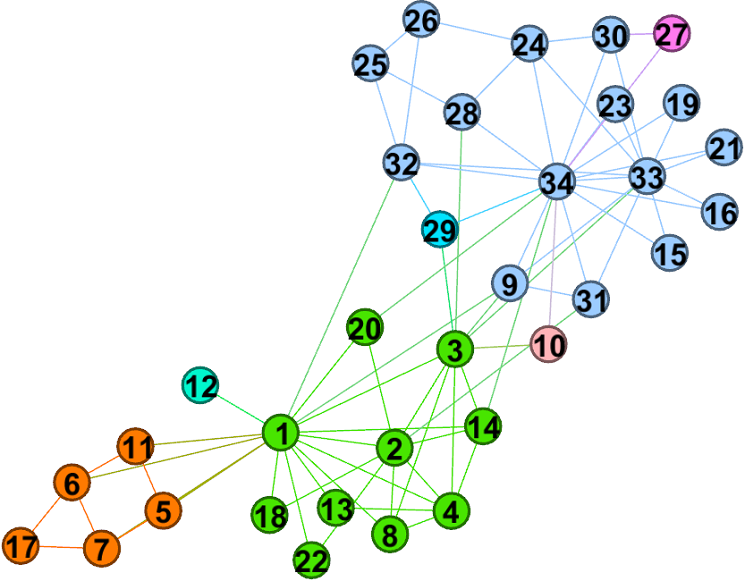}}
        ~
	\subfloat[Initial clusters (stage 1).]{\label{fig:kcgstep1} 
		\includegraphics[width=0.3\textwidth]{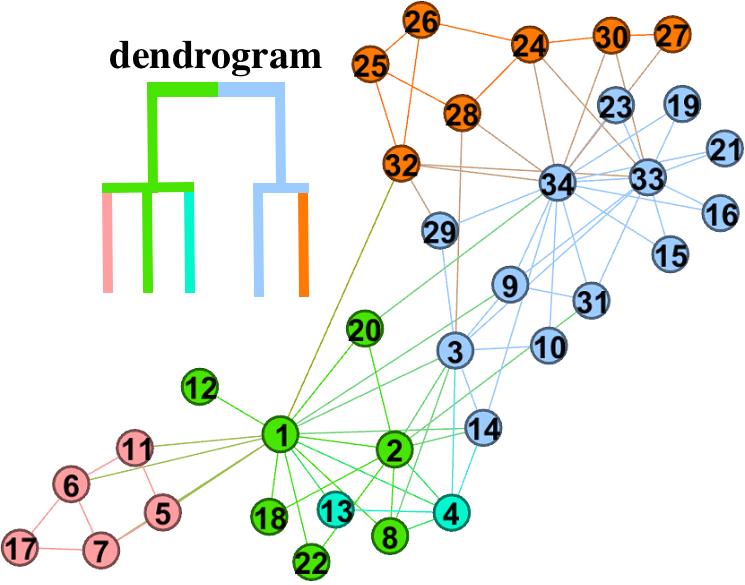}}
	~
	\subfloat[The final 2-cluster result (stage 2).]{\label{fig:kcgstep2}
		\includegraphics[width=0.3\textwidth]{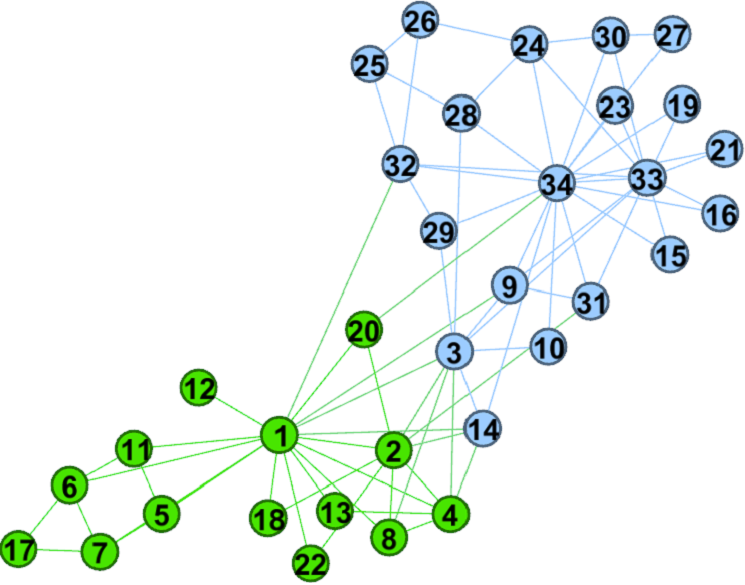}}

	\caption{\label{fig:cluskcg} Clustering of the karate club network: using the edge removal clustering of \cite{NRK} on the left, and the proposed algorithms in the middle (stage 1) and on the right (stage 2, with agglomeration).
	}
\end{figure*}
\subsection{Clustering of the Karate Club Network}

\begin{figure}
	\centering
	\includegraphics[scale=0.45]{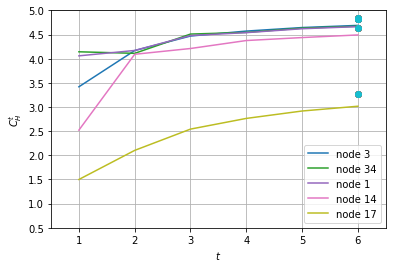}
	\caption{\label{fig:morekaratecentrality} Time evolution of entropic centrality of the three nodes with the highest centralities (nodes 3, 34, 1 respectively), node 14, which visually appears in the graph to be at the border of the clusters (and has, in fact, the seventh highest centrality) and the node with the lowest entropic centrality (node 17).}
\end{figure}

We consider the karate club network \cite{KarateClubOriginal} (see Figure \ref{fig:cluskcg} for the cluster results), to illustrate and analyze the workings of the clustering algorithm with a toy example with known baseline, before studying larger graphs. In Figure \ref{fig:kcgstep1}, the initial set $S_{ini,v_q}$ of clusters is shown along with the dendrogram for agglomeration, and Figure \ref{fig:kcgstep2} shows the final clusters. Clustering obtained using the edge removal technique (20 iterations) from \cite{NRK} is shown in Figure \ref{fig:karateremoval} for comparison. We also show the time evolution of some of the nodes with highest asymptotic entropic centrality, and the node with the lowest asymptotic entropic centrality in Figure \ref{fig:morekaratecentrality} (this is in addition to Figure \ref{fig:kcgplot} where we demonstrated the time evolution of certain nodes too), to illustrate the behavior of nodes which are hubs, at the interface of the clusters and at the periphery of the network.
This top-down clustering approach follows the idea of edge removal from \cite{girvan2002community}, but using the reduction of average Markov entropic centrality to determine which edge to remove. 

While it is visually apparent that our approach yields better clustering, we quantify this based on the ground truth \cite{KarateClubOriginal} using F-score \cite{F-score}. The result obtained with our approach has a F-score of 0.884 while the one obtained with \cite{NRK} has a F-score of 0.737.

We furthermore benchmark the computation time: the edge removal based clustering approach \cite{NRK} took 14.154 seconds, while our final result of 2-clusters were computed in 0.026 seconds. These experiments were run on a 64-bit PC with x64-based Intel(R) Xeon(R) CPU E5-1650 0 @3.20GHz processor and 16 GB RAM. This is easily explained: in our algorithm, the transition probability matrix $\hat{P}$ needs to be computed only once. In contrast, even within a single iteration, the edge removal algorithm \cite{NRK} needs to recompute the said matrix for every graph instance created by removal of each possible edge, to determine which edge to remove, and this exercise is then repeated in every iteration. That accounts for the huge discrepancy in the computation time, and demonstrates the computational efficacy of our approach. 

\subsection{Clustering of Bitcoin Subgraphs}
\label{sec:bitcoin178clusters}

\begin{figure*}
	\centering
	\subfloat[Initial clusters (Algo. \ref{alg:querynodecentric} output)]{\label{fig:undirstep1} 
		\includegraphics[width=0.3\textwidth]{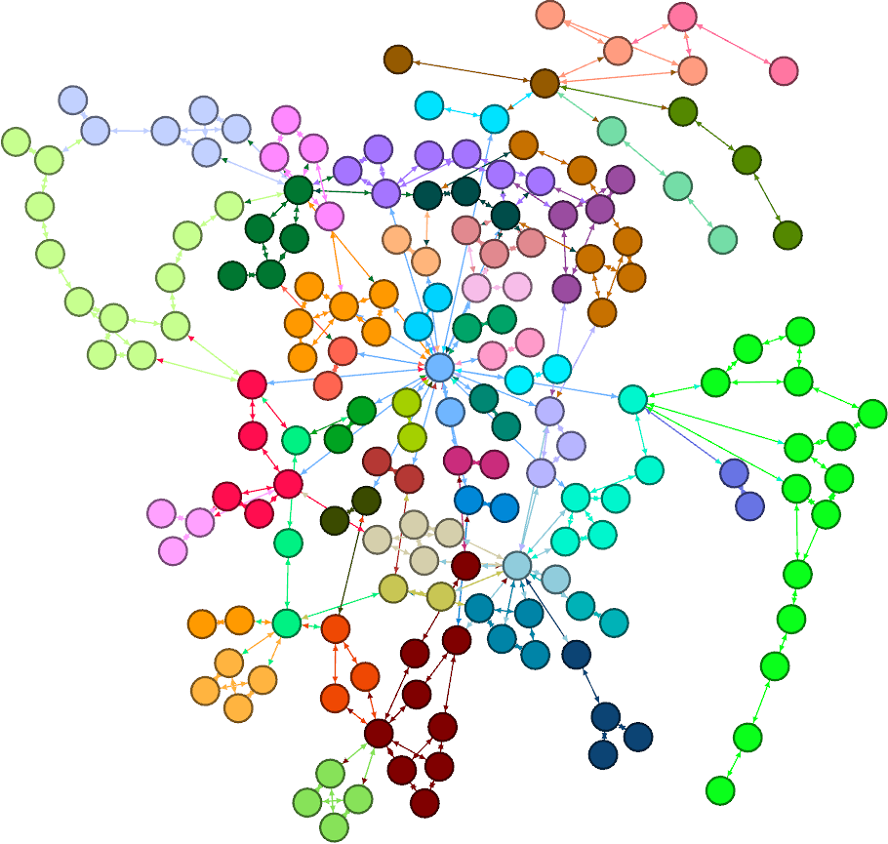}}
	~
	\subfloat[Two rounds of agglomeration.]{\label{fig:undirstep2} 
		\includegraphics[width=0.3\textwidth]{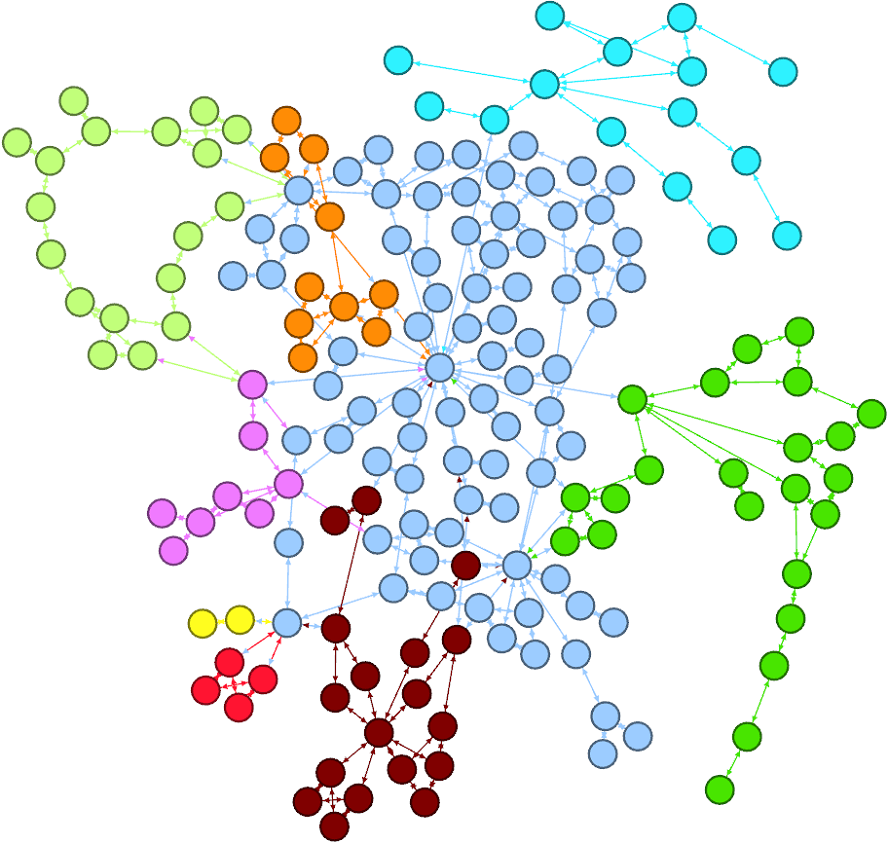}}
        ~
	\subfloat[Agglomeration dendrogram]{\label{fig:undirdendo} 
		\includegraphics[width=0.3\textwidth]{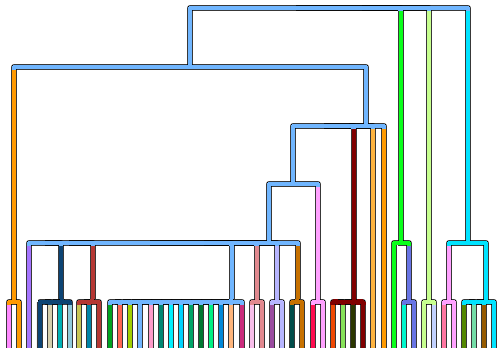}}

	\subfloat[Initial clusters (Algo. \ref{alg:querynodecentric} output)]{\label{fig:dirstep1}
		\includegraphics[width=0.3\textwidth]{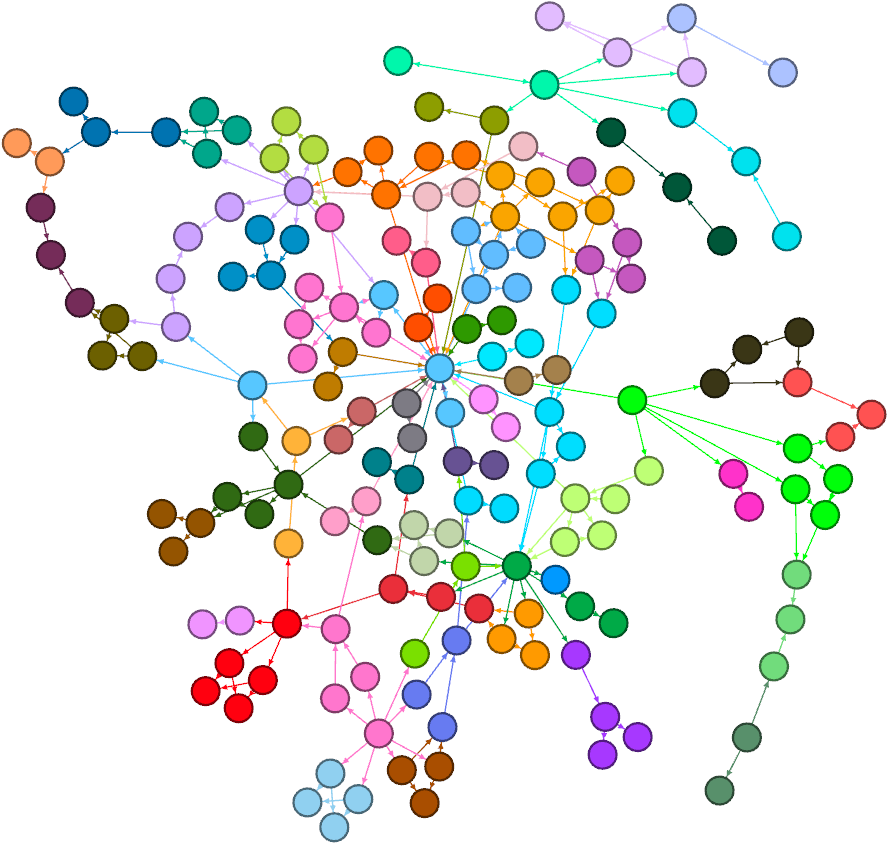}}
	~
	\subfloat[One round of agglomeration]{\label{fig:dirstep2}
		\includegraphics[width=0.3\textwidth]{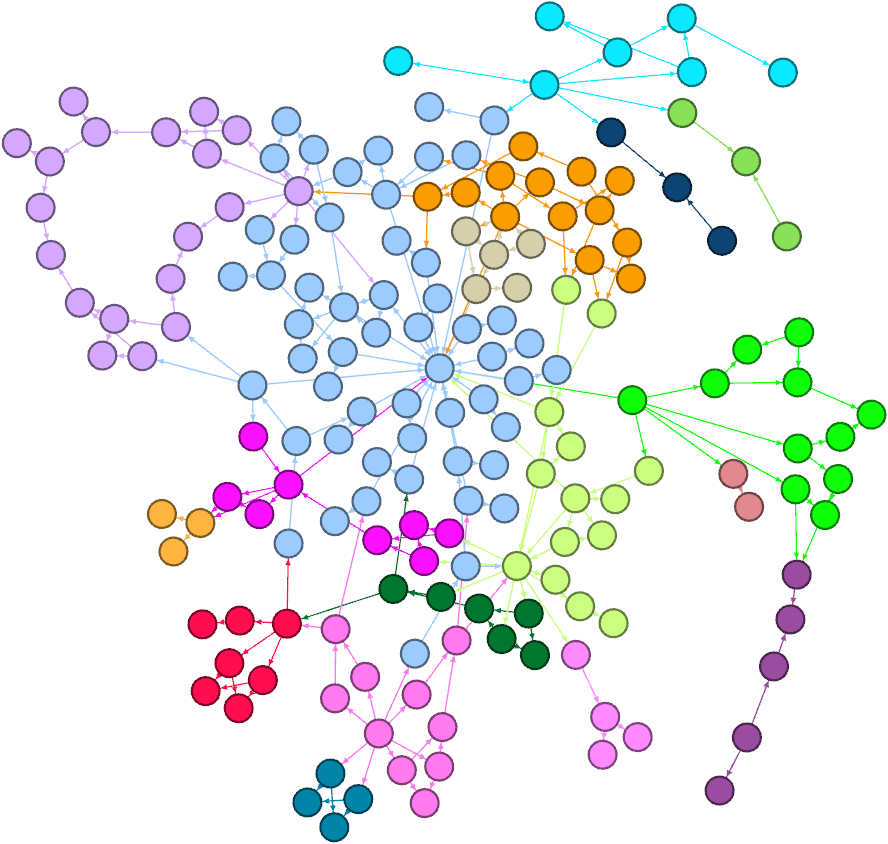}}
	~	 
	\subfloat[Two rounds of agglomeration]{\label{fig:dirstep3} 
		\includegraphics[width=0.3\textwidth]{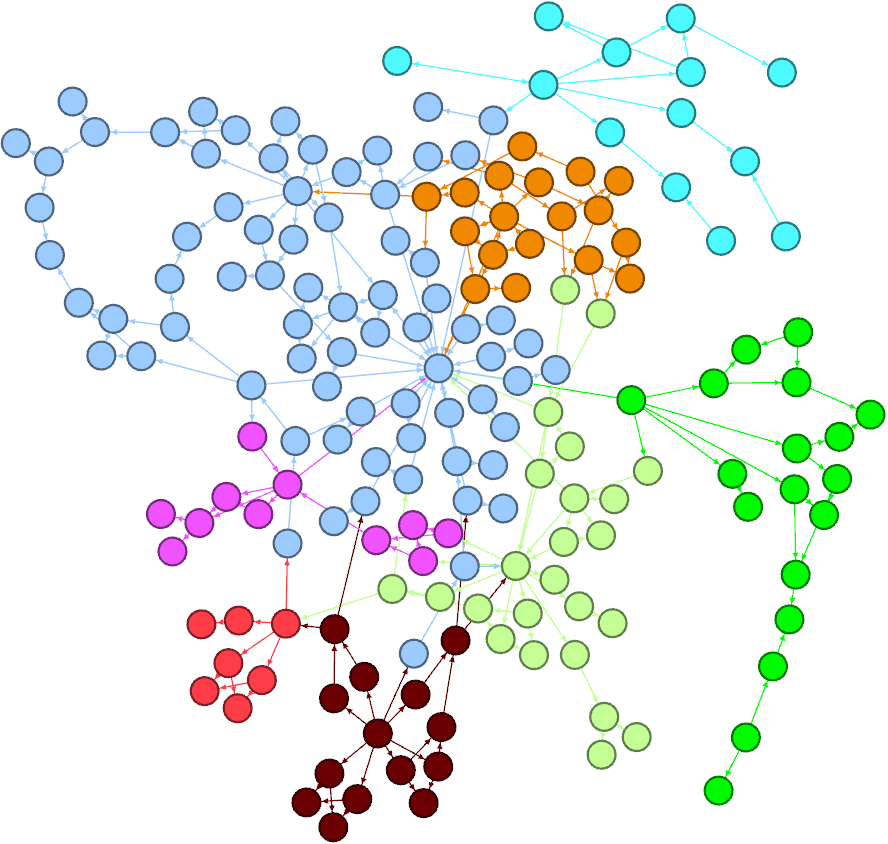}}

	\subfloat[Initial clusters (Algo. \ref{alg:querynodecentric} output)]{\label{fig:weightstep1} 
		\includegraphics[width=0.3\textwidth]{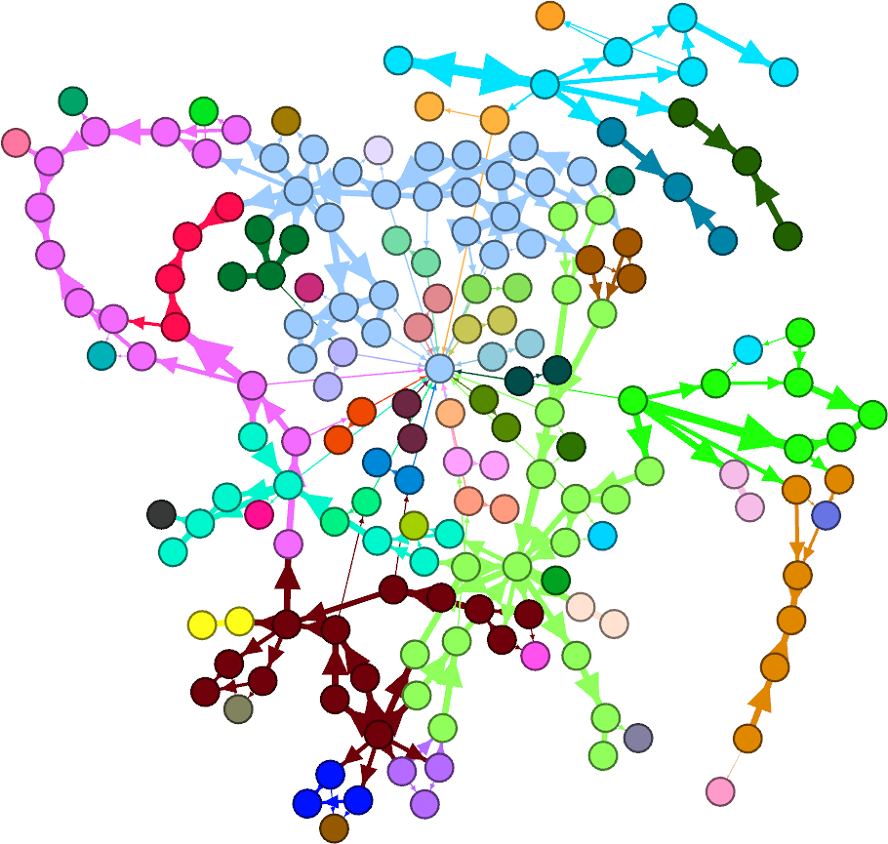}}
	~
	\subfloat[Four rounds of agglomeration]{\label{fig:weightstep2} 
		\includegraphics[width=0.3\textwidth]{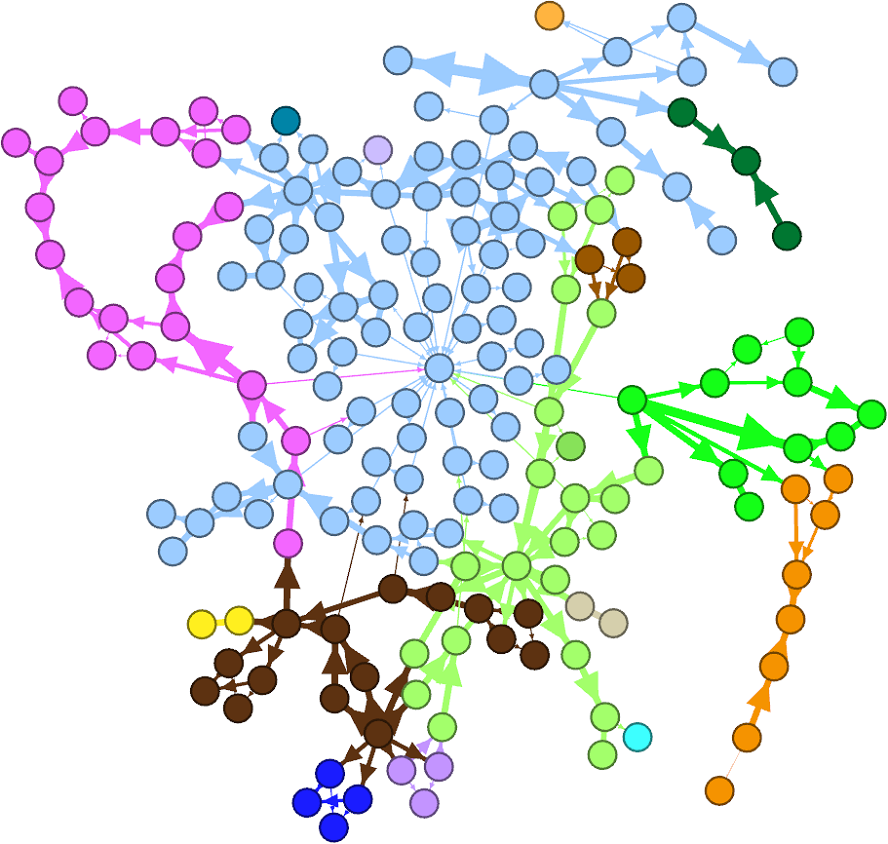}}
        ~
	\subfloat[Agglomeration dendrogram]{\label{fig:weightdendo} 
		\includegraphics[width=0.3\textwidth]{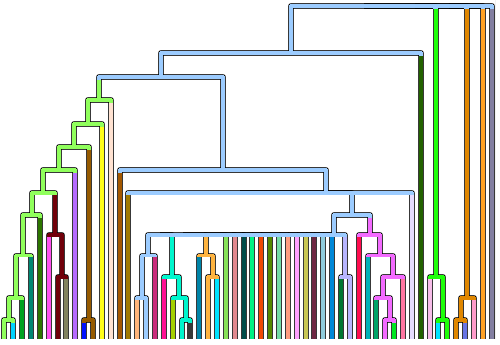}}

	\caption{\label{fig:178}
Clustering of the 178 node bitcoin subgraph (results obtained in 1.072, 1.077 and 1.123 seconds respectively): \textbf{1st row} - undirected unweighted graph, \textbf{2nd row} -  directed unweighted graph, \textbf{3rd row} -  directed weighted graph ($\alpha=1,~\mu=\tfrac{d_{w,out}(v)}{d_{out}(v)}$).}	
\end{figure*}

\begin{figure}
\centering
	\includegraphics[scale=0.14]{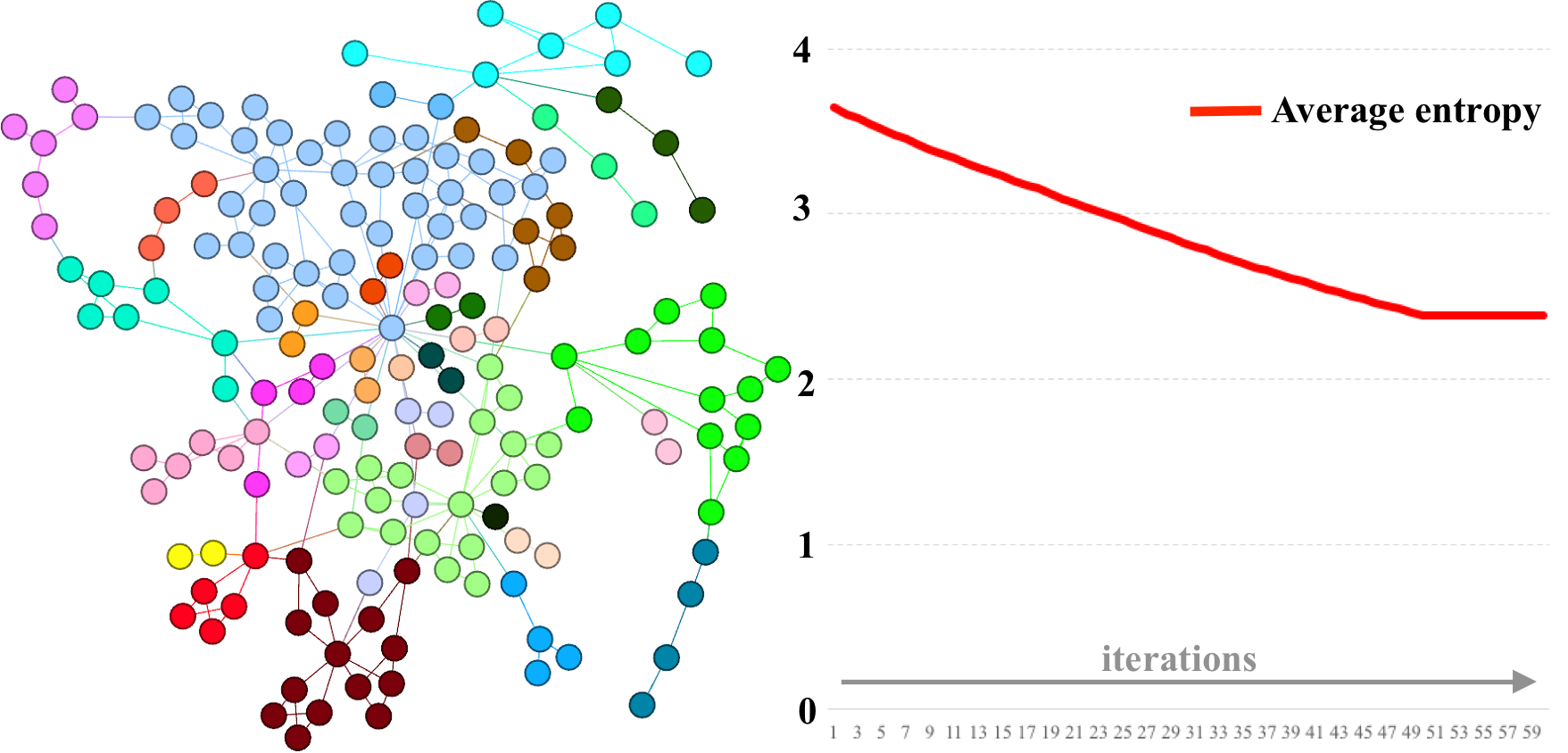}
	\caption{\label{fig:178removal}
		Clustering by iterative edge removal \cite{NRK}.}
\end{figure}

We apply our proposed clustering algorithm to variations (un/directed, un/weighted) of the 178 node Bitcoin network subgraph \cite{data178} and report the results in Figure~\ref{fig:178}. The results obtained using the edge removal algorithm \cite{NRK} on the undirected unweighted graph variant is shown on Figure~\ref{fig:178removal}. Unless otherwise stated, we use as a parameter $N=53$, essentially considering $S_{HE}$ to comprise the top 30\% entropic centrality nodes (see the sensitivity analysis below).

While it is visually clear from Figure~\ref{fig:178} that we obtain different clusters depending on the scenario considered, Table \ref{tab:differentgraphtypes} confirms this by reporting F-scores \cite{F-score} across the graph variants. Also, the effect of the parameter $\mu$ (without the transition probabilities being altered by edge weights) is rather low. This reinforces an underlying motivation of our work, namely that which graph variant (un/directed and/or weighted) to study is application dependent, and hence having one graph clustering algorithm that works across all variants is beneficial.

The clusterings of the undirected unweighted graph found by our algorithm in 1.072 seconds (Figure \ref{fig:178}) and by the edge removal algorithm in 3196.07 seconds (Figure \ref{fig:178removal}) are very different (F-score of 0.125). Our algorithm visually yields (more) meaningful results, though in the absence of a ground truth, this assertion remains subjective. 
The agglomeration process in our algorithm could have been continued beyond the final result shown here, as indicated by the associated dendrogram. The edge removal based clustering was stopped at the 50th iteration, after which no further average entropy reduction was observed with any single edge removal.

\begin{table}[]
	\centering
	\begin{tabular}{@{}lllllll@{}}
		\cellcolor{AnotherGray} graph
		& \cellcolor{ashgrey} $UU$ & \cellcolor{ashgrey} $D_{1,1}$ & \cellcolor{ashgrey} $D_{w,1}$ & \cellcolor{ashgrey} $D_{1,M}$ & \cellcolor{ashgrey} $D_{w,M}$ & \cellcolor{ashgrey} $UU_{er}$
		\\ 
		\cellcolor{ashgrey} $UU$
		&  1                        &  0.416                         &  0.506                        &   0.372                       & 0.271                         &   0.125                       \\
		\cellcolor{ashgrey} $D_{1,1}$&   0.416                       &    1                      &      0.473                    &     0.806                     &   0.364                       &    0.146                      \\
		\cellcolor{ashgrey} $D_{w,1}$&  0.506                        &    0.473                      &    1                      &  0.408                        &   0.416                       &   0.122                       \\
		\cellcolor{ashgrey} $D_{1,M}$&   0.372                       &   0.806                       &   0.408                       & 1                         &   0.413                       &  0.145                        \\
		\cellcolor{ashgrey} $D_{w,M}$&   0.271                       &  0.346                        &  0.416                        &   0.413                       &   1                       &  0.131                        \\
		\cellcolor{ashgrey} $UU_{er}$ &   0.125                       & 0.146                         &     0.122                     & 0.145                         &   0.131                       &   1                       \\ 
	\end{tabular}
\caption{\label{tab:differentgraphtypes}
Pairwise F-score among clusterings achieved for different graph variants. Legend --- UU: undirected \& unweighted; $D_{x,y}$: Directed with $\alpha=x$, and $\mu=1$ if $y=1$, else $\mu=\tfrac{d_{w,out}(v)}{d_{out}(v)}$ if $y=M$; $UU_{er}$: edge removal algo \cite{NRK}.} 
\end{table}

When to stop agglomerating is typically purpose dependent. The second row of Figure~\ref{fig:178} shows various stages of agglomeration, the dendrogram for agglomeration is given for the 1st and 3rd row. There are  many community structures that repeat across the considered scenarios, and most of the cluster boundaries can be traced back to the high entropy nodes (Figure \ref{fig:entropybitcoin}), yet there are also subtle differences, e.g., in the weighted directed graph, there are instances of single isolated nodes, which stay isolated for several interactions of agglomeration because of weak (low weight) connections.

{\bf Sensitivity analysis.}
For each graph variant, the size of $S_{HE}$ was varied to comprise 10\%, 20\%, 30\% and 40\% of the top entropic centrality nodes. For the unweighted and weighted directed graph, the F-score between clusterings obtained with 10\% and the others are 0.824 and 0.989 respectively, while all the other pairs have F-score of 1. This suggests very consistent results in these cases, irrespectively of the choice of $|S_{HE}|$. However for the unweighted undirected graph, $|S_{HE}|$ has a significant impact, with the F-score between 20\% and 40\% being the lowest at 0.626, while the best score of 0.912 is obtained between 30\% and 40\%. This justifies the use of the top 30\% entropic centrality nodes for the reported results.
{Looking back at Figure \ref{scatter}, we observe that considering only 20\% for the size of $S_{HE}$ means including a number of query nodes with relatively low highest probabilities, while between 30\% and 40\%, these highest probabilities are not changing significantly, which is consistent with the observed variations in sensitivity.} 

\begin{figure}
	\centering
	\includegraphics[scale=0.4]{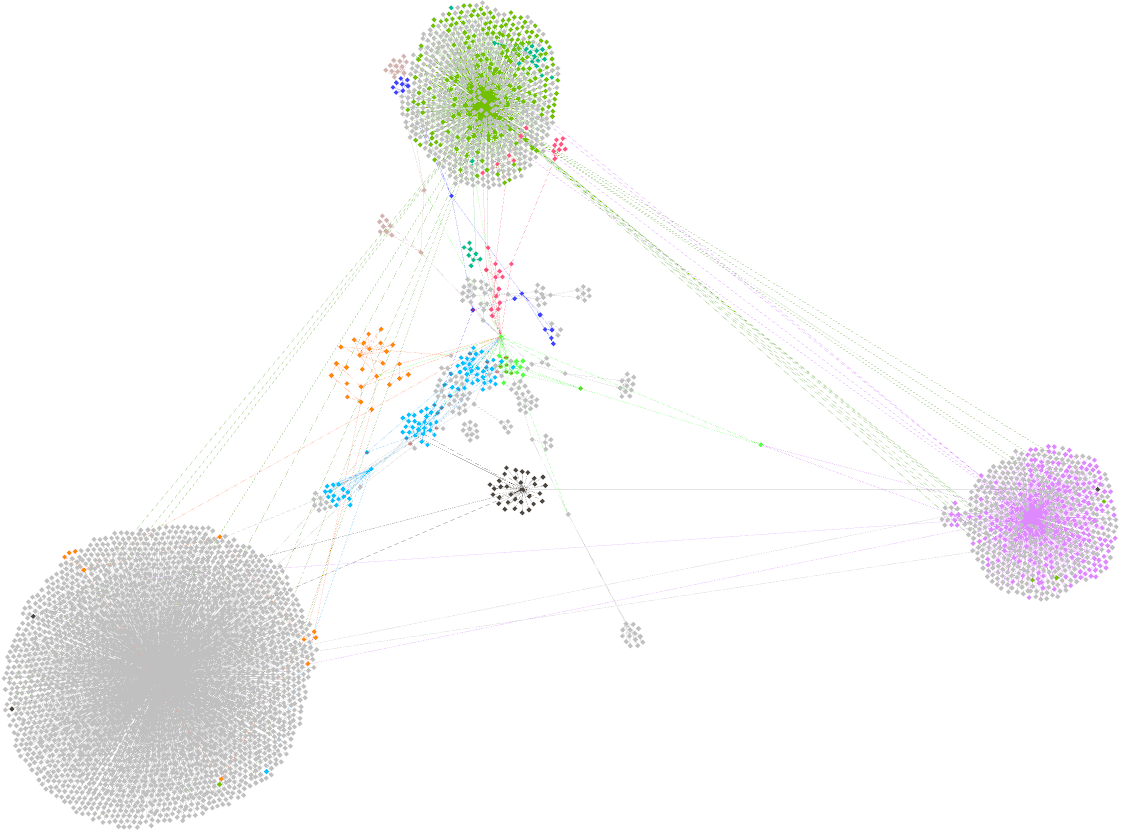}
	\caption{\label{fig:4kclustering_Bigpicture}
		Clustering of a Bitcoin subgraph involved in the Ashley-Madison extortion scam \cite{Egret-paper}.}
\end{figure}

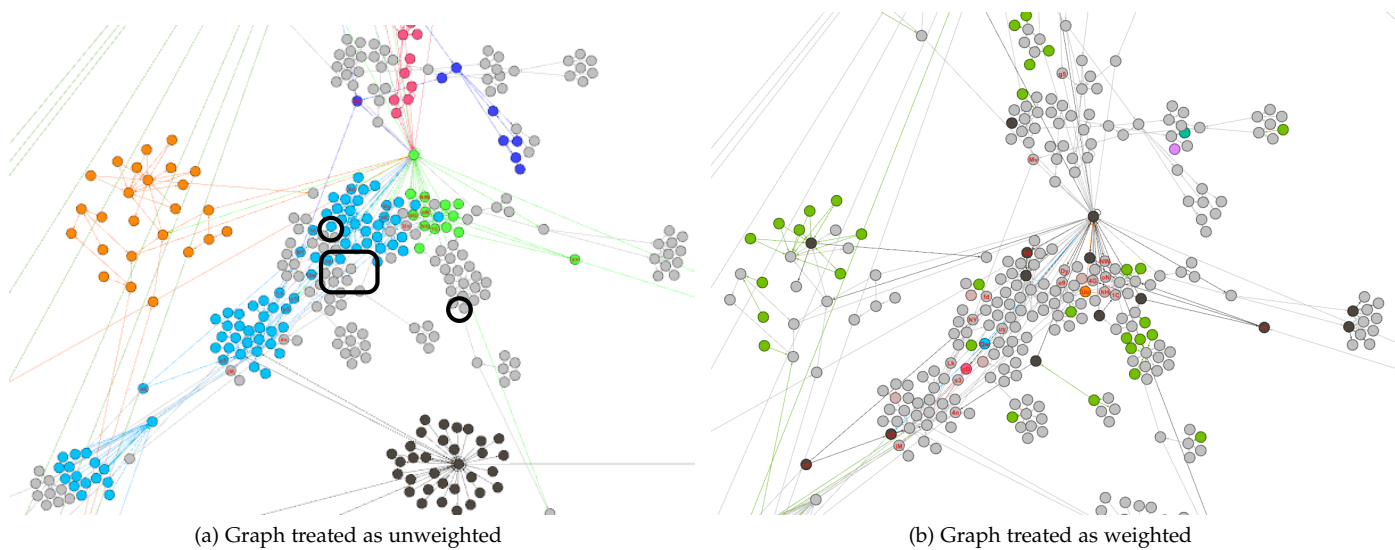
\begin{figure*}
\centering
\subfloat[Graph treated as unweighted]{\label{fig:unweightedzoom}
\begin{tikzpicture}
\node[anchor=south west,inner sep=0] at (0,0){
\includegraphics[width=0.5\textwidth]{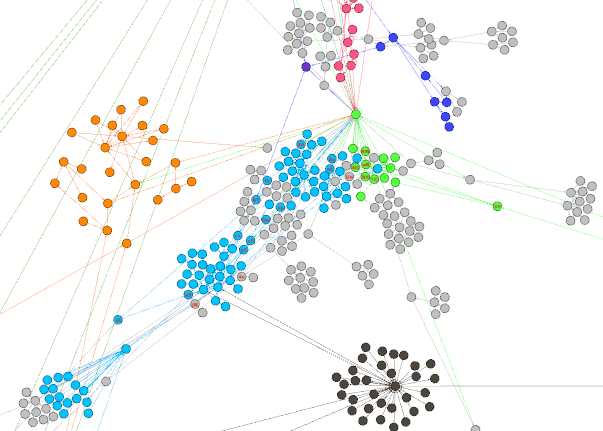}};
\draw[black,ultra thick] (4.28,3.8) circle (0.15cm);
\draw[black,ultra thick] (5.99,2.73) circle (0.15cm);	
\draw[black,ultra thick, rounded corners] (4.15,2.96) rectangle ++(0.75,0.54);
\end{tikzpicture}}
~
        \subfloat[Graph treated as weighted]{\label{fig:weightedzoom}
        \includegraphics[width=0.5\textwidth]{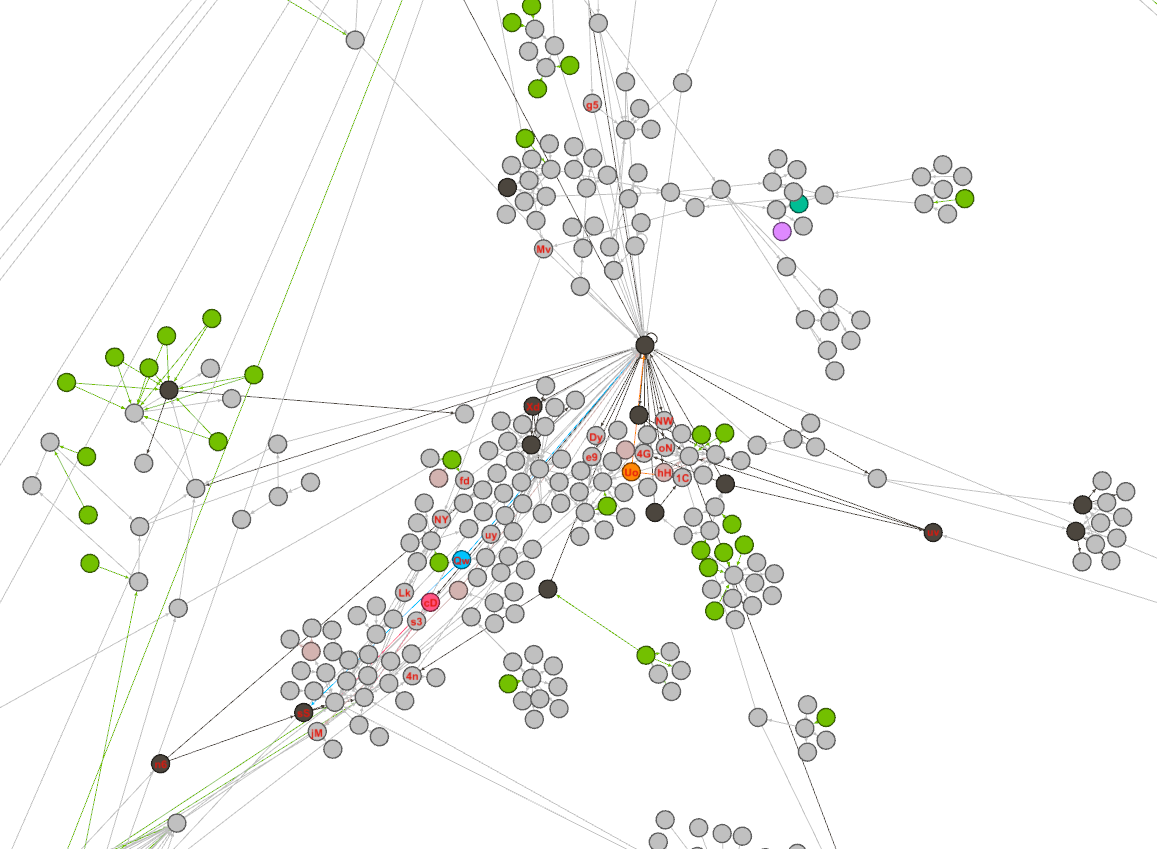}}
       
	\caption{Zoom in of the Ashley-Madison extortion scam \AD{(Bitcoin transactions induced)} graph.\label{fig:4kcluster-zoom}}
\end{figure*}

We next consider another Bitcoin subgraph, but this time, we use a specific subgraph \cite{data4571} of 4571 nodes, constructed around Bitcoin addresses suspected to be involved in the Ashley-Madison extortion scam \cite{Egret-paper}. The result of the clustering algorithm is shown on Figure~\ref{fig:4kclustering_Bigpicture} {and Figure \ref{fig:4kcluster-zoom}}: (1) the graph was considered {once unweighted (the emphasis is thus on node connections), once weighted (to capture the amount of Bitcoins involved)}, (2) the asymptotic Markov entropic centrality was used (we have no specific diameter of interest), (3) the top 30\% nodes with the highest centrality are chosen as high centrality nodes $S_{HE}$, (4) five agglomeration iterations were performed for clustering {the unweighted graph, and one for the weighted variant}.

{On Figure~\ref{fig:4kclustering_Bigpicture}, showing the overall unweighted graph}, there are three visually obvious main clusters: the upper green cluster, the purple cluster on the right, and the grey cluster on the left. The first observation is that the grey color here only represents nodes whose cluster size is too small to be significant (only 5 iterations were performed), they are thus kept in grey so as to make the other clusters more visible. The green and purple clusters are easily interpreted: each contains one Bitcoin address that is a hub for all its neighbors.

We then zoom into the central clusters, shown on Figure~\ref{fig:unweightedzoom}. The actual relationship among the constituent wallet addresses in a cluster can be determined e.g. by using \url{blockchain.com/explorer}. We observe a green cluster near the middle (boxed). In our layout, we have isolated one of the constituent nodes (on the right, encircled), to show that the nodes in this collection have multiple mutual connections, as expected among nodes within a cluster. We see that the encircled node above the boxed group has also been assigned to the same cluster. This node is in fact connected to several of the other clusters that have been identified with our algorithm, and is one of the high centrality nodes, which lies at the interface of clusters. It happens to have been assigned to the green cluster, since each node is assigned to at most a single cluster. Some of the nodes in the (boxed) cluster were previously identified to be suspect addresses involved in the Ashley-Madison data breach extortion scam \cite{Egret-paper}. The resulting clusters thus help draw our attention to the other nodes which have been grouped together, since it indicates that Bitcoin flows have circulated among them, for their relationship with the already known suspected nodes to be investigated further.

Zooming into the weighted graph gives a very different picture: since the amounts of Bitcoin involved drive the clustering in this case, we prominently see two clusters highlighting nodes dealing with high volume of Bitcoins. This confirms an expected behavior from scammers, which consists of collecting few Bitcoins from many addresses to avoid attention. Combining both clustering results however correlate nodes that are likely to be involved in the scam, together with nodes dealing with high volume of Bitcoins. For example, this could be a direction to consider for Bitcoin forensics: nodes appearing in clusters by interpreting the graph in both manners could possibly be involved in aggregating scam money, since they stand out both in terms of the volume of Bitcoin they handle, and in terms of the frequency of interactions with multiple suspected wallet addresses.

\subsection{\AD{Benchmarking With Synthetic Graphs}}\label{sec:synthnetwork}

In the previous subsection, we looked at the clusterings that the algorithm provides with Bitcoin subgraphs, for which no ground truth is available. In order to benchmark our proposed algorithm rigorously, we thus further experiment with graphs for which some form of ground truth is assumed.

An acknowledged concern in the research community is that a unique objective benchmark to compare graph clustering algorithms is not feasible \cite{LWG,letopeel}. Different algorithms may yield different clusters for a given graph, that may each be meaningfully interpreted based on distinct contexts. Conversely, in the real world, connections may have been induced as a consequence of multiple causes (contexts), and using the meta-data representing a subset of these contexts to determine a `ground truth' for the resulting graph may not be accurate. Furthermore, there may be implicit or explicit hierarchical community structures in the graph, or the communities may be fuzzy, and a clustering algorithm may find clusters at a coarser or finer granularity than the one considered as the ground truth. 

Synthetic graphs (for example \cite{LFR}) are considered to alleviate the issue of the lack of a unique and objective ground truth. Yet such synthetic graphs may not carry all the characteristics and associated complications of a real network.

Considering the merits of benchmarking with graphs with ground truth, but also the inherent limitations associated with any particular instance(s) of real or synthetic (family of) graph(s), we study a wide range of graphs with assumed ground truth. We next discuss our experiments with synthetic graphs, before studying some real graphs in the following subsection.

The principal idea of generating synthetic graphs with a known ground truth is to first create isolated subgraphs (with certain properties such as a given node degree distribution) that represent the ground truth communities. Then, rewiring of a fraction of the connections is carried out to establish cross community links such that, probabilistically, a $1-\mu$ fraction of links are with nodes within the same community, while a fraction $\mu$ (mixing probability) of connections are with other nodes. Though this rewiring process itself might affect the neighborhood of individual nodes to an extent that it materially changes the community it actually belongs to (particularly for high values of the mixing probability $\mu$), the original allocation of the communities is considered to continue to hold, and is treated as the ground truth. For the reported experiments below, we used synthetic benchmark graph instances randomly generated using NetworkX. 


\begin{figure}
\centering
\subfloat[100 nodes graph: 4 clusters detected with F-score 0.900]{\label{fig:synth100} 
\includegraphics[width=0.3\textwidth]{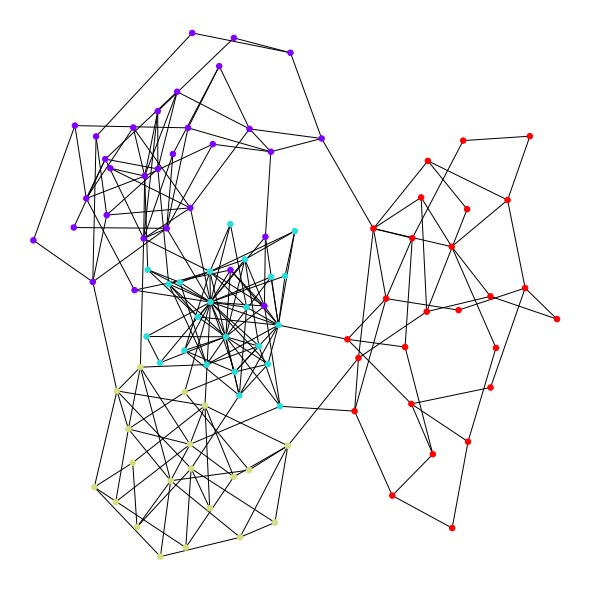}}
~
\subfloat[300 nodes graph: 15 clusters detected with F-score 0.973]{\label{fig:synth300} 
\includegraphics[width=0.3\textwidth]{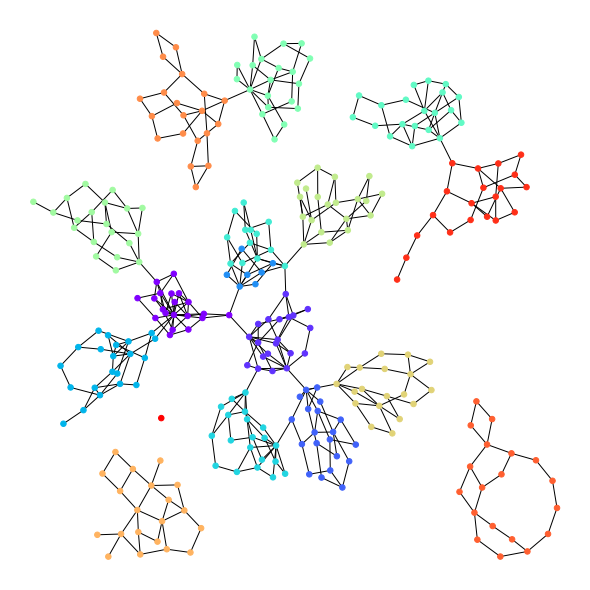}}
~
\centering
\subfloat[500 nodes graph: 21 clusters detected with F-score 0.909]{\label{fig:synth500} 
\includegraphics[width=0.3\textwidth]{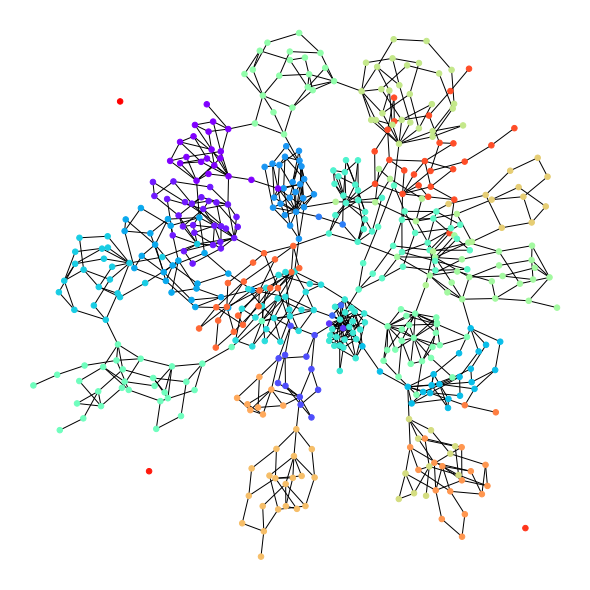}}

\caption{\label{fig:synthnets}
\AD{Clusterings of synthetically generated networks with mixing probability $\mu=0.1$}.}
\end{figure}

\AD{For sanity check and to manually (visually) interpret the results, we start the benchmarking with small graphs and a small value of $\mu=0.1$. }
The clusters identified by our algorithm for synthetic graphs with 100, 300 and 500 nodes are shown in Figure \ref{fig:synthnets} for $\mu=0.1$. Visually, we see that the algorithm yields meaningful clusters. We also determine the F-score with respect to the ground truth as determined by the network generation process, and across the different networks we observe very good (0.9 or above) F-score values. For the 100 nodes graph, one can visually see certain nodes, particularly in the middle group being allocated cluster labels that mismatch, explaining the relatively lower score of 0.9 among these graph instances. For the 500 nodes graph instance, the isolated nodes had distinct labels in the ground truth. Some other misattributions can also be seen visually, explaining the relatively lower score of 0.909. The 300 nodes graph instance had disconnected components, which might also have made it easier for the clustering algorithm to find relevant communities, yielding a noticeably high score of 0.973 was obtained.

\begin{figure}
	\centering
		\includegraphics[scale=0.35]{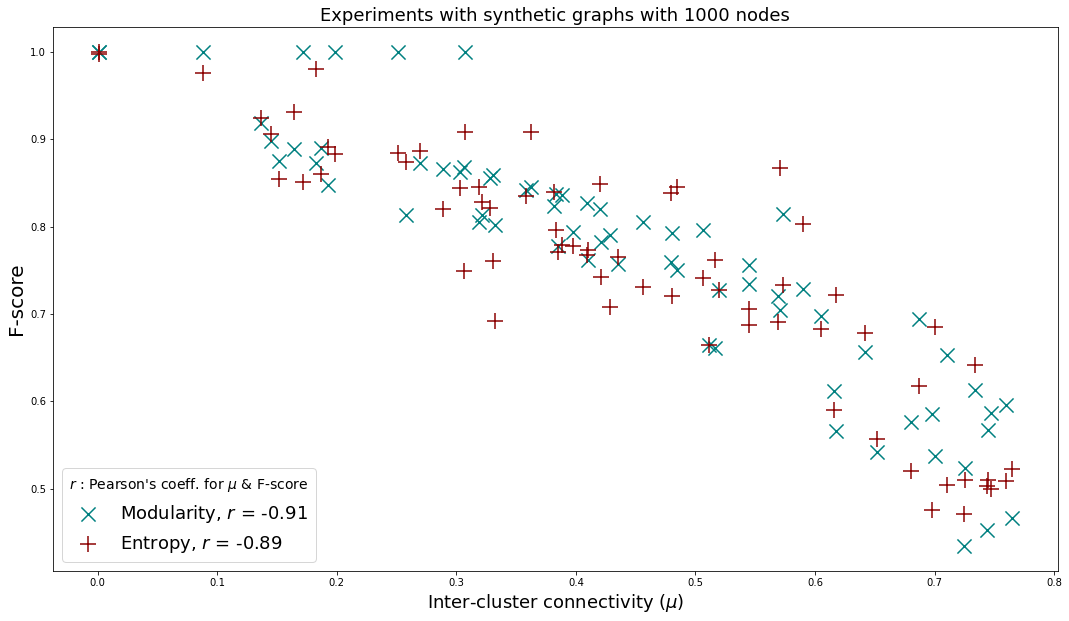}
	\centering
		\includegraphics[scale=0.35]{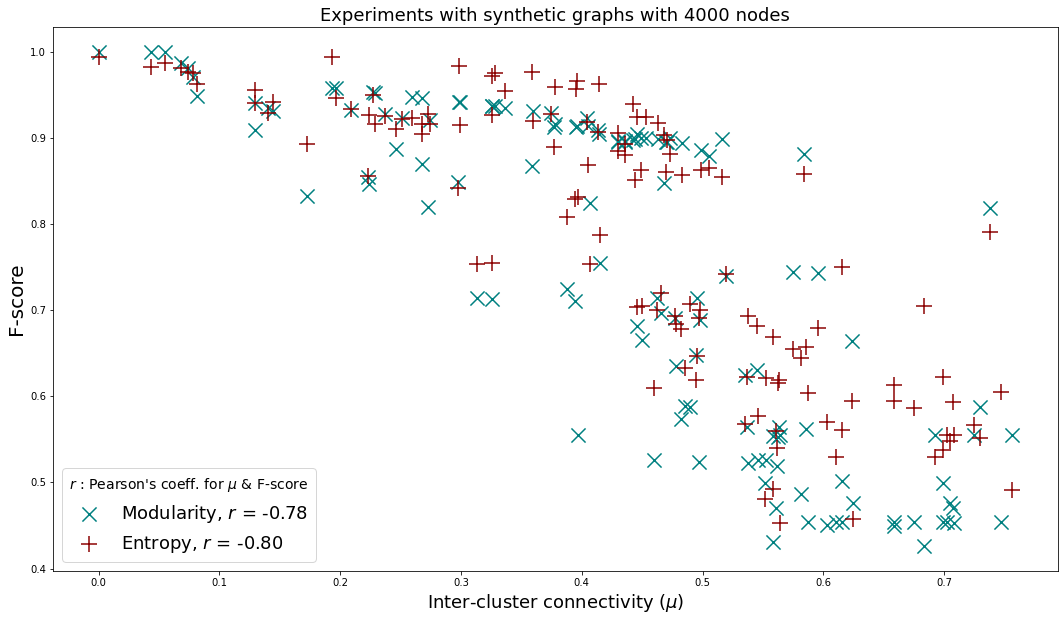}
	\caption{\label{fig:synthetic-scatter}
		\AD{Scatter plots of F-scores for large-scale benchmarking of our \textbf{Entropy} based graph clustering approach with randomly generated synthetic graphs comprising 1000 and 4000 nodes respectively for a wide range of inter-cluster linkage $\mu$ characteristics are shown. \textbf{Modularity} optimization based Louvain \cite{blondel2008fast} community detection algorithm is also shown to provide a point of comparison.}} 
\end{figure}

\AD{
We next extend our study with larger scale experiments both (i) in terms of graph size (1000 and 4000 nodes) and (ii) in terms of range of $\mu$ values representing different extents of cross-community linkages. We show the scatter plots of observed F-scores for our \textbf{Entropy} based graph clustering algorithm. For a point of reference, we also provide the results observed for clustering with Modularity optimization \cite{blondel2008fast}. For relatively smaller values of $\mu$, perfect clustering is obtained by \cite{blondel2008fast} while near perfect clustering is also obtained by our approach. For the rest of the spectrum of $\mu$ values, performance of both the approaches varies to certain extent (more so in the larger graphs with 4000 nodes), but overall both algorithms yield good results, e.g., F-score is consistently higher than 0.8 for $\mu<0.3$, and the deterioration of the F-score with increasing $\mu$ is gradual, rather than sharp. Furthermore, while the performance varies across different graph instances, very high (absolute values of) Pearson's correlation coefficients between F-score and $\mu$ (precise correlation coefficient $r$ values are indicated in the figures) indicate a good degree of consistency in the behaviour for both the algorithms. From individual data points, we observe that among the two algorithms, there is no clear winner, and each outperforms the other for some graph instances across most of the $\mu$ value ranges. These large-scale benchmarking experiments with randomly generated synthetic graphs  demonstrate the efficacy of our proposed approach on its own. While the original objective of our proposal was to investigate a new way to carry out graph clustering rather than to necessarily compete with existing approaches, the comparison using synthetic graphs with one of the popular existing approaches further demonstrates that the quality of clustering results obtained by our approach is in fact comparable.}

 
\subsection{\AD{Benchmarking With Real World Graphs}}\label{sec:groundtrutheva}

\AD{Since synthetic graphs may not exhibit all the nuances of communities occurring in the real world, we complement our study with benchmarking experiments} using networks with known ground truth, namely, the dolphin network \cite{lusseau2003bottlenose} and the American college football network \cite{girvan2002community} which were previously used in the work \cite{NRK} that we extend. \AD{Moreover, we extend the comparative aspect of our evaluation, and to that end} we compare our approach with other popular community detection algorithms, namely, InfoMap \cite{RB}, label propagation \cite{andersen2006local} and Louvain clustering \cite{blondel2008fast}. 

\subsubsection{Clustering of the Dolphin Network}

We consider the dolphin network \cite{lusseau2003bottlenose}, an undirected unweighted social network where bottlenose dolphins are represented as nodes and association between dolphin pairs are represented as links. The network comprises 62 nodes and 159 links, and it was noticed that the dolphin community splits into two communities \cite{lusseau2003bottlenose}  comprising 20 nodes and 42 nodes. We use this as the ground truth.

\begin{figure}
	\centering
	\subfloat[The asymptotic Markov entropic centralities (darker the color, higher the relative entropic centrality score).]{\label{fig:dolnet} 
		\includegraphics[scale=0.3]{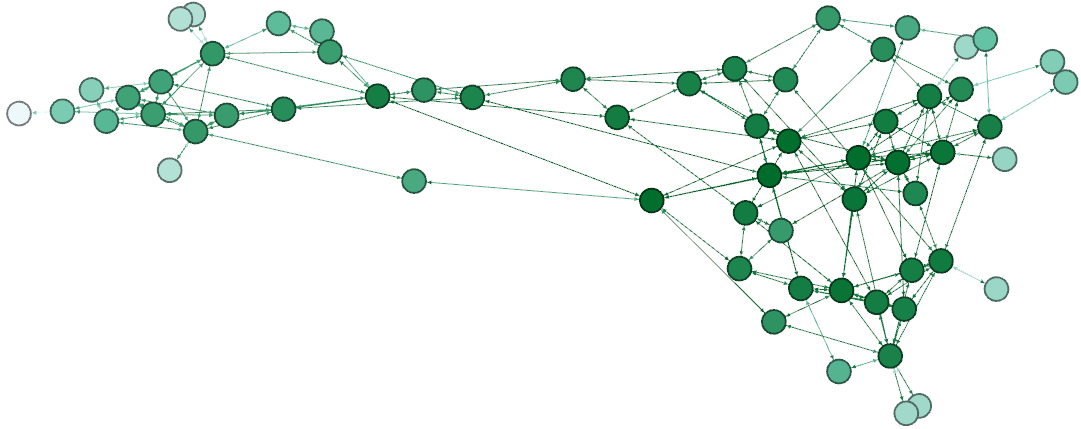}}
        ~
	\subfloat[Histogram of the normalized (by the maximum) centralities.]{\label{fig:dolhist} 
		\includegraphics[scale=0.37]{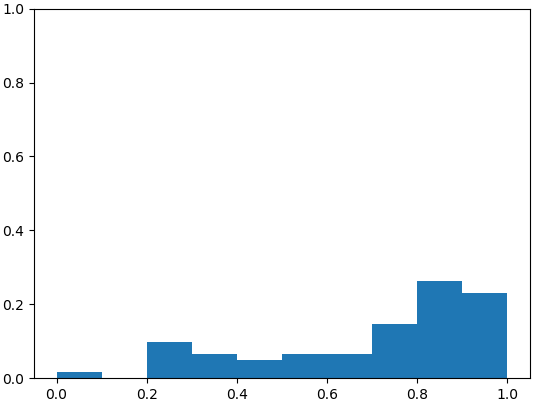}}
		
	\caption{\label{fig:dolphin_gradient}
		The dolphin network} 
\end{figure}

\begin{figure}
\centering

\subfloat[Proposed algorithm (2 iterations, top 60\%).]{\label{fig:dolphours} 
\includegraphics[width=0.3\textwidth]{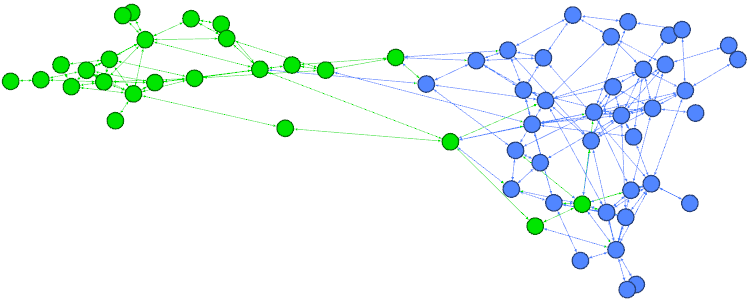}}
~
\subfloat[InfoMap.]{\label{fig:dolphinfomap} 
\includegraphics[width=0.3\textwidth]{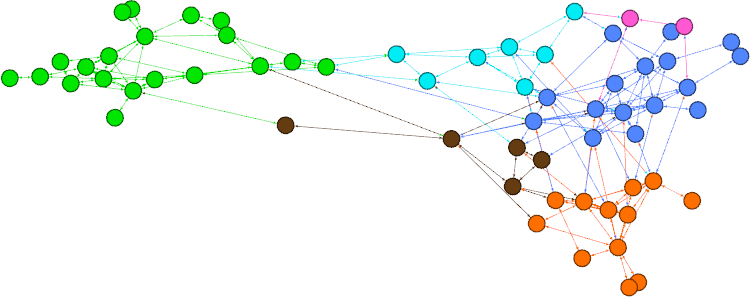}}
~
\subfloat[Louvain.]{\label{fig:dolphlouvain} 
\includegraphics[width=0.3\textwidth]{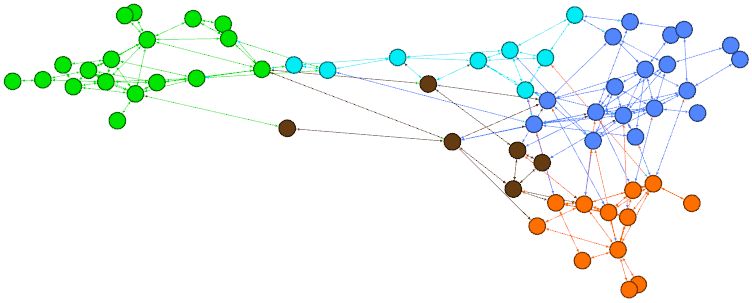}}

\subfloat[Label propagation.]{\label{fig:dolphlabel} 
\includegraphics[width=0.3\textwidth]{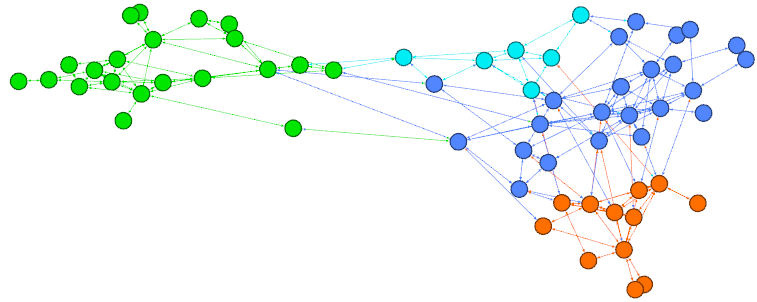}}
~
\subfloat[The ground truth \cite{lusseau2003bottlenose}.]{\label{fig:dolphgt}
\includegraphics[width=0.3\textwidth]{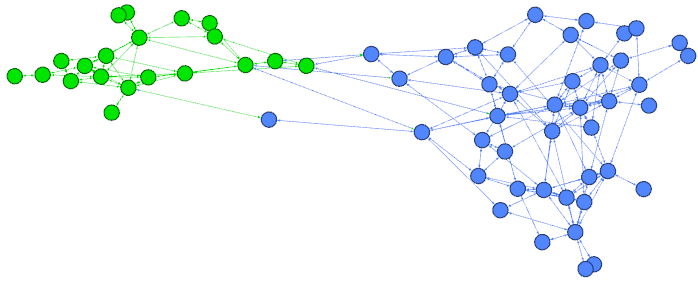}}
~
\subfloat[Maximum absorption probability as a function of the entropic centrality.]{\label{fig:dolphsp} 
\includegraphics[width=0.3\textwidth]{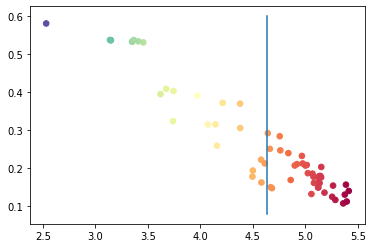}}
	
\caption{\label{fig:dolphin_compares}
Clusterings of the dolphin network.}
\end{figure}

In Figure \ref{fig:dolnet}, we show the network structure and the corresponding relative (that is, normalized by the maximum observed value) Markov entropic centralities: the darker the node color, the higher the relative centrality. 
	Figures \ref{fig:dolhist} and \ref{fig:dolphsp} respectively depict the relative Markov entropic centrality (fractional, bucketed) distribution and the scatter diagram of the absolute entropic centrality score ($x$-axis), versus the maximum absorption probability at any node ($y$-axis) for a random walker starting from the given node.

The histogram indicates that it would be more meaningful to choose the clustering parameter $S_{HE}$ to include the top 50\%-70\% (instead of $\approx$ 30\%, as used in the previous experiments) since more than 60\% of the nodes have normalized entropy value above 0.7. The scatter diagram helps us see that, indeed, taking the top 30\% nodes for $S_{HE}$ would mean including many nodes whose highest probability is relatively small, while there are few such a node by fixing the choice threshold at 60\% (the vertical line demarcates the top 60\% nodes on the right). The result obtained with two iterations of the algorithm is shown on Figure~\ref{fig:dolphours}, next to clustering results obtained using InfoMap \cite{RB}, Louvain \cite{blondel2008fast} and label propagation \cite{andersen2006local} (with their default parameters). We observe visually that the proposed clustering produces a better result compared to other clusterings. This is confirmed by computing the F-score \cite{F-score} for each clustering result against the ground truth: the F-score of the proposed algorithm is 0.858, in contrast, it is 0.545 for InfoMap, 0.565 for Louvain, and 0.657 for label propagation. These three community detection techniques also find more than two clusters. From Figure \ref{fig:dolphin_compares}, we also visually infer that the other clustering results could be improved if agglomeration techniques were applied to the smaller communities located on right hand side of the dolphin network. In fact, it could be argued that even though the group of dolphins split in two groups (which is the basis of the ground truth), it does not preclude the existence of further smaller communities within those two split groups, which could then be what is being detected by these algorithms.
	
\subsubsection{Clustering of the American College Football Network}

We next consider the American college football network \cite{girvan2002community}, an undirected unweighted network representing the Division-I football games from Fall 2000. A team is represented as a node, and a game between two teams is represented as a link between two nodes. There were in total 115 teams and 613 games. Teams were divided into 12 conferences, and teams in the same conference frequently had games against others. We treat the 12 conferences as the network's ground truth, comprising 12 clusters.

\begin{figure}
	\centering
	\includegraphics[scale=0.25]{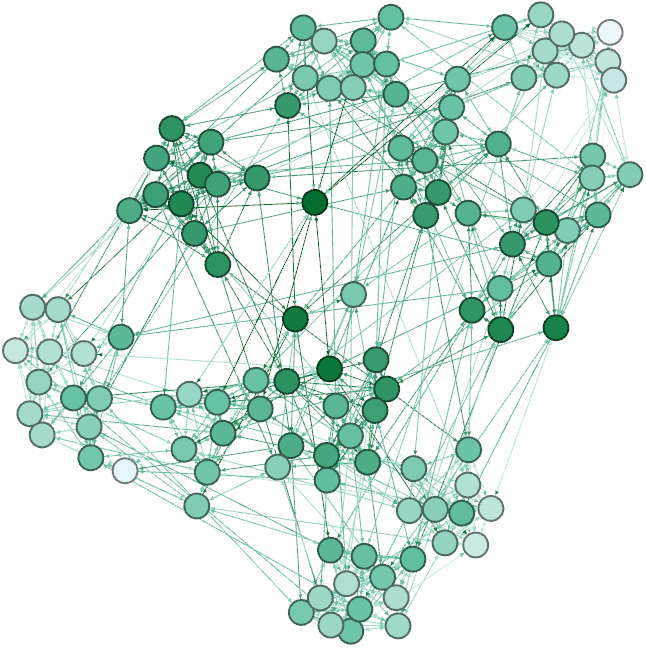}
        \includegraphics[scale=0.4]{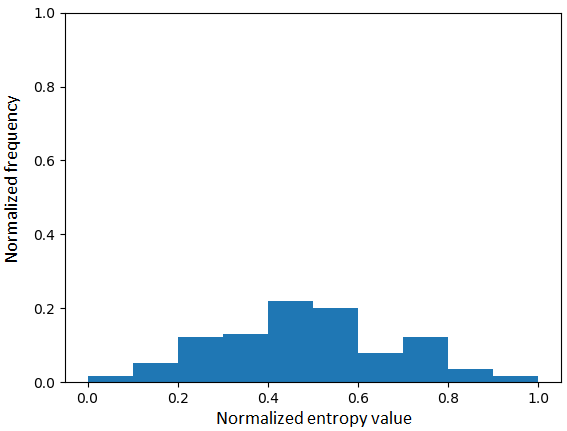}
	\caption{\label{fig:footballentropygradient}
		The American College football network \cite{girvan2002community} is shown (darker the color, higher the relative entropic centrality score). The histogram of the normalized (by the maximum) centralities distribution for the same network is also shown.}
\end{figure}

\begin{figure}
\centering
\subfloat[Proposed technique (one iteration).]{\label{fig:afour} 
\includegraphics[width=0.3\textwidth]{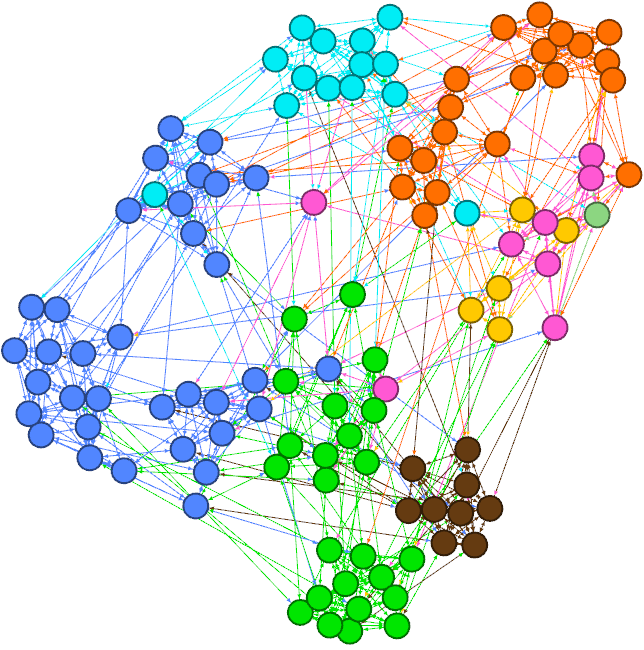}}
~
\subfloat[Proposed technique (applied  on the 3 largest subgraphs in Figure~\ref{fig:afour}).]{\label{fig:afour3} 
\includegraphics[width=0.3\textwidth]{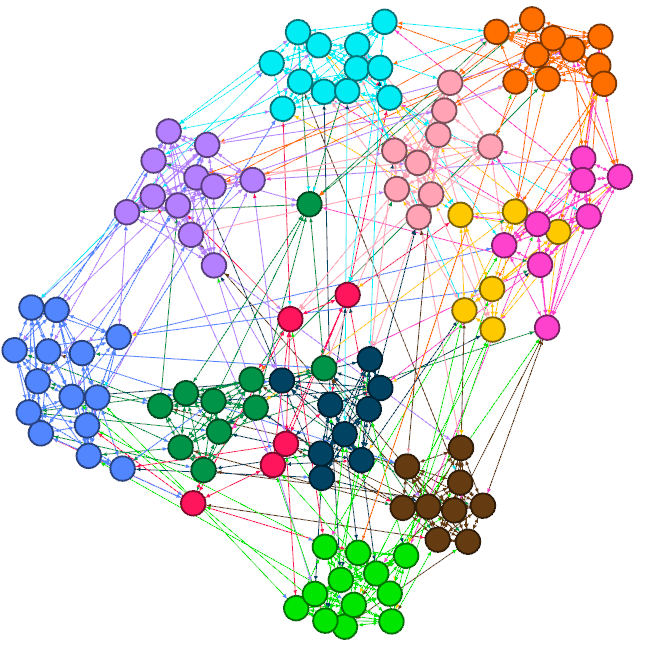}}
~
\centering
\subfloat[InfoMap.]{\label{fig:afinfomap} 
\includegraphics[width=0.3\textwidth]{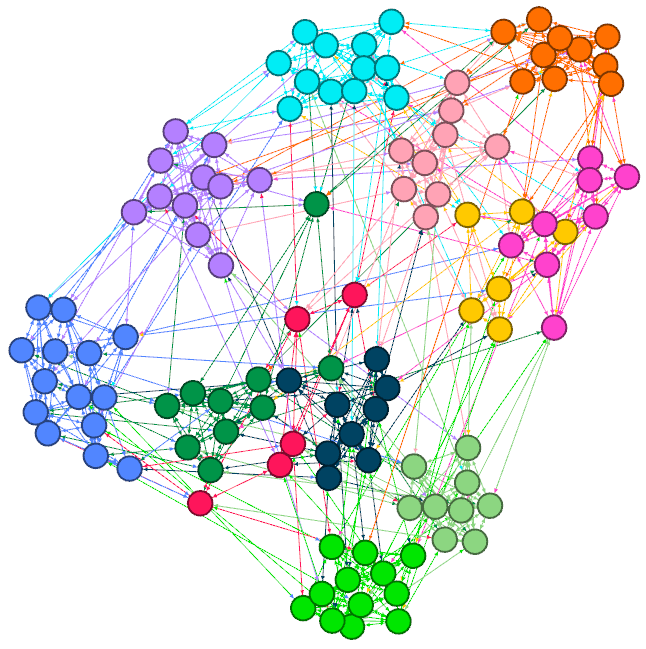}}

\subfloat[Louvain.]{\label{fig:aflouvain}
\includegraphics[width=0.3\textwidth]{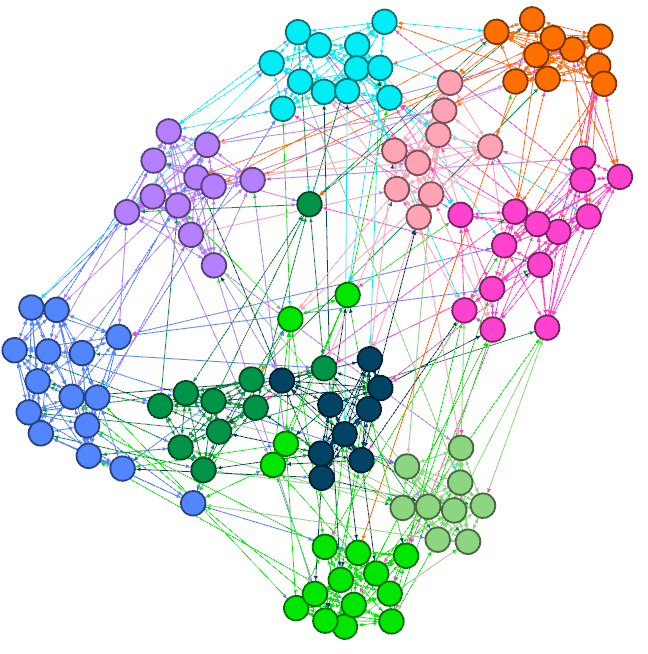}}
~
\centering
\subfloat[Label propagation.]{\label{fig:aflabel} 
\includegraphics[width=0.3\textwidth]{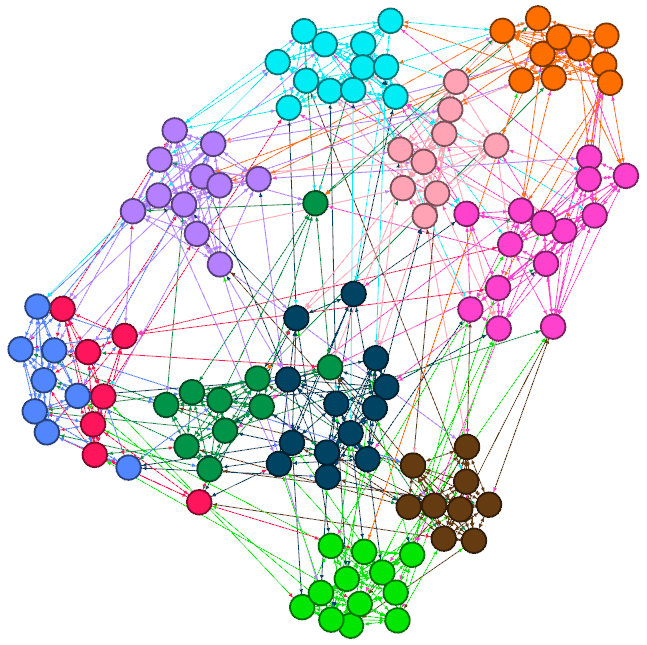}}
~
\subfloat[The ground truth \cite{girvan2002community}.]{\label{fig:afgt} 
\includegraphics[width=0.3\textwidth]{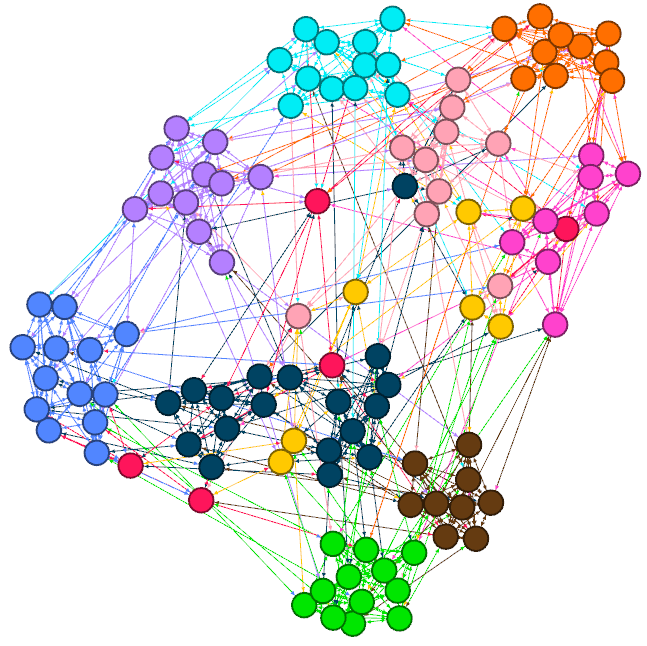}}

\caption{\label{fig:football_compares}
Clusterings of the American College football network.}
\end{figure}

\begin{figure}
\centering
\includegraphics[scale=0.45]{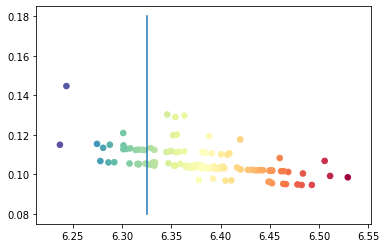}
\includegraphics[scale=0.45]{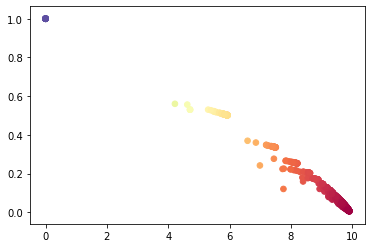}
\caption{\label{fig:scatter-threegraphs}
	{Scatter diagram with the entropic centrality of nodes on the $x$-axis, and on the $y$-axis, the maximum absorption probability at any node for a random walker starting at that node: the American college football network (left), and the EU mail network (right)}.}
\end{figure}

{Figure \ref{fig:footballentropygradient} shows the network and the corresponding relative (that is, normalized by the maximum entropic centrality value) Markov entropic centralities: darker the color, higher the relative entropic centrality score. We also see in the distribution that a majority of nodes have normalized entropic centrality between 0.2 to 0.4. This helps us to identify our clustering parameter $S_{HE}$ to identify the set of high entropy nodes deemed as center/border of a cluster. Accordingly, we chose $S_{HE}$ to comprise the top 50/60/70/80\% entropic centrality nodes and results obtained had F-scores against ground truth as 0.273, 0.406, 0.409, and 0.517 respectively. 
	The scatter diagram showing the (absolute) entropic centrality and the maximum absorption probability at any node for a random walker starting at corresponding nodes is shown in Figure \ref{fig:scatter-threegraphs} (left). The vertical line in the diagram shows the demarcation for $S_{HE}$ for 80\%. Unlike for the dolphin network, there is barely any node with distinctively high probability. The threshold of 80\% separates a few nodes with both slightly highest entropic centrality and highest probability.
	The result with F-score 0.517 is shown in Figure \ref{fig:afour}. We stop at the first iteration since our clustering technique is a bottom-up approach, which means that second iteration will produce fewer number of clusters. Based on the ground truth (Fig.~\ref{fig:afgt}), we notice for our algorithm a similar behavior as was observed with the other algorithms for the dolphin network, namely: the algorithm coalesced several of
	the ground truth communities to create larger communities. By extracting the largest three subgraphs of these larger communities and re-applying the algorithm on each of these subgraphs (again with parameter $S_{HE}$ consisting of the top 80\% entropic centrality nodes), we obtained a new group of clusters, shown on Fig.~\ref{fig:afour3}, with a significantly improved F-score of 0.811. The overall computation time was a total of 0.221 seconds. We compare this with results obtained with InfoMap \cite{RB}   (F-score of 0.904 and total time of 0.013 seconds), Louvain \cite{blondel2008fast} (F-score of 0.823 and total time of 0.002 seconds), and label propagation \cite{andersen2006local} (F-score of 0.796 and total time of 0.001 seconds) as shown in Figure \ref{fig:football_compares} along with ground truth. In the ground truth \cite{girvan2002community}, a cluster with yellow color and another cluster with crimson color spread their members out over the network. Considering this anomalous `ground truth', ours as well as other clustering techniques produce very good results. InfoMap has the highest F-score, Louvain has similar F-score to our clustering technique, and label propagation has the lowest F-score. 
}

Finally, we considered the European email network \cite{euemail} representing email communication in a large European research institution, among members that belong to 42 departments (thus 42 clusters). The corresponding scatter diagram is shown in on the right of Figure \ref{fig:scatter-threegraphs}. Looking at the scatter diagram from left to right, we observe (actually) a few nodes with entropy 0, these are isolated nodes with no edge. Then there is a small group with entropies varying between 4 and 6, and finally on the right the bulk of nodes have both entropies more than 6 and highest probability less than 0.3. This suggests that the proposed algorithm will have difficulties in identifying clusters, since choosing $S_{HE}$ to include the large right group gives too many clusters, but either iterating or taking smaller $S_{HE}$ leads to too few clusters. This demonstrates how the scatter diagram informs whether and when our approach is suitable for clustering a given graph instance.

\section{Concluding Remarks}
In this paper, we investigated the entropic centrality of a graph, using the spread/uncertainty of a random walker's eventual destination as a measure, that is applicable for all families of graphs: un/weighted, un/directed. Studying the probability distribution of a random walker to be absorbed at any given node when initiated at a node of a given entropic centrality, we established principled insights on how to choose query nodes for random walkers, and how to exploit said probability distribution to identify local community structures. We utilized these mechanisms to realize heuristic bottom-up clustering algorithms, relying on the centrality informed choice of query nodes, which inherit the universality of the entropic centrality model, and hence are also applicable across families of graphs. We benchmarked the proposed clustering mechanism using a variety of data sets and by comparing it with other popular algorithms. Given the principles that guided the design of our heuristics, we also explored how the underlying analysis could be leveraged to reason about when and whether our algorithm is suitable to cluster a given graph.

Given the bottom-up and localized nature of the most relevant information that are used in the decision making process, in the future we also want to explore the trade-offs in the quality of results obtained against computational scalability and possible distribution/parallelization, if partial information is used to compute approximate centrality scores and random walker distributions.

Our model naturally fits applications such as the flow of money and its confined circulation among subsets of users, where the volume or frequency of interactions can be mapped to edge weights, and the direction of the flow is vital. Money laundering and cryptocurrency forensics are thus application areas of interest which we want to explore with the designed tools in the immediate future.

\section*{Funding} This work was not supported by any funding.


\newpage
\ifx\undefined\BySame
\newcommand{\BySame}{\leavevmode\rule[.5ex]{3em}{.5pt}\ }
\fi
\ifx\undefined\textsc
\newcommand{\textsc}[1]{{\sc #1}}
\newcommand{\emph}[1]{{\em #1\/}}
\let\tmpsmall\small
\renewcommand{\small}{\tmpsmall\sc}
\fi

\end{document}